\documentclass[twoside,11pt]{article}
\def\forjournal{1}

\usepackage{amsmath}
\usepackage{nccmath}
\usepackage{hyperref}
\usepackage{amsthm}

\usepackage{cleveref}

\usepackage[abbrvbib,nohyperref]{jmlr2e}
\newtheorem{assumption}{Assumption} 

\newcommand{\rev}[1]{#1}

\usepackage{graphicx}

\usepackage[utf8]{inputenc}

\usepackage{shortcuts}
\usepackage{outlines} 
\usepackage{tabulary}
\usepackage{booktabs}
\usepackage{float}
\usepackage{nicefrac}
\usepackage{algpseudocode}
\usepackage{algorithm}
\newcommand\dif{\mathop{}\!\mathrm{d}}
\usepackage{wrapfig}
\newcommand{\EE}[2]{\mathbb{E}_{#1}\left[#2\right]}

\usepackage{subcaption}
\usepackage{cancel}

\newcommand{\yw}[1]{}
\usepackage{enumitem} 
\newcommand\norm[1]{\lVert#1\rVert}
\usepackage{mathtools}
\newcommand{\flag}[1]{}
\crefname{assumption}{assumption}{assumptions}
\usepackage{xcolor}
\newcommand{\new}[1]{#1}

\newcommand{\az}[1]{}

\newcommand{\hsmoothdelta}{\lambda}

\newcommand{\tepspyax}[1]{\tilde p_\eps(Y, A=1)}
\newcommand{\tepspnloweryax}[1]{\tilde p_\eps(y, A=1,x)}

\newcommand{\tepspax}[1]{\tilde p_\eps(A=1,{#1})}
\newcommand{\tpnloweryax}[1]{\tilde p(y, A=1, x)}

\newcommand{\tpnlowerx}[1]{\td{p}(x)}

\newcommand{\tpnalowerx}[1]{\tilde p(A=1, x)}

\newcommand{\tepspnax}[1]{\tilde p_\eps(A=1, X_{#1})}
\newcommand{\tepspnalowerx}[1]{\tilde p_\eps(A=1, x)}

\newcommand{\tepspnamidx}[1]{\tilde p_\eps(A=1\mid X_{#1})}
\newcommand{\tpnx}[1]{\tilde p(X_{#1})}

\newcommand{\sumn}[1]{\sum_{{#1}=1}^n}
\newcommand{\tp}{\tilde{p}}
\newcommand{\htdel}{{{\tilde{\phi} }}}

\newcommand{\eps}{\epsilon}
\newcommand{\est}{\Psi} % estimand functional
\newcommand{\tP}{\tilde{P}}

\newcommand{\td}[1]{\tilde{#1}}
\newcommand{\paren}[1]{\left(#1\right)}
\newcommand{\bces}[1]{\left\{#1\right\}}
\newcommand{\bkts}[1]{\left[#1\right]}

\newcommand{\fnl}{\Psi}
\newcommand{\ptbdir}{{i}}

\newcommand{\smoothdel}[1]{\tilde{\delta}_{#1}^\lambda}
\makeatletter
\newcommand*{\defeq}{\mathrel{\rlap{%
                     \raisebox{0.3ex}{$\m@th\cdot$}}%
                     \raisebox{-0.3ex}{$\m@th\cdot$}}%
                     =}
\makeatother

\newcommand{\intd}{\mathrm{d}}

\usepackage{lastpage}
\jmlrheading{}{}{1-\pageref{LastPage}}{Sep. 24}{}{21-0000}{}

\ShortHeadings{Data-Driven Influence Functions}{Jordan, Wang, Zhou}
\firstpageno{1}

\begin{document}

% This file includes all changes from the resubmitted manuscript to the initial conference version. Changes from the conference version are in purple; furthest changes in the newest resubmitted version are in blue. 
% \clearpage 

\title{Data-Driven Influence Functions for Optimization-Based Causal Inference}

\author{\textbf{Michael Jordan} \\
       \texttt{jordan@cs.berkeley.edu} \\
       \addr Department of EECS and Statistics\\
       \addr University of California, Berkeley
       \AND
       \textbf{Yixin Wang} \\
       \texttt{yixinw@umich.edu} \\
       \addr Department of Statistics\\
       \addr University of Michigan
       \AND
       \textbf{Angela Zhou\footnotemark[1]} \\
       \texttt{zhoua@usc.edu} \\
       \addr Department of Data Sciences and Operations\\
       \addr University of Southern California, Marshall School of Business}
\renewcommand{\thefootnote}{\fnsymbol{footnote}}
\footnotetext[1]{Authors listed in alphabetical order.}
\renewcommand{\thefootnote}{\arabic{footnote}}
% \author{\name Author One \email %one@stat.washington.edu 
% \\
%        \addr Department of Statistics\\
%        University of Washington\\
%        Seattle, WA 98195-4322, USA
%        % \AND
%        % \name Author Two \email two@cs.berkeley.edu \\
%        % \addr Division of Computer Science\\
%        % University of California\\
%        % Berkeley, CA 94720-1776, USA
%        }

\editor{.}

\maketitle

 % \tableofcontents

\begin{abstract}%   <- trailing '%' for backward compatibility of .sty file
We study a constructive algorithm
    that approximates Gateaux derivatives for statistical functionals by finite differencing, with a focus on functionals that arise in causal inference. We study the case where probability distributions are not known \textit{a priori} but need to be estimated from data. These estimated distributions lead to empirical Gateaux derivatives, and we develop a new analytical framework to study precisely how finite-differences approximates known estimators, starting with a case study of the interventional mean (average potential outcome), which gives analytical tools to analyze more complicated functionals. We then derive requirements on the rates of numerical approximation in perturbation and smoothing that preserve statistical benefits  such as rate double robustness. We then study more complicated functionals such as dynamic treatment regimes and the linear-programming formulation for policy optimization in infinite-horizon Markov decision processes. We give arguments that can generalize to the case of arbitrary constraints,  illustrating the expressivity of constructive approaches for Gateaux derivatives. %\new{For causal inference estimators that can be expressed as optimization programs, we find that recovering orthogonality adjustment requires perturbing the optimal solution, in contrast to typical envelope theorem arguments in deterministic optimization.
    % }
\end{abstract}

\begin{keywords}
causal inference, 
influence functions, 
stochastic optimization, 
algorithmic statistics,
semiparametric statistics \end{keywords}

\section{Introduction} 

\looseness=-1
Inferential targets in causal machine learning often take the form of statistical functionals of the data distribution. Examples include average treatment effects or average policy values for infinite-horizon offline reinforcement learning. Estimation of these statistical functionals can be vulnerable to the first-stage bias introduced by the estimation of nuisance functions such as outcome regressions or transition probabilities. By leveraging the causal structure, however, it is possible to derive bias-adjusted estimators that weaken the need for accurate estimation of these nuisance functions.
% because of the causal structure, estimation of the functional can be improved by deriving bias adjustments. 
% Indeed, 
The celebrated doubly robust estimator is one such bias adjustment.  there are general frameworks for deriving such estimators, including semiparametric efficiency, double robustness, Neyman orthogonality, and ``debiased/double" machine learning~\citep{bickel1993efficient,robins1994estimation,newey1994asymptotic,chernozhukov2016locally,chernozhukov2018double,hines2022demystifying,kennedy2016semiparametric,kennedy2022semiparametric}. 
% The celebrated doubly robust estimator is one such bias adjustment \citep{robins1994estimation}. 
A common framework is to derive an influence-function–based correction \citep{newey1994asymptotic,hines2022demystifying,kennedy2016semiparametric,kennedy2022semiparametric}. Debiased/double machine learning provides a practical recipe for implementing this construction with flexible nuisance estimates \citep{chernozhukov2018double}. Such influence-function adjusted estimators achieve desirable properties such as semiparametric efficiency, double robustness, and/or Neyman orthogonality ~\citep{bickel1993efficient,chernozhukov2016locally}. 
These general frameworks can also be applied to more complicated functionals in longitudinal causal inference and offline reinforcement learning (also known as offline policy evaluation/learning) \citep{zhao2015new,chakraborty2013statistical,van2003unified,jl16,pb2016}.

A drawback of these general frameworks is that significant analytic effort is often necessary to obtain concrete estimators that are appropriate for particular situations.  Indeed, a practitioner may be interested in novel variants of an estimand, or be working within a constrained class of functionals that are appropriate for a particular class of problems. For example, in \textit{constrained} Markov Decision Processes (MDPs) \citep{altman1999constrained}, or optimization-based estimators more broadly, a practitioner can easily avail themselves of custom convex-optimization solvers that are computationally efficient and make deploying additional constraints straightforward.  Unfortunately, such choices may require case-specific re-analysis to establish the desired statistical properties. It is therefore important to develop a suite of constructive, numerical or algorithmic methods that yield desired forms of bias adjustment.  Such a suite would be complementary to analytic derivations.

% In this work, we develop an algorithmic approach to bias adjustments for statistical functionals.
To this end, we study procedures for estimating the statistical functionals used for bias adjustment in causal inference via numerical finite-difference approximation. %develop off-the-shelf procedures for estimating the statistical functionals used for bias adjustment in causal inference.  We do so via numerical approximation of the Gateaux derivatives that underlie the general analytic frameworks, building on prior work in this area. 
% We focus on interventional effects.
% \cite{newey1994kernel} suggested numerical differentiation to estimate Gateaux derivatives of functionals unavailable in closed form but approximated by computational procedures, such as the solution to a system of differential equations; this approach is later used in \cite{hausman1995nonparametric}. 
% Recent pedagogical surveys of \cite{hines2022demystifying,kennedy2022semiparametric} also emphasize the usefulness of Gateaux derivatives for deriving influence functions, an approach which is directly applicable for discrete data and generalizes to continuous distributions via a smoothing argument as discussed in \cite{ichimura2015influence,carone2018toward}.
We build specifically on a line of work studying finite differences for approximating influence functions \citep{newey1994kernel,frangakis2015deductive,carone2018toward,ichimura2015influence}. To the best of our knowledge, the first mention appears in \citet{newey1994kernel,hausman1995nonparametric} for computing influence functions of differential equation systems --- where re-using the outputs of black-box solvers is favorable. 
\rev{\citet{frangakis2015deductive} propose a constructive procedure that approximates an influence function (in the sense of \citet{hampel1974influence,huber2004robust}) by finite differences. \citet{luedtke2015discussion,ichimura2015influence} propose an additional smoothing of the perturbation to preserve validity under semiparametric models. \citet{carone2018toward} also studies the smoothed finite difference} and generic smooth statistical functionals of the observational distribution, including causal plug-in estimands (e.g., regression adjustment and g-formula functionals for the ATE and longitudinal effects) and non-causal examples as well. 
%The point of departure of \citet{carone2018toward} is a causally-identified plug-in functional of the observational joint distribution, such as regression adjustment or the G-formula for the average treatment effect or longitudinal treatment effects. 

Numerical differentiation can be computationally favorable. Given one causal identification result, \rev{evaluating the one-step estimator requires a weaker computational oracle of black-box functional evaluations, rather than a stronger first-order computational oracle of gradient information of the functional}. In other areas such as optimization, numerical differentiation enables analysts to optimize, given only function evaluation access to the objective, without specialized training in a particular modeling language or paradigm. {For example, many data scientists use optimization through the {\tt scipy.optimize} or {\tt r.optim} packages: if they do not provide a gradient function callback, in some configurations, it is by default approximated by numerical differentiation. } %Although more advanced paradigms can compute exact gradients, they can require specialized training to use or implement. 

In this paper, we develop new analytical frameworks that advance theoretical studies of methods that use
numerical derivatives to approximate influence functions, focusing on the implications for \textit{statistical} estimation rather than computational approximation alone.
In more detail, we provide exact characterizations of data-driven influence functions, in contrast to numerical derivatives based on known probability distributions. 
% Although we expect that some of our exact characterizations may be apparent to experts, we also hope that this level of concreteness can be helpful to non-experts. 
\new{In the nascent and rapidly growing area of automatic causal inference, our exact characterizations illustrate ``analytical statistical unit tests". 
%Existing numerical studies indeed verify approximation computationally, but we show how 
\textit{Analytical equivalences and approximations} can refine our understanding of estimation properties and help practitioners more concretely understand these approaches and their tradeoffs. We start with simple examples, illustrate how our analytical framework extends to more complicated examples, and give arguments that would generalize our stable approximation certificates to variants of known estimators without known EIFs (so long as these variants satisfy regularity conditions). %Our goals are to more deeply characterize finite-difference approximation in causal inference, rather than to compete with other methodological proposals. 
Further empirical studies and improvements are important but beyond the scope of this work.}

\rev{Our paper is organized as follows}.\footnote{An earlier version of this manuscript, with a subset of results, appeared in the proceedings of Neurips 2022 with the title ``Empirical Gateaux Derivatives for Causal Inference". This manuscript contains expanded results: Section 5 and Section 6 regarding stochastic optimization functionals including sensitivity analysis.} \Cref{sec-problemsetup} introduces problem setup and background, including our simple running example of the average potential outcome, while \Cref{sec-relatedwork} introduces related work. \rev{Our key novel contributions are in \Cref{sec-aipw,sec-stochopt,sec-application}. In \Cref{sec-aipw} we introduce our analytical framework for studying finite-difference approximation stability based on simple product and quotient rules. We illustrate these rules with the canonical example of AIPW, where by specializing the analysis to a causal functional, we can obtain finer analysis of the rates of \textit{numerical} approximation that can preserve \textit{statistical} performance. We also illustrate how these rules can be used to study other functionals, like a dynamic treatment regime, obtained via compositions of products, quotients, compositions of probability densities. In \Cref{sec-stochopt}, we apply our new analytical framework to study more complex functionals, marginalizations of conditional linear optimization problems, including infinite-horizon off-policy evaluation and sensitivity analysis in causal inference, and establish finite-difference approximation thereof. %Our arguments require taking total derivatives of optimization problems, in contrast to non-causal envelope theorems which drop zero-mean terms. 
In \Cref{sec-application} we conduct a brief proof-of concept empirical study, including finite-differences for stochastic optimization for the sensitivity analysis functional.
} 
% Throughout we emphasize concretizations and fine-grained comparisons to compare sources of estimation and numerical error to highlight tradeoffs. 
% \new{Further, along the way, we derive product and quotient rules in a ``finite-difference calculus.", which simplifies verification of favorable finite-difference approximation properties for more complicated functionals. We summarize this new analytical framework in \Cref{sec-fdcalc} and apply it to a more complex functional, policy evaluation in a dynamic treatment regime. %Our exact approximations via product/quotient/matrix inversion rules can certify stable approximation for compositions thereof. 
%\new{()}

% \vspace{-5pt}

\section{Problem Setup: From Numerical to Empirical Gateaux Derivatives }\label{sec-problemsetup}
% We use the treated mean as a running example to fix notation. The average treatment effect follows by differencing treated and control means, and the broader framework extends to more general identified functionals.
We begin by introducing our first example, the mean potential outcome, and the canonical doubly robust estimator. We then introduce the key objects in the more general framework of orthogonalizing or debiasing statistical functionals by Gateaux derivative adjustments via the one-step estimator. We also describe the numerical approximation of these adjustments via perturbed black-box plug-in evaluations of the statistical functional. After, we clarify what we mean by data-driven influence functions, and \textit{empirical} rather than \textit{numerical} Gateaux derivatives, and delineate our specific research questions.

% We let $O\sim P$ denote a draw of observed data following the distribution $P$ belonging to a statistical model $\mathcal{M}$. We will focus on the estimation of statistical functionals $\fnl(P)$, where $\Psi:\mathcal{M}\to \mathbb{R}^q$ is pathwise differentiable. 
\rev{Let $\mathcal M_{\mathrm{NP}}$ denote the collection of all distributions on $\mathcal O$.
We consider statistical functionals $\Psi:\mathcal M_{\mathrm{NP}}\to\mathbb R^q$ that are pathwise differentiable at $P$.}
%Throughout, we take as given the assumptions that grant causal identifiability (e.g., overlap/positivity, unconfoundedness/ignorability). 
\rev{In this paper, we focus on causal applications, although the method is more broadly applicable. 
}
% Our starting point for estimation is plug-in estimation of a statistical functional, which presumes causal identification of the statistical functional. This in itself requires some background knowledge in general. 

% \begin{assumption}[Causal identification of plug-in functional]\label{asn-identification}
%     The causal estimand is identified in $\Psi(P)$. 
% \end{assumption}
Our first example is the mean potential outcome: we start from regression adjustment identification. There are other more complex estimands, such as longitudinal estimands, which readily admit causal identification without expert knowledge (such as G-computation for longitudinal causal inference), but which may admit many variations whose nonparametric influence functions may require expert knowledge analysis. In these settings, with general identification approaches but more case-by-case estimation bias adjustments, a starting identification is readily available and computerization can still be useful. 
We very briefly overview the celebrated doubly robust estimate of the mean potential outcome. The average treatment effect is the difference of these doubly robust estimators for the treated and control means; for simplicity, we discuss the treated mean only. \rev{The observation $O=(X,A,Y)$ includes covariate $X$, treatment $A\in\{0,1\}$, and outcome $Y\in \mathbb{R}.$ Under the Neyman-Rubin potential outcomes framework \citep{rubin2005causal}, an individual is endowed with a vector of potential outcomes $Y(a)$, one for each level of (binary) treatment $a \in \{0,1\}$, where $Y(a)$ denotes the outcome that would obtain under intervention $A=a$. For causal estimands, we apply conventional identifiability assumptions, e.g. consistency ($Y(A)=Y$), ignorability ($Y(0), Y(1) \perp A \mid X$) and positivity ($P(A=1 \mid X) \in(0,1)$).}
\begin{example}[Mean potential outcome]\label{ex-cfactualmean} The statistical functional corresponding to the mean potential outcome is: 
$$\textstyle \fnl(P) =\E[Y(1)]= \E[\E[Y\mid A=1,X]].$$
\end{example}

\paragraph{Influence function and Gateaux derivative.}

A functional $\fnl$ is \emph{Gateaux differentiable at $P$ relative to $\mathcal{M}_{\mathrm{NP}}$} if $\fnl'(H;P) \coloneqq \frac{d \fnl (P+\eps H)}{d\eps}\vert_{\eps=0}$ exists and is linear and continuous in $H$, \rev{where $H$ ranges over the tangent space $T_P(\mathcal{M}_{\mathrm{NP}})$, i.e., the closure in
$L^2_0(P)$ of score functions of regular parametric submodels
$\{P_\epsilon:\epsilon\in(-\delta,\delta)\}\subset\mathcal{M}_{\mathrm{NP}}$ with $P_0=P$.
Thus perturbations are taken only along such submodels, ensuring that
$P_\epsilon \in \mathcal{M}_{\mathrm{NP}}$ for sufficiently small $\epsilon$.}
% \footnote
Another way to state the definition is that $\fnl$ is Gateaux differentiable at $P$ if there exists $\phi_P(H)$ such that $\fnl(P+\eps H) - \fnl(P)=\eps \phi_P(H)+o(\eps),$ as $\eps \to 0$ \cite[p.296]{van2000asymptotic}. This statement is equivalent when the functional maps to the real line. \citet{hampel1974influence} and \citet{huber2004robust} define the \textit{influence function} $\phi(O;P)$ as a Gateaux derivative with respect to perturbations $H=\delta_o-P$, where $\delta_o$ is a degenerate Dirac measure satisfying the identity $\int o \delta_{o'}(o) = o'$.  That is, \rev{we have \(\phi_{P}\left(\delta_{o}-P\right)=\phi(o ; P)\) with}:%\useshortskip
 $$ \phi(o;P) \coloneqq \frac{\mathrm{d} \Psi \left(P + \epsilon(\delta_{o} - P)\right)}{\mathrm{d} \epsilon}\Big\vert_{\epsilon=0}.$$ 
\rev{
Influence functions are valuable because they provide the canonical \emph{first-order linear approximation} to how a smooth functional $\Psi(P)$ changes under perturbations of $P$. This von Mises expansion underlies one-step (debiased) estimators. %It then remains to control the second-order remainder in a von Mises expansion. %Accordingly, we work with functionals that admit a von Mises expansion around $P$:
}

\rev{Throughout, let $\tilde P$ denote a generic plug-in estimate of $P$ (e.g., induced by estimated probability distributions). We assume that $\Psi:\mathcal{M}_{\mathrm{NP}}\to\mathbb{R}$ is pathwise differentiable at $P$ under the nonparametric model. That is, for every regular parametric submodel $\{P_\epsilon:\epsilon\in(-\delta,\delta)\}\subset\mathcal M_{NP}$
with $P_0=P$ and score
$
s(u)\ :=\ \left.\frac{\partial}{\partial \epsilon}\log p_\epsilon(u)\right|_{\epsilon=0}\in L^2_0(P),
$
the map $\epsilon\mapsto \Psi(P_\epsilon)$ is differentiable at $\epsilon=0$ and there exists
$\phi(\cdot;P)\in L^2_0(P)$ such that
$
\left.\frac{d}{d\epsilon}\Psi(P_\epsilon)\right|_{\epsilon=0}
\;=\;
\E_P\!\big[\phi(O;P)\,s(O)\big]
$ \citet{bickel1993efficient,pfanzagl2012contributions,newey1990semiparametric}.
Under standard regularity conditions, this implies the von Mises expansion
\[
\Psi(\tilde P)-\Psi(P)
=
\int \phi(u;P)\,d(\tilde P-P)(u)
+
R_2(\tilde P,P),
\qquad
R_2(\tilde P,P)=o(\|\tilde P-P\|).
\]
See 
\citet{kennedy2022semiparametric} for discussion of the connection to the von Mises expansion.
}
% We assume $\Psi$ is pathwise differentiable under the nonparametric model so that there exists an influence
% function $\phi(\cdot;P)\in L^2_0(P)$ satisfying
% \[
% \Psi(\tilde P)-\Psi(P)=\int \phi(u;P)\,d(\tilde P-P)(u)+R_2(\tilde P,P),
% \]
% where $R_2(\tilde P,P)$ is a second-order remainder term. See \citep{bickel1993efficient,pfanzagl1982,newey1990semiparametric} for formal definitions of pathwise differentiability, and \citep{kennedy2022semiparametric} for discussion on connections between these definitions of pathwise differentiability and the von-Mises expansion above where the influence function ``represents" the Gateaux derivative. Often pathwise differentiability is proved by showing the remainder term is second-order.
\looseness=-1
\paragraph{Numerical Gateaux derivatives.}
 \rev{
    For continuous data, the naive perturbation $(1-\varepsilon)P+\varepsilon\delta_o$ is not $P$-dominated and therefore does not correspond to a regular $P$-dominated submodel/tangent direction. Therefore, these delta-function perturbations are not alone sufficient for semiparametric inference. \citet{carone2018toward,luedtke2015discussion} and \citet{ichimura2015influence} restore absolute continuity in semiparametric models} by taking an additional limit as a \textit{smoothed perturbation} converges to the point mass. 
That is, they replace the Dirac delta measure $\delta_o$ with a smoothed distribution $\smoothdel{{o}'}(o)$, a function satisfying that it integrates to 1, $\int  \smoothdel{{o}'}(u)\mathrm{d} u= 1,$ and is dominated by $P$. A common choice for $\smoothdel{o'}$ has the form $\smoothdel{o'}(o) = K_{\hsmoothdelta}(o-o')$, where $K_\lambda$ is a bandwidth-normalized kernel function and notation is overloaded to also designate product kernels. We then define a perturbation in the direction of an observation $o_i$ as follows: \useshortskip
\if \forjournal 0
$P_{\eps,\hsmoothdelta}^{o_\ptbdir}=P_{\eps,\hsmoothdelta}^{\ptbdir}  \coloneqq (1-\eps) P + \eps \smoothdel{o_{\ptbdir}} %\label{defn-smoothing}
$.
\fi 
\if \forjournal 1
\begin{align*}%\SwapAboveDisplaySkip 
P_{\eps,\hsmoothdelta}^{o_\ptbdir}=P_{\eps,\hsmoothdelta}^{\ptbdir}  \coloneqq (1-\eps) P + \eps \smoothdel{o_{\ptbdir}}. %\label{defn-smoothing}
\end{align*}
\fi

\citet{frangakis2015deductive,carone2018toward} discuss \textit{numerical} approximation by finite differences for a fixed $\eps, \hsmoothdelta$ with $\eps \ll\hsmoothdelta$, to evaluate the numerical derivative $\eps^{-1}( {\est(P_{\eps,\hsmoothdelta}) - \est( P )) }{}.$
%deriving implications on approximation error in $\eps,\lambda$ from generic finite-difference results. 
Further, \cite{hines2022demystifying,kennedy2022semiparametric} highlight influence-function derivations based on the analytical Gateaux derivative. \citet{carone2018toward} establish that the limit of the finite-difference quotient when $\eps$ and $\lambda$ tend to zero is precisely the Gateaux derivative, and they provided generic approximation rates in $(\epsilon,\lambda)$ for a multi-point finite-difference scheme. 
% , where $\Psi(P_{\epsilon, \hsmoothdelta}^{*})$ is the projection of $\Psi(P_{\epsilon, \hsmoothdelta})$ onto the tangent space of the semiparametric model:  \begin{equation*} \textstyle 
% \phi_{P}(o;P) =\underset{\lambda \rightarrow 0}{\lim}\left[\frac{d}{d \epsilon} \Psi\left(P_{\epsilon,\lambda}\right) \right] =\underset{{\hsmoothdelta \rightarrow 0}}{\lim } \;\underset{\epsilon \rightarrow 0}{\lim} \;\; {\epsilon}^{-1} \left( {\Psi(P_{\epsilon, \hsmoothdelta}^{*})-\Psi(P)} \right).
% \end{equation*}
These rates establish sufficient conditions for approximation error that ensure validity of the numerical approximation, but can be conservative. For a two-point approximation with a uniform kernel, the rate from \citet[Thm. 5]{carone2018toward} is $\left(\epsilon \lambda^{-d}\right)+\lambda^{2}$. We will strengthen this result later in the paper.
% , which can be useful for expanding the regime of numerical approximation without running into numerical precision issues.

\paragraph{Empirical Gateaux derivatives (data-driven influence functions).} 

In the context of statistical estimation, the underlying probability distributions also need to be estimated. We consider \textit{plug-in estimation} of the statistical functional via \rev{density estimation of the joint probability estimates $\tilde{P}(O)$}. We assume there already exists a causal identification argument such that the estimand can be written as a functional of the observational joint distribution. This can be the case when considering variations of a class of identified functionals, such as the G-formula. For example, a plug-in \rev{estimator} of the mean potential outcome is as follows:
% \if\forjournal 1
$$ \fnl(\tP)= \int \int y \frac{\td{p}(y, A=1,x)}{\td{p}(A=1,x)} \tp(x) \mathrm{d}y\mathrm{d}x.$$
% \fi
We study the \textit{empirical Gateaux derivative} obtained by numerical approximation with estimated probability densities (e.g., $\tilde P(Y,A=1,X)$),% the \textit{numerical Gateaux derivative} obtained by finite differencing with the true probabilities, 
and how it approximates the analytical Gateaux derivative:
\begin{align*}
    \htdel(O_i) &
    \if\forjournal 0
\textstyle 
\fi
    % \textstyle
    = \eps^{-1}\left( {\est(\tP_{\eps}^i) - \est( \tP ) }{} \right) && \text{ empirical derivative at smoothed and estimated distributions,}\\
%         \hdel(O_i) &
%         \if\forjournal 0
% \textstyle 
% \fi
%         % \textstyle
%         = \eps^{-1}\left( {\est(P_{\eps}^i) - \est( P ) }{}\right)  && \text{ numerical derivative at smoothed and true distributions,}\\
        \phi(O_i) &
        \if\forjournal 0
\textstyle 
\fi
        % \textstyle
        = \frac{\mathrm{d}}{\mathrm{d}\epsilon} \est(P_{\eps}^i) \Big\vert_{\eps=0} && \text{ analytical Gateaux derivative.} 
\end{align*}

% It is the first of these representations that is appropriate for serving as a statistical estimator of influence functions. 
\rev{Obtaining influence functions computationally or algorithmically, rather than analytically, in order to improve statistical estimation will necessitate starting from \textit{probability estimates} in practice.}
In particular, the \textit{one-step estimator} \rev{has the following} asymptotically linear representation~\citep{pfanzagl1990estimation}:
\begin{equation}
 \if\forjournal 0
\textstyle 
\fi
 \fnl_n = \fnl(\tilde{P}) + \frac 1n \sumn{i} \htdel(O_i), \qquad \text{ where } \htdel_i(O_i) = \frac{1}{\epsilon}(\fnl(\tP^{i}_\epsilon) - \fnl(\tP)). \label{eqn-onestepestimator}
\end{equation}
See \Cref{apx-if} for further discussion of this expansion. % Therefore, the empirical procedure resembles estimation of a functional based on plug-in estimation of probability densities. If we use complex machine learning estimators for $\tilde P$ then we may incur additional estimation bias from evaluating estimated probability densities upon the same data used to obtain these estimates. We may either assume Donsker conditions (classically typical in the semiparametric literature); or consider data-splitting based estimators, where we separately learn estimated densities from a subset of the data $\tP^{n_1}$ and evaluate on another split of the data to obtain conditional unbiasedness (conditional on the other fold). (Both approaches are standard in the literature). Since computing the empirical Gateaux derivative resembles estimating functionals of probability densities, advances in leave-one-out estimation that preclude the finite-sample data loss inherent in sample-splitting may also be used in this setting \citep{kandasamy2015nonparametric}. 
Given these definitions, we now turn to the major mathematical and statistical questions that we address in this paper:
\begin{description}
\item \textbf{Q1}: How exactly does the empirical Gateaux derivative approximate the analytical derivative? 
\item \textbf{Q2}: What are the required rates of \textit{numerical approximation} in perturbation and smoothing based on $(\eps, \lambda)$ that preserve the beneficial \textit{statistical} properties of the constructed estimator?
\item \textbf{Q3}: How does the finite-difference representation approximate the influence functions of optimization-based estimators? 
\end{description} 
We begin in \Cref{sec-aipw} by studying these questions in the example of the mean potential outcome. 
% \begin{example}[\Cref{ex-cfactualmean}, continued]\label{eqn-aipw-if}
The canonical doubly robust estimator (augmented inverse propensity weighting, or AIPW) is a one-step adjustment.
The influence function \citep{robins1994estimation,hahn1998role} for a mean potential outcome (\Cref{ex-cfactualmean}) is: 
\begin{equation}
\if\forjournal 0
\textstyle 
\fi
{\phi(O) = \frac{\indic{A=1}}{p(A=1\mid X)} (Y - \E[Y\mid A=1,X])+\E[Y\mid A=1,X]-\fnl(P)}.\label{eqn-AIPW-IF}
\end{equation}

We provide a term-by-term exact characterization of how finite-differences approximates the simplest and most canonical IF in causal inference. \new{In \Cref{sec-stochopt}, we build on this initial exact characterization and study more complicated functionals to highlight more general applicability.}

% \end{example}
%Recent pedagogical reviews  \cite{fisher2021visually,hines2022demystifying} emphasize the Gateaux derivative definition of the influence function; this intuition is justified by the smoothing device developed in these works. 
We conclude our introductory remarks by noting limitations of our work. 
  We omit discussion of semiparametric efficiency since we focus on computing influence functions without the consideration of model restrictions and tangent spaces.
That is, we focus on influence functions under a nonparametric model. %Note that \citet{carone2018toward} suggest that one can handle the semiparametric case by evaluating a Gateaux derivative and projecting onto the model space directly; such an investigation is out of scope of the current paper.
 Throughout, we assume the functional is pathwise differentiable. This assumption does warrant caution, because this condition can pose a fundamental barrier to efforts for ``automation"; hence
%  for a completely general approach, since it may be difficult to verify pathwise differentiability generically without analytically analyzing/deriving the influence function,, and establishing that the remainder term is second-order is typically a problem-specific exercise. 
 empirical Gateaux derivatives are intended to augment but not replace analytical expertise. See \Cref{apx-discussion-pathwisediff} for discussion on generic conditions.
 \if\forjournal 0 
\footnote{However, under the (stronger) assumption of Fr\'echet rather than Gateaux differentiability, previous work gives sufficient conditions that imply the sufficiency of our assumption. \citet[sec. 4]{ichimura2015influence} provide a generic sufficient condition for asymptotic linearity assuming that the Fr\'echet derivative is Lipschitz, while \cite{kandasamy2015nonparametric} show the remainder term is second order; both results are obtained under convergence rate conditions on the input densities. %See \Cref{apx-discussion-pathwisediff} for more discussion, including a class of problems where general guarantees are possible, i.e., strongly convex stochastic programs.
}
\fi

\section{Related Work}\label{sec-relatedwork}
% \vspace{-5pt}

We discuss the most closely related directions but do not provide a survey or overview of influence functions or semiparametric statistics. See \citet{bickel1993efficient,tsiatis2006semiparametric,fisher2021visually,kennedy2022semiparametric} for a broader overview and further details. See \Cref{apx-relwork-numdiff} for additional discussion on numerical derivatives in optimization and machine learning. 

\paragraph{Numerical approximation of Gateaux derivatives.}
\cite{carone2018toward} derive high-level sufficient conditions for approximation error via general results for finite-difference approximations. Relative to those general sufficient conditions, we focus on specific improvements in approximation error for various functionals. 
For the specific case of the mean potential outcome/average treatment effect, \citet{frangakis2015deductive} exploit additional structural knowledge. In particular, they interchange integration and differentiation as a result of an analytical insight and obtain functional evaluations in terms of the \textit{conditional expectation}. For more complex functionals, making these determinations is similar to analytically deriving the influence function; i.e., it is less ``automatic." In general, it may be difficult to deduce the form of the nuisance functions without using knowledge of the estimand or working with specific graphical or algorithmic structure. The approach in Appendix B of \cite{bravo2020two} considers the specific case of debiasing moment conditions with respect to first stages but also implicitly leverages this change in order of integration. 
Recent pedagogical reviews  \citep{fisher2021visually,hines2022demystifying,kennedy2022semiparametric} emphasize the Gateaux derivative and its use  for deriving influence functions. While our point of departure is a finite-difference approach to the former, we also consider more complicated optimization functionals. 
% Appendix B of \cite{bravo2020two}, which had studied the debiasing moment correction of \cite{ichimura2015influence} for the case of empirical likelihood and preserving asymptotically pivotal properties (Wilk's phenomenon), studies a numerical derivative estimation approach and provides conditions for a valid asymptotically linear representation. 
% They focus on debiasing moment conditions with respect to first stages. %For example, in their framework with $(\theta, \eta)$ identified parameter and nuisance first stage, $\theta$ is the ATE and $\eta$ is the propensity score in an IPW estimator. 

\rev{One limitation of our work is that we focus on the non-parametric influence function, rather than the semi-parametric setting. \citet{carone2018toward} discusses model projection to obtain efficient influence functions, and we expect similar approaches to work here. For the optimization-type functionals we study, we conjecture that tools from Riemannian manifold optimization could be helpful in projection, but such tools are theoretically and practically complicated and outside of the scope of the current study.}

\paragraph{Other work on automating causal inference.} Recent work in causal inference develops more algorithmic or computerized semiparametric inference via a variety of methods~\citep{kandasamy2015nonparametric,bhattacharya2020semiparametric,chernozhukov2021automatic,chernozhukov2020adversarial,chernozhukov2022riesznet,farrell2021deep,jung2021estimating,alaa2019validating}. \rev{To summarize differences briefly, we focus on numerical derivatives that only require black-box access to plug-in evaluation of the functional. In the language of optimization, this is a weaker \textit{zeroth-order computation oracle}. %This approach can be particularly helpful in cases like optimization or simulation where black-box evaluations of the functional are available, but obtaining analytical derivatives can be difficult. On the other hand, other approaches may achieve better performance when such restrictions are not binding.
} 

\rev{Subsequent to the initial conference version of this paper, \citet{luedtke2026simplifying} develops the theory for an auto-differentiation framework to obtain influence functions. When auto-differentiation is available for a particular functional, it could lead to overall more numerically stable performance that is not subject to finite-difference approximation error. We expect that as more research continues in this area, other approaches could further become available for practitioners to use, and we urge practitioners to compare to the latest improvements. }

\rev{Finite-difference approximations to influence functions for statistical estimation could play an analogous role as they do in optimization. Optimization distinguishes different classes of algorithms based on the information needed about the function: first-order (gradient-based) vs. zero-th order (function-evaluation based, such as finite-differences). Generally methods that leverage more gradient information have stronger convergence guarantees. However sometimes such stronger gradient information (i.e. auto-differentiation) is not available, and so studying zeroth-order methods can be valuable in complementary ways\footnote{\rev{For example, for the stochastic optimization functionals we study later in the paper, auto-differentiation through an optimization problem remains an active research area and is far less mature than for other elementary functionals. It is ultimately algorithmically available (e.g. due to solving large systems of equations from KKT conditions), rather than symbolically available. There are a number of different algorithmic schemes \citep{agrawal2019differentiable,blondel2022efficient,bolte2022automatic}, whose algorithmic details would be expected to change statistical implications/properties and approximation guarantees. We leave a thorough comparison for future work.}}.
}

In more detail, \cite{kandasamy2015nonparametric} outline a general approach for Gateaux derivative estimation based on leave-one-out techniques. \cite{chernozhukov2021automatic} build on the variational characterizations of \cite{ichimura2022influence} via Riesz representation for generalized regression residuals. \cite{chernozhukov2020adversarial,chernozhukov2022riesznet} and other works further develop ``automatic'' estimation based on this approach, including for more complex functionals as well. However, deriving a Riesz characterization for our later optimization examples could be complex. We refer to \citet{ichimura2015influence} for the smoothing arguments required in a semiparametric setting. \cite{farrell2021deep} develop a framework similar to a functional coefficient model and compute a parametric IF. %develop a framework for modeling heterogeneity with deep networks, similar to a functional coefficient model that admits an influence function computable by automatic differentiation (or numerical differentiation of Hessians/Jacobians for a given parametrization). 
\cite{bhattacharya2020semiparametric} focus on graphical algorithms and characterizations of the efficient influence function for identified latent variable models, for ATE-like functionals, given a causal graph. %\cite{alaa2019validating} develop approaches for validating causal machine learning methods with influence functions.
\paragraph{Influence functions or Gateaux derivatives in machine learning.} 
More broadly in machine learning, influence functions have been used for qualitative sensitivity analysis or bias correction \citep{koh2017understanding,giordano2019higher,giordano2019swiss,wilson2020approximate}. However, the use of influence functions in the causal setting is different because one computes influence functions of the statistical functional with respect to perturbations in the \textit{distribution}, rather than perturbing the identifying moment function.
\paragraph{Influence functions for stochastic optimization.}
% To the best of our knowledge, the first use of influence functions in stochastic optimization, albeit for substantive robustness analysis, is in \citet{dupavcova1990stability}. 
% \cite{tercca2021envelope} study multi-stage linear stochastic programs without causal structure. 
% Our later arguments show that orthogonality adjustments are zero-mean terms that would be omitted from conventional envelope theorem arguments that are not tailored for statistical purposes.  
Recently, IFs have also been used for debiasing: \citet{gupta2021debiasing} debiases the in-sample bias of (unstructured) stochastic optimization via finite-differences. \citet{guo2022off} build on the finite-difference method of \citet{ito2018unbiased} for debiasing optimization with causal outcome coefficients but without characterization of the adjustment.

\section{A General Framework for Finite-Difference Approximation Stability via Calculus Rules}\label{sec-aipw}
\rev{In \Cref{sec-aipw}, we develop calculus rules for finite-difference expansions and introduce our framework. Then we illustrate its use by applying it first to AIPW, then to dynamic treatment regimes.
% In this section, first in \Cref{sec-framework-fdcalc} we introduce our general framework for verifying finite-difference approximation stability for functionals that are the compositions of simple calculus rules, such as product, quotient, chain rules, etc. We demonstrate how these rules can illustrate the mechanical outputs of the finite-difference algorithm and recover term-by-term approximations to the components of the analytical Gateaux derivative. In \Cref{sec-framework-ate-estimation} we revisit the familiar AIPW estimator with these analysis tools and study implications for statistical estimation. In \Cref{sec-framework-dtr}, we use these rules to deduce approximation stability of more complicated functionals, such as a dynamic treatment regime. 
}

% \rev{Our framework is extensible to other functionals via deriving additional finite difference calculus rules. We do so in the next section to investigate stochastic optimization functionals of research interest to causal inference. }
% \vspace{-5pt}

\subsection{Finite-difference approximation error calculus}\label{sec-framework-fdcalc}
\rev{First, we give product and quotient rules that can be used to establish finite difference approximation error for functionals that are compositions of these rules.
\rev{\cite{van1991efficiency,kennedy2022semiparametric} suggest an influence-function calculus, to derive influence functions via calculus rules like product and quotient rules}\footnote{We thank a referee for pointing to \citet{van1991efficiency}. Analogously, our proofs later on repeatedly apply these rules and repeatedly apply various simplifications to establish the finite-difference approximation error. } 
}

% \az{Stable finite-difference estimation under product, quote and chain rules }
\rev{
\begin{proposition}[Product and quotient rules.]\label{prop-prodquotchainrule}
For any $f: \mathbb{R}^{d_u} \to \mathbb R,\; g: \mathbb{R}^{d_v} \to \mathbb R, $
    \begin{align}%\SwapAboveDisplaySkip
\eps^{-1} (f_\eps(u) g_\eps(v) - f(u) g(v)) &= f (u) 
\cdot \eps^{-1}(g_\eps(v)-g(v)) + g_\eps(v) \cdot \eps^{-1}(f_\eps(u)-f(u))  && \text{ product rule} \label{eq-productrule}\\
\eps^{-1} \left( \frac{f_\eps (u)}{g_\eps(v)} - \frac{f (u)}{g(v)} \right) &= 
\frac{g(v) \cdot \eps^{-1} (f_\eps(u) -f(u)) - f (u)\cdot \eps^{-1} (g_\eps(v) -g(v)) }{g(v) g_\eps(v)}  
&& \text{ quotient rule}  \label{eq-quotient-rule}%\\
% \eps^{-1} (f(g_\eps(v)) - f(g(v))) &= \left( \frac{f(g_\epsilon(v)) - f(g(v))}{g_\epsilon(v) - g(v)} \right) \cdot \left( \epsilon^{-1} (g_\epsilon(v) - g(v)) \right) && \text{chain rule} \label{eq-chain-rule}
% check
\end{align}
\end{proposition}
}

% \az{Next talk about how this is relevant in terms of perturbed nuisances. Then specialize to probability densities? }
\new{
% Finite-difference calculus is studied more generically in numerical analysis. This suggests a general recipe for analyzing functionals that can be decomposed into such operations, provided the relevant primitive perturbation bounds and local stability conditions can be verified. 
\rev{Additional composition rules arise for more complex functionals. For example in \Cref{sec-stochopt}, we develop an analogous rule for matrix inverses that extends the analysis to optimization-based functionals.}
}

\rev{Although we state these generic rules, we will apply them to the probability densities that make up the plug-in functional, which we can then trace to the final terms that appear in the influence function and estimator.  }

\paragraph{Smoothed and perturbed nuisance functions induced by the perturbed probability estimates.}
% We will show how the evaluation of the integrals with respect to smoothed degenerate distributions leads to expressions that recover the analytical derivative with \textit{smoothed} density-induced nuisances that we define in this subsection. 
\rev{We walk through how plugging-in probability densities into the finite-differences approach approximates an AIPW estimator, albeit with nuisance functions that include perturbations.} For statistical estimation, approximating Gateaux derivatives requires plug-in estimation of the statistical functional with \textit{estimated} probabilities. To resolve Q1, we observe that the perturbed probability densities induce the functional form of nuisance functions in the analytical derivative. This differs from typical use, which describes the estimator in terms of, e.g., conditional expectations, allowing the analyst to specify the functional form of nuisance functions. 

First we describe some notational conventions we will use to describe these induced nuisances in general. We will denote the $\tP$-induced conditional expectation as%\useshortskip
\begin{equation*}{\E_{\tP}[Y\mid X=x] \coloneqq  \int y {\td{p}(y\mid x)} \mathrm{d}y}.\end{equation*} 
For example, the density-induced conditional expectation of kernel density estimates is exactly Nadaraya-Watson regression. %\footnote{For example, $$\textstyle \hat\mu^{NW}(x)=\int \frac{y \sum_{i=1}^{n} K_{h}\left(x-X_{i}\right) K_{h}\left(y-Y_{i}\right)}{\sum_{j=1}^{n} K_{h}\left(x-X_{j}\right)} \mathrm{d} y =\frac{\sum_{i=1}^{n} K_{h}\left(x-X_{i}\right) \int y K_{h}\left(y-Y_{i}\right) \mathrm{d} y}{\sum_{j=1}^{n} K_{h}\left(x-X_{j}\right)} =\frac{\sum_{i=1}^{n} K_{h}\left(x-X_{i}\right) y_{i}}{\sum_{j=1}^{n} K_{h}\left(x-\lambda}\right)}.$$}
% \else
% That is, $\textstyle \hat\mu^{NW}(x)=\int \frac{y \sum_{i=1}^{n} K_{h}\left(x-X_{i}\right) K_{h}\left(y-Y_{i}\right)}{\sum_{j=1}^{n} K_{h}\left(x-X_{j}\right)} \mathrm{d} y %{=\frac{\sum_{i=1}^{n} K_{h}\left(x-X_{i}\right) \int y K_{h}\left(y-Y_{i}\right) \mathrm{d} y}{\sum_{j=1}^{n} K_{h}\left(x-X_{j}\right)} 
% {=\frac{\sum_{i=1}^{n} K_{h}\left(x-X_{i}\right) y_{i}}{\sum_{j=1}^{n} K_{h}\left(x-X_{j}\right)}}.$
% \fi 
\rev{Similarly, for other nuisance functions, we denote those induced by $\tilde{P}$ as $\E_{\td{P}}$, etc.} Given this, we study empirical derivatives with kernel density estimation (KDE) as a classic example of density estimates that induce conditional expectation estimators. Other density estimation approaches can be used, simply substituting their rates of convergence in \Cref{lemma-rates}. We include more discussion in \Cref{apx-practicalconsiderations}.
% : {any ``modern" generative modeling approach/density estimation approach can be used for plug-in evaluation of the statistical functional.} 
Of course, nonparametric estimation approaches suffer from the curse of dimensionality in general, so additional structure is required for product-rate conditions. %\added{Our analysis is will be modular, so that it would only remain to verify convergence rates of unperturbed density estimates.}

\sloppy
% Our characterization will show exactly how plug-in evaluation with finite differences is nearly equivalent to evaluating the doubly robust estimator with a Nadaraya-Watson regressor as the conditional outcome regression for $\E[Y\mid A=1,X]$ and a kernel density estimate of the propensity score. It is equivalent up to an additive bias from the smoothed perturbation distribution%, which induces what we will term a \textit{smoothed nuisance evaluation}, that we define next.% relative to estimation error of the induced nuisance. 
\paragraph{Integrating the smoothed perturbation induces smoothed nuisance functions.}
Integrating with respect to the smoothed perturbation distribution can be interpreted as a smoothed evaluation of the resulting nuisance functions. We define the \textit{$\smoothdel{x_0}$}-smoothed conditional expectation 
%$\tilde{\mathbb{E}}[Y\mid A=1,X=x_{\ptbdir}]$ 
which smooths the evaluation point, i.e., smooths evaluation around $x_0$ rather than precisely at $x_0$. 
\begin{definition}[\textit{$\smoothdel{x_0}$}-smoothed conditional expectation]%\useshortskip
\begin{equation}
\textstyle \tilde{\mathbb{E}}_P[Y\mid X=x_{0}] \coloneqq \int  \E_P[Y\mid X=u] \;\smoothdel{x_0}(u) \;\mathrm{d}u .
\end{equation}
\end{definition}

A similar definition appears in the literature on nonregular inference or the localization of a global functional \citep{newey1994kernel}. For example, \citet{van2018cv} propose smoothing a nonregular target estimand, such as a dose-response curve or density evaluated at a point $x_0$, by conducting a smoothed evaluation locally around $x_0$. In that work, a function evaluated at $x_0,$ $g(x_0)$,
%$\fnl(P)(x)$ (such as $\fnl(P)(x) = \E[Y\mid A=1,X=x]$) 
is approximated with a kernel smooth over $x$ in a neighborhood of $x_0$ with bandwidth $h$; so 
${\fnl_\hsmoothdelta(x_0) = \int K_{\hsmoothdelta}(u-x_0) g(u) \mathrm{d}u}.$ \citet{jung2021double} use a similar smoothing/localization device but for a different problem in general. 
\rev{Combining our earlier rules with this new notation to describe simplifications of the resulting induced nuisances, we can obtain useful finite-difference identities for common operations like marginalizations and conditional expectations, stated generically.}
\rev{
\begin{lemma}[Useful finite-difference identities]\label{eq-cndlexp-fd}
    Let $p_\epsilon(o) = (1-\epsilon)p(o) + \epsilon \tilde\delta_\lambda(o)$, where $\tilde\delta_\lambda(o)$ is a product kernel centered at $o_i=(x_i,y_i)$ that integrates to one. 
(i) For any integrable function $h(X)$, 
\begin{equation*}
{\epsilon}^{-1}\big(\E_{P_\epsilon}[h(X)] - \E_P[h(X)]\big) = \int h(x)\smoothdel{(x_i)}(x)\,dx - \E_P[h(X)].
\end{equation*}
(ii) For the conditional mean, 
\begin{equation*}
{\epsilon}^{-1}
\big(\E_{P_\epsilon}[Y \mid X = x] - \E_P[Y \mid X = x]\big) = \frac{\smoothdel{(x_i)}(x)}{p_\epsilon(x)}\Big(\int y\,\smoothdel{(y_i)}(y)\,dy - \E_P[Y \mid X = x]\Big).
\end{equation*}
\end{lemma}
The first identity generates the familiar centering/regression terms. The second generates the IPW/correction terms.
}
\rev{
The identities in \Cref{prop-prodquotchainrule} are exact and can be applied recursively to propagate first-order finite-difference expansions through compositions. In particular, if the primitive components of a composite functional admit first-order finite-difference expansions, then any functional $\Psi$ built from them by sums, products, quotients, and smooth maps admits a corresponding first-order expansion as well. The resulting approximation error is obtained by combining the primitive remainder terms with higher-order cross terms generated by the composition rules.
}

\rev{
The constants controlling this approximation error depend on local stability properties of the operations involved. In our later analysis, \Cref{lemma-rates} gives norm bounds for perturbed nuisance functions, while \Cref{asn-matrixentriesarewellapproxfnls} imposes a first-order perturbation expansion for matrix-valued primitives in the optimization examples. We control these remainder terms at two levels: pointwise deterministic expansions, which isolate smoothing bias and other approximation error, and integrated or sample-average bounds used for statistical analysis, such as $L_2(P)$ control of perturbed nuisance functions.
}

\rev{The following theorem isolates the approximation error requirements on finite-differences in order to preserve the desired statistical improvements from the standard one-step estimator. It suffices to verify that the empirical Gateaux correction $\tilde{\phi}$ is $o_p(n^{-1/2})$-close, in sample average, to the corresponding plug-in influence-function term $\phi_{\tilde{\eta}}$.}
\rev{
\begin{theorem}[Sufficient condition for asymptotic linearity: empirical Gateaux vs.\ plug-in influence correction]\label{thm-one-step-sufficient-condition}
Let $\eta$ denote the (true) nuisance functions and $\tilde{\eta}$ denote the nuisance functions implied by plug-in estimates of $\tilde{P}$, and $\phi_{\tilde{\eta}}$ the non-parametric influence function evaluated at these $\tilde{\eta}$ nuisance functions. 
Suppose the traditional one-step estimator
$
\Psi_{\text {OS }}\coloneqq \Psi(\tilde{P})+\frac{1}{n} \sum_{i=1}^n \phi_{\tilde{\eta}}\left(O_i\right)
$
is asymptotically linear for $\Psi(P)$ with influence function $\phi$, i.e.
$
\Psi_{\text {OS }}-\Psi(P)=\frac{1}{n} \sum_{i=1}^n \phi\left(O_i\right)+o_p(n^{-1 / 2}) .
$
If, in addition,
\begin{equation}\textstyle 
\frac{1}{n} \sum_{i=1}^n(\tilde{\phi}\left(O_i\right)-\phi_{\tilde{\eta}}(O_i))=o_p(n^{-1 / 2}), \label{eqn-one-step-decomposition-empiricalappx-vanishes-fast}
\end{equation}
then the empirical Gateaux one-step estimator
$
\Psi_n \coloneqq \Psi(\tilde{P})+\frac{1}{n} \sum_{i=1}^n \tilde{\phi}\left(O_i\right)
$
is asymptotically linear with the same influence function $\phi$.\footnote{We thank a reviewer for this suggested meta-theorem.}
\end{theorem}
}

\rev{In summary, our framework consists of: (i) exact finite-difference calculus rules to propagate approximation error for compositions, (ii) identities that interpret the resulting perturbations as familiar nuisance-function terms, and (iii) a meta-theorem showing that asymptotic linearity follows once the empirical Gateaux correction is sufficiently close, in sample average, to the corresponding plug-in influence correction. }

\rev{Therefore, for a given functional, it remains to verify primitive perturbation bounds and local stability conditions for the functional of interest, and to verify \cref{eqn-one-step-decomposition-empiricalappx-vanishes-fast}, that the approximation error of the data-driven influence function is lower order. Such analysis is analogous to showing remainder terms are second-order in proving pathwise differentiability. 
% In the examples in this paper that follow, we do so by applying \Cref{prop-prodquotchainrule} and \Cref{eq-cndlexp-fd} to identify how the $P_\eps$-induced nuisances would induce approximation error $
% \Psi_n \coloneqq \Psi(\tilde{P})+\frac{1}{n} \sum_{i=1}^n \tilde{\phi}\left(O_i\right)
% $, and controlling an additional second-order remainder-term. 
In the next subsection, we illustrate the framework by discussing its application to the average treatment effect. }
% \az{Supposition of the theorem says, for example, that the plug-in nuisances satisfy product error rates. }
\subsection{Recap and illustration: How exactly does finite differences approximate AIPW?}\label{sec-framework-ate-estimation}

\paragraph{Setup.}
 We target the estimation of a mean potential outcome as in \Cref{ex-cfactualmean}. For the smoothed perturbation, we specialize to kernel functions $K_{\hsmoothdelta}(u)$
from kernel density estimation satisfying 
$\int_{-\infty}^\infty K(u)du =1$ (normalization to a probability density) and $K(-u)=K(u)$, for all $u$ (symmetry). %\footnote
{Examples include the Gaussian kernel with $K(u)={({2\pi})^{-\frac 12}} e^{-u^2/2},$ or uniform with $K(u) = \nicefrac{1}{2}\indic{\abs{u}\leq 1}.$ We generally consider product kernels.
%We generally consider $K_{\hsmoothdelta}(u)=\hsmoothdelta^{-d}K(u/\hsmoothdelta),  K(x)=K\left(x_{1}, \ldots, x_{d}\right)$.})

% and $\fnl(\tP),\fnl(\tP_\eps)$ by 
% plugging-in the perturbed estimates of the density functions. 
% The derivation of Nadaraya-Watson kernel regression from the integral representation with respect to kernel estimates of the probability densities is analogous to the change in order of integration that yields conditional expectation representations of the Gateaux derivative.
\paragraph{Recap: Augmented IPW estimator} \rev{To provide a self-contained discussion for non-expert readers, we overview well-known properties of the augmented IPW estimator.} \rev{Under conventional assumptions of consistency $(Y(A)=Y)$, ignorability $( Y(0), Y(1) \perp A \mid X)$ and positivity $(P(A=1 \mid X) \in(0,1))$,} the mean potential outcome is identified as:%\useshortskip
\begin{equation}\label{eqn-fnl-cfactualmean}
\textstyle \fnl(P) = \E[Y(1)]  %\E[ \E[Y(1) \mid X] ] 
\textstyle 
= \E[ \E[Y\mid A=1,X]]= \int \int y p(y \mid A=1,x) \mathrm{d}y\mathrm{d}x 
 = \int \int y \frac{p(y, A=1,x)}{p(A=1,x)} p(x) \mathrm{d}y\mathrm{d}x.
\end{equation}
We obtain $\fnl(\tP_\eps)$ by plugging in the perturbed $\td{p}_\eps$ probability density estimates. %Note this is distinct from the plug-in estimator in empirical process theory, which plugs in the empirical CDF; we plug in statistical estimates. 
See \Cref{alg-empiricalif} for a summary of the procedure. Such plug-in estimation of the conditional expectation is termed \textit{regression adjustment}. %The overall functional is also referred to as \textit{inverse propensity weighting,} $\textstyle \fnl(P) =\E[Y(1)]= \E \left[Y \frac{\mathbb{I}[A=1]}{e(X)}\right]$, where $e(x) = P(A=1\mid X=x)$ is the \emph{propensity score}. 
% $$ \textstyle \fnl(P) =\E[Y(1)]= \E \left[Y \frac{\mathbb{I}[A=1]}{e(X)}\right].$$
% Given two identifying functionals for the mean potential outcome, a natural question is: can these identification approaches be combined in some way to improve estimation? 
The canonical \textit{doubly robust} estimator of \citet{robins1994estimation} is:
$$ \textstyle \E[Y(1)]=  \E \left[ \frac{\indic{A=1}}{p(A=1\mid X)} (Y - \E[Y\mid A=1,X])+\E[Y\mid A=1,X] \right].$$
Appealing properties such as mixed-bias and rate double robustness can be verified easily. This is one example of the broader framework of one-step estimation using influence functions.%This provides a canonical example of one-step adjustment by influence functions that applies more broadly. 

% \subsection{Application to the ATE}\label{sec-framework-ate-estimation}
\paragraph{Characterizing the finite-differences output for AIPW.}
Our characterization will show exactly how plug-in evaluation with finite differences is nearly equivalent to evaluating the doubly robust estimator with a Nadaraya-Watson regressor as the conditional outcome regression for $\E[Y\mid A=1,X]$ and a kernel density estimate of the propensity score. It is equivalent up to an additive bias from the smoothed perturbation distribution. The following lemma, similar to \citet[Lemma 25.1]{van2018cv}, summarizes convergence rates of these smoothed nuisances. 
\begin{lemma}\label{lemma-rates}
Let $\td{\mu}_\eps(X)=\td{E}_{\tP_\eps}[Y\mid A=1,X]$ and $\td{e}_\eps(X) = \td{p}_{\eps}(A=1, X)/\td{p}(X)$ denote the nuisances induced by plug-in estimation of probability distributions $\td{p}$. \rev{Assume that, for each $\epsilon$ in a neighborhood of $0$, the function
$\tilde{\mu}_\epsilon(\cdot)$ is $\beta$-H\"older (uniformly in $\epsilon$),
so that $\int \tilde{\delta}^\lambda_x(u)\tilde{\mu}_\epsilon(u)\,du-\tilde{\mu}_\epsilon(x)=O(\lambda^\beta)$.
}%For any bounded function $g(x)$ in a Hölder class of degree $\beta$,
% $\textstyle (\int \smoothdel{x}(u) g(u)\dif u - g(x))^2 \dif x = O(\hsmoothdelta^\beta).$ 
Assume that $\smoothdel{}$ is a product kernel in a Hölder class of degree $\beta$, and $\textstyle \int \vert{u}\vert^\beta \vert{\smoothdel{u}}\vert \dif u < \infty$ and $\int u^s \smoothdel{u} \dif u= 0$ for $s \leq \beta.$ Assume $Y$ is bounded. %$\textstyle \int  \left\{\int \smoothdel{x}(u) g(u)\dif u \right\}- g(x))^2 \dif x = O(h^4 + \frac{1}{nh^d}).$ 
Then, if the unperturbed nuisance estimation satisfies the rates $\|\tilde{\mu}-\mu\|=O_p\left(a_n\right)$ and $\|\tilde{e}-e\|=O_p\left(b_n\right),$ the perturbed nuisances satisfy the following rates: %\useshortskip
    %         \begin{align*}
    % &\E[(\td{\mu}_\eps(X) - \mu(X))^2]= \E[(\td{\mu}(X) - \mu(X))^2] + O(\eps^2 \hsmoothdelta^{-d}),\\
    % &\E[(\td{e}_\eps(X) - e(X))^2] = \E[(\td{e}(X) - e(X))^2]  + O(\eps^2\hsmoothdelta^{-d}).
    % \end{align*} 
    % RMSE
                \begin{align*}%\SwapAboveDisplaySkip
    &
    \norm{
    % \sqrt{\E[(
    \td{\mu}_\eps(X) - \mu(X)
    }
    % })^2]}
    = 
    % \norm{
    % % \sqrt{\E[(
    % \td{\mu}(X) - \mu(X)
    % }
    % )^2]} 
    % + 
    O_p(a_n+\nu^{-1} \eps \hsmoothdelta^{-d/2}),
    \\
    &
    % \sqrt{\E[(
    \norm{
    \td{e}_\eps(X) - e(X)
    }
    % )^2]} 
    = 
    % \sqrt{\E[(
    % \norm{
    % \td{e}(X) - e(X)
    % }
    % )^2]}
    % + 
    O_p( b_n+\nu^{-1} \eps \hsmoothdelta^{-d/2}).
    \end{align*} 
\end{lemma}
The perturbed nuisances are asymptotically consistent, so long as $\td{p}$ induces asymptotically consistent nuisances and we fix $\eps$ as a sequence vanishing in $n$ at a rate appropriate to counteract the growth in $\lambda$ due to ``roughness"; i.e., the variability of $\smoothdel{}$. The difference between the function evaluation and its smoothed nuisance arises from the bias analysis of kernel density estimators and the dimension-dependence of ensuring absolute continuity, but it is a non-stochastic, entirely deterministic argument. While the analysis improves mildly in the dimension dependence on $\hsmoothdelta$ compared to the  general analysis of \citet[Thm. 5]{carone2018toward}, this still highlights the generally unfavorable dependence of the numerical approach on the dimension due to smoothing. 
\rev{
\begin{remark}
\Cref{lemma-rates} provides the link between deterministic finite-difference perturbations and the statistical condition \Cref{eqn-one-step-decomposition-empiricalappx-vanishes-fast}. It gives $L_2(P)$ control of the perturbed nuisance functions $\tilde{\mu}_\eps$ and $\tilde{e}_\eps$, which can then be used to bound $\frac{1}{n} \sum_{i=1}^n(\tilde{\phi}(O_i)-\phi_{\tilde{\eta}}(O_i))$ to satisfy \Cref{thm-one-step-sufficient-condition}. %More generally, for functionals whose finite-difference expansions depend on perturbed nuisance functions, analogous integrated bounds play the same role.
\end{remark}
}
\begin{algorithm}[t!]
\caption{Empirical Gateaux derivatives }
\label{alg-empiricalif}
\begin{algorithmic}[1]
% \Procedure{Empirical Gateaux Derivatives}{}  
    \State Inputs: $\tP$ probability density estimates, data $\{O_i\}_{1:n},$ perturbation and smoothing $(\eps,\lambda)$ parameters, causally identified functional of the observational joint distribution $\fnl$
    \For{$i=1,\dots n$}
    \State $\tP_\eps^i = (1-\eps) \tP + \eps \smoothdel{o_i}$ and  $\htdel(O_i) \gets \eps^{-1}(\fnl(\tP_\eps^i)-\fnl(\tP))$
    % \State
    \EndFor 
\State Output: one-step adjusted estimate from empirical Gateaux derivatives, $\fnl(\tP) + \frac{1}{n} \sum_i \htdel(o_i)$
% \EndProcedure

\end{algorithmic}
\end{algorithm}

Our first proposition establishes the exact pointwise approximation error induced by finite differencing, where $\tilde{\mu}_{\lambda} (x_i)= \int \smoothdel{(x_i)}(x) \mu(x) \mathrm{d}x$.%Denote for brevity ${\mu}(x) = \int y\,\frac{p(y,A=1,x)}{p(A=1,x)}\,\mathrm{d}y, {\mu}_\eps(x) = \int y\,\frac{p_\eps(y,A=1,x)}{p_\eps(A=1,x)}\,\mathrm{d}y.$ \az{too many notation collisions with $\tilde{E}_{P}$ etc and can we redefine something smoother like $\tilde{\mu}_{\lambda} (x_i)= \int \smoothdel{(x_i)}(x) \mu(x) \mathrm{d}x $ }
% \az{give example for AIPW}
\rev{
\begin{proposition}[Empirical Gateaux Derivatives approximate AIPW estimators]\label{cor-aipw-additivebias}
Suppose the smoothed perturbation kernel $\smoothdel{y_i}(y)$ is uniform so that $ \int y \; \smoothdel{y_i}(y) \mathrm{d}y =y_i$. Consider the mean potential outcome functional $\fnl(P) = \E[Y(1)].$ When the perturbation observation is the observation datapoint $O_i = (X_i,A_i,Y_i)$,
\begin{align*}%\SwapAboveDisplaySkip
 \htdel(O_i)
% \textstyle \epsilon^{-1}( {\Psi(\tilde{P}_\epsilon^{\ptbdir} ) - \Psi(\tilde P)}{})
&= 
\int \tilde{p}(x)\frac{\indic{A_i = 1} }{  \tepspnalowerx{j}} \smoothdel{(x_i)}(x)
        ( Y_i - 
 \td{\mu}(x)) \mathrm{d} x
      + (\tilde{\mu}_\lambda(X_i) - \fnl(\tilde{P}))
    \\
    & \qquad +   \int \{\td{\mu}_\eps(x) - \td{\mu}(x)\}\{\smoothdel{(x_i)}(x)-p(x)\}\,\mathrm{d}x
\\&
=
\frac{\indic{A_i=1} }{   \tepspnamidx{i}} 
   \left( Y_i - 
  \tilde{\mu}(X_i)\right)
        % \\&\textstyle\qquad 
+\left(\tilde{\mu}(X_i)- \Psi(\td{P}) \right) + O(\hsmoothdelta^\beta) + O(\eps).
\end{align*}
\end{proposition}
}
\rev{Although the form of the AIPW influence function is itself very well known, in \cref{cor-aipw-additivebias} we apply the framework to AIPW as a starting example. We do so in order to inspect the differences in individual terms, which reveals where the finite-differences approximation error arises from in general. }First, observe that convolution with the smoothing kernel introduces additive bias at the observation $o_i$. \rev{Next, we outline how we proved the decomposition/approximation error by applying our more general rules in turn. %Similar approximation stability results hold for other functionals that are not the mean potential outcome, but arise from similar compositions of product/quotients of the input probability distributions.
}
\rev{
\paragraph{Applying the framework: applying the finite difference rules to the mean potential outcome, $E[Y(1)]$.}
We give a proof outline for \Cref{cor-aipw-additivebias} by leveraging our more general product/quotient rules and simplifications. 
\begin{align*}
&\eps^{-1}(\Psi(P_\eps)-\Psi(P))
= \eps^{-1}\left(\int p_\eps(x)\,\mu_\eps(x)\,\mathrm{d}x-\int p(x)\,\mu(x)\,\mathrm{d}x\right) \\
& {}{=} \int p(x){\{\mu_\eps(x)-\mu(x)\}}\,\mathrm{d}x 
      + \int \mu(x)\{p_\eps(x)-p(x)\}\,\mathrm{d}x
      +   \int \{\mu_\eps(x) - \mu(x)\}\{p_\eps(x)-p(x)\}\,\mathrm{d}x \tag{\cref{eq-productrule}, \text{product rule}}
      \\
& 
{=} 
\int p(x)\frac{\indic{A_i = 1} }{  \tepspnalowerx{j}} \smoothdel{(x_i)}(x)
        ( Y_i - 
 \mu(x)) \mathrm{d} x
      + (\mu_\lambda(X_i) - \fnl(\tilde{P}))
      +   \int \{\mu_\eps(x) - \mu(x)\}\{p_\eps(x)-p(x)\}\,\mathrm{d}x \tag{\cref{eq-quotient-rule}, quotient rule and simplifying}
\end{align*}
\Cref{cor-aipw-additivebias} follows by applying \Cref{eq-cndlexp-fd} to the first term. 
}
% \az{need to write up formal proof of this?}
% \begin{corollary}[Empirical Gateaux Derivatives approximate AIPW estimators]\label{cor-aipw-additivebias}
% Suppose the smoothed perturbation kernel $\smoothdel{y_i}(y)$ is uniform so that $ \int y \; \smoothdel{y_i}(y) \mathrm{d}y =y_i$. When the perturbation observation is the observation datapoint $O_i = (X_i,A_i,Y_i)$,
% \begin{align*}%\SwapAboveDisplaySkip
%  \htdel(O_i)
% % \textstyle \epsilon^{-1}( {\Psi(\tilde{P}_\epsilon^{\ptbdir} ) - \Psi(\tilde P)}{})
% &
% \if\forjournal 0
% \textstyle
% \fi
% =
% \frac{\indic{A_i=1} }{   \tepspnamidx{i}} 
%    \left( Y_i - 
%   {\E_{\tP}[Y\mid A=1,X_i]} \right)
%         % \\&\textstyle\qquad 
% +\left( {\mathbb{E}}_{\tP_\eps}[Y\mid A=1,X_{\ptbdir}] - \Psi(\td{P}) \right) + O(\hsmoothdelta^\beta).
% \end{align*}
% \end{corollary}
\Cref{lemma-rates} allows us to further simplify and conclude that the decomposition of \Cref{prop-decomposition-gateaux-derivative-pdf-rep} is close to the canonical AIPW estimator up to the $O(\hsmoothdelta^\beta)$ bias induced by smoothed evaluation.

\paragraph{Implications for estimation}
Next, we use our characterization in \Cref{prop-decomposition-gateaux-derivative-pdf-rep}, \Cref{lemma-rates}, and \Cref{cor-aipw-additivebias} to study the \textit{statistical} properties of our \textit{computational/numerical} approximation. \Cref{cor-aipw-additivebias} is a deterministic equivalence and allows us to study the relationship to the typical ``oracle" AIPW estimator, albeit with induced $\epsilon$-perturbed nuisances. An appealing property of bias-adjusted treatment effect estimation is \textit{rate double robustness}: because we incur the product of convergence rates of the nuisances, we may enjoy parametric $n^{-\frac 12}$ rate convergence of the target functional while nuisances converge at slower rates, for example at $n^{-\frac 14};$ see \cite{chernozhukov2018double,rotnitzky2021characterization}. We use the rate double robustness property of the target and \Cref{lemma-rates} to infer the required rate conditions on $(\epsilon, \hsmoothdelta)$ that  retain the beneficial statistical properties of the canonical AIPW estimator. The next result combines our previous characterizations with the standard analysis of AIPW. 

\rev{Next, we verify asymptotic normality for AIPW by giving sufficient conditions and characterizing the statistical and approximation error rate conditions that would be required. }
\begin{assumption}[Regularity conditions]\label{asn-regularityconditions}
Assume the following regularity conditions hold:
\begin{enumerate}[label=(\roman*)]
\item $Y$ is bounded. 
\item    $p(y\mid A=1,x), p(A=1\mid x)$ are in Hölder classes of minimum degree $\beta$.
\item The estimates $p(y,a,x), p(a,x),p(x)$ belong to a Donsker function class. 
\item Assume $\td{\mu},\td{e}$ satisfy the product-rate condition: 
$%\sqrt{\E[(
\norm{\td{\mu}(X) - \mu(X)}
%)^2}
% } 
\times %\sqrt{\E[(
\norm{\td{e}(X) - e(X)
% )^2] 
}
= O_p(n^{-\frac 12}).$
Assume $\td{\mu},\td{e}$ are RMSE-consistent with $r_\mu, r_e$, respectively, 
so that $r_\mu+r_e\geq \frac 12$.
\end{enumerate}
\end{assumption}
\rev{
\begin{theorem}[AIPW rate double-robustness]\label{prop-rate-double-robustness} 
    Under \Cref{asn-regularityconditions}, if the approximation error rates further satisfy that $\eps \lambda^{-d/2}=o(n^{-\max(r_\mu, r_e)}),$ $\eps \lambda^{-d/2}\|\tilde{p}(X)-p(X)\|=o_p(n^{-\frac 12})$, and $\lambda^\beta = o(n^{-\frac 12}),$ then $
\frac{1}{n} \sum_{i=1}^n\left(\tilde{\phi}\left(O_i\right)-\phi_{\tilde{P}}\left(O_i\right)\right)=o_p\left(\nu^{-1} n^{-1 / 2}\right).
$
\end{theorem}
}
% \az{is this dependent on numerical approximation error alone?}
% \az{fixup}
% \begin{theorem}[Rate double robustness]\label{prop-rate-double-robustness} 
% Consider the one-step estimator with the empirical Gateaux derivative. 
% Under \Cref{asn-regularityconditions}, when $\eps \lambda^{-d/2}=o(n^{-\max(r_\mu, r_e)}),$ and $\lambda^\beta = o(n^{-\frac 12})$:%\useshortskip
%  $$
%  \textstyle \left(
% \if\forjournal 0
% \textstyle
% \fi
% \textstyle \fnl(\tilde{P}) + \frac 1n \sumn{i} \htdel(O_i)\right) - \fnl(P) = o_p(\nu^{-1} n^{-\frac 12}). $$ 
%  \end{theorem}
% The requirement that $\xi >0$ enforces $\eps \ll \hsmoothdelta$. 
\rev{Here $O_p\left(\nu^{-1}\right)$ reflects the dependence on the overlap constant $\nu^{-1}$ as we connect the finite difference analysis with causal analysis. We omit other fixed regularity constants, such as bounded outcomes from the $O(\cdot)$ notation. }
\Cref{prop-rate-double-robustness} states the rate conditions required of \textit{numerical approximation} in $\eps, \hsmoothdelta$ to preserve the \textit{statistical property} of rate double robustness. Assuming that the nuisances induced by the probability density estimates satisfy a product-rate condition that the product of their RMSEs is faster than the parametric $O(n^{-\frac 12})$ rate, the form of \Cref{prop-decomposition-gateaux-derivative-pdf-rep} suggests that the error induced by finite-differences must be faster than the rates of the unperturbed nuisance functions, $\max(r_\mu, r_e)$, which can be a slower rate than the $\epsilon=o(n^{-\frac 12})$ implied by the generic analysis of finite differences. In general, the dimension dependence on $\lambda$ in \Cref{lemma-rates} will prevent setting a slower rate for $\eps$ for nontrivial dimensionality. \rev{In total, if the marginal density estimator satisfies
$\|\tilde p(X)-p(X)\|=O_p(n^{-r_x})$, then a sufficient rate condition is
$\epsilon \lambda^{-d/2}=o_p(n^{-\max\{r_\mu,r_e,1/2-r_x\}})$,
together with $\lambda^\beta=o(n^{-1/2})$.}

\rev{For a direct comparison of these requirements to prior work, note that for a uniform kernel and our two-point approximation scheme, \citet[Thm. 5]{carone2018toward} suggests a rate requirement of 
% $O(\epsilon \lambda^{-d} + \lambda^{2})$
$\epsilon \lambda^{-d}= o(n^{-\frac 12})$ and $\lambda^{2} = o(n^{-\frac 12}).$\footnote{
\rev{
\citet{carone2018toward} obtained a general rate that reflects potential improvements in approximation error due to higher-order finite difference schemes. Under a uniform kernel, they obtain a generic approximation error of $O\!\left((\epsilon \lambda^{-d})^{s} + \lambda^{2}\right)$ where $s=\min(m,m_0)$, $m$ denotes the order of the forward finite-difference
scheme (i.e., an $(m+1)$-point scheme), and $m_0$ denotes the degree of pathwise
differentiability of the projected functional along the smoothed path. In our paper we use the standard two-point forward
difference, corresponding to $m=1$ and $s=1$, so the generic discretization term reduces to
$O(\epsilon \lambda^{-d} + \lambda^{2})$.}
}
% \footnote{
% \rev{
% }
% }
}
% It may be possible to choose $\eps$ on the order of the faster nuisance which could be a rate \textit{slower} than the $n^{-\frac 12}$ (in RMSE) that arises from a generic finite-difference analysis. 
\rev{\citet{carone2018toward} provide a general bound for \textit{multi-point finite-difference schemes} which reduce approximation error requirements via additional function evaluations; these also additionally require higher-order pathwise differentiability of the functional. }
Note that our tighter analysis based on rate-double robustness can improve the rate term involving $\eps$ on the order of the improvements from \rev{multi-point difference schemes discussed in \citep{carone2018toward}.  
}
% \footnote{
\rev{
For a concrete comparison, consider a higher-order $m=2$ multi-point scheme, which would give a weaker rate requirement of 
$(\epsilon \lambda^{-d})^2 + \lambda^{2} = o(n^{-\frac 12})$. %, where $d$ is the dimension of smoothing at the cost of additional evaluations of $\Psi$.
% However, our refined analysis exploits the specific statistical structure of the
% interventional mean and shows that the $\epsilon$ portion of the numerical requirement is instead governed by the
 Our 
weaker conditions in \Cref{prop-rate-double-robustness} indicate that if $r_\mu=r_e=1/4$, then our two-point scheme only requires $\epsilon \lambda^{-d/2}=o_p(n^{-1/4})$,
together with $\lambda^\beta=o(n^{-1/2})$.
This matches the $n^{-1/4}$ exponent in the $\epsilon$-term of a generic
second-order scheme ($s=2$), but without algorithmically requiring higher-order finite
differences.%; moreover, the dependence on $\lambda$ is milder here
($\lambda^{-d/2}$ rather than $\lambda^{-d}$).
Thus, when nuisance or density estimation rates are slower than parametric,
the statistical structure can yield improvements comparable to those of
multi-point schemes while retaining a simple two-point implementation.
}
% }
% That is, our analysis shows we can attain the approximation benefits of algorithmically improved multi-point finite-difference schemes via the specialized \textit{statistical} structure of the target adjustments. 

\rev{Our analysis indicates that the specific structure of the function gives an opportunity to improve upon prior general approximation bounds, which raises the question: for what other functionals are such improvements possible? As we discuss in our later analysis of more complicated functionals, it depends on the structure of the functional; later examples, e.g. with inverse function operations do not admit such sharpenings. Following \citep{carone2018toward}, we still recommend $(\epsilon, \lambda)$ plots in practice to assess approximation quality since the approximation theory is not tight.}

Our analysis does make use of our characterization in \Cref{cor-aipw-additivebias}; i.e., a deterministic approximation error from AIPW with induced nuisance functions. While our refined \textit{analysis} requires quantification via smoothness parameters of the probability densities, the \textit{algorithm} generally does not require knowledge of these parameters. %, so that our discussion of  statistical empirical process terms and higher order remainder terms is not completely general but rather uses the known structure of the functional.

\subsection{More complex estimands: dynamic treatment regimes}\label{sec-framework-dtr}

We study estimation properties of the empirical Gateaux derivative for multi-stage DTR. 
\begin{example}[Dynamic Treatment Regime]
    In $T$-stage dynamic treatment regimes, the causal quantity of interest is the mean potential outcome
$\EE{}{Y({\bar{a}})}$, where $\bar{a} = (a_0, \ldots, a_{T})$ is the (deterministic) treatment strategy and $\mathcal{A}$ is the action space. \rev{Assume $Y({\bar{a}})$ is sequentially ignorable given the treatment and covariate history, $(\bar{A}_t, \bar{X}_t)$, at each time $t$:
$A_t \perp Y(\bar{a}) \mid \bar{A}_{t-1}, \bar{X}_t,$ for all $\bar{a}.$}
The $g$-formula identifies the estimand: \useshortskip
\begin{equation*} %\textstyle 
\EE{}{Y({\bar{a}})} = \int \EE{}{Y \,|\, \bar{A}=\bar{a}, \bar{X}=\bar{x}}\prod_{t=1}^T \tilde{p}(x_t \,|\, \bar{a}_{t-1}, \bar{x}_{t-1})\dif \bar{x},
\end{equation*} 
where $\bar{A} = (A_0, \ldots, A_{t}), \bar{X} = (X_0, \ldots, X_{t})$. To derive its influence function, we take the empirical Gateaux derivative in the direction of $o_i=(\bar{x}_i, \bar{a}_i, y_i)$, where $\bar{x}_i = (x_{i0}, \ldots, x_{it})$ and $\bar{a}_i = (a_{i0}, \ldots, a_{it})$,
\begin{align}\SwapAboveDisplaySkip
    \Psi(\tilde{P}^i_\epsilon) 
    &\textstyle =\int \left(\int y  \frac{\tilde{p}_\epsilon(y , \bar{A}=\bar{a}, \bar{X}=\bar{x})}{\tilde{p}_\epsilon(\bar{A}=\bar{a}, \bar{X}=\bar{x})} \dif y\right)\prod_{t=1}^T \frac{\tilde{p}_\epsilon(x_t , \bar{a}_{t-1}, \bar{x}_{t-1})}{\tilde{p}_\epsilon(\bar{a}_{t-1}, \bar{x}_{t-1}))}  \tilde{p}_\epsilon(x_0)\dif \bar{x}.
\end{align}
\end{example}
Below we characterize how this empirical Gateaux derivative at the smoothed distribution differs from the one at the (unsmoothed) estimated distribution. The following result is analogous to \Cref{prop-decomposition-gateaux-derivative-pdf-rep}
and \Cref{cor-aipw-additivebias} but is extended to the dynamic treatment regime.
\begin{proposition}[Dynamic treatment regime]\label{prop-fd-dtr}
\begin{align}\SwapAboveDisplaySkip
    & \eps^{-1} ({\Psi(\tilde{P}^i_\epsilon)-\Psi(\tilde{P})})
    \nonumber
    % \\
    %     & 
        = 
        \left(\E_{\tilde{p}(\bar{x}_{1:T})}\left[ \EE{\tilde{P}_\epsilon}{Y \,|\, 
        \bar{A},\bar{X}} \mid \bar{x}_0
        \right]  - \Psi(\tilde{P})\right) \\
        & \qquad 
    +
    \frac{
    \indic{\bar{A}=\bar{a}_i}
    }{\tilde{p}_\epsilon(
    \bar{a}\,|\,
    \bar{x})}{}\left[ Y_i - \EE{\tilde{p}(\bar{x})}{\EE{\tilde{P}}{Y\,|\, 
    \bar{A}, 
    \bar{X} }}\right]  \nonumber\\
    &
    \qquad+   \sum_{s=1}^T\frac{
        \indic{\bar{A}=\bar{a}_{i,T-s}}
    }{\tilde{p}_\epsilon(\bar{a}_{T-s}\,|\, \bar{x}_{T-s})}
    \left\{  \EE{\tilde{P}}{ \EE{\tilde{P}_\epsilon}{Y\,|\, 
    \bar{A}, 
    \bar{X}} \mid \bar{a}_{i,T-s}, \bar{x}_{i,T-s+1}  } 
    \right. \nonumber
    \\
    &
    \quad \quad\quad\quad 
    \left.- \EE{\tilde{P}}{ \EE{\tilde{P}_\epsilon}{Y\,|\, 
    \bar{A}, 
    \bar{X}} \mid \bar{a}_{i,T-s}, \bar{x}_{i,T-s}  }
    \right\}  + O(\eps^2)  + O(\lambda^{\beta}). \nonumber
\end{align}
\end{proposition}
The expansion verifies that the requirements in $\eps$ for numerical approximation are analogous to \Cref{ex-cfactualmean}, despite the complexity of the nested expectation. Again, smoothing incurs overall unfavorable dependence in the dimension, as the extension of \Cref{lemma-rates} to $\E_{\td{P}_\eps}[Y\mid \bar{A},\bar{X}]$ also incurs unfavorable dependence on the dimension. %The history-dependent nuisances imply that the dimension additionally grows in the time horizon. 
\new{
% \paragraph{Simplifying verification of approximation properties via finite-difference calculus.}
}
\rev{
\paragraph{Applying the framework: dynamic-treatment regime}
In the appendix, we provide an alternate, much shorter derivation via the finite-difference calculus, which allows deducing the main insight without cumbersome manual analysis: the approximation error from finite differences accumulates linearly. 
\begin{align*}
\MoveEqLeft{\frac{\Psi(\tilde{P}^i_\epsilon)-\Psi(\tilde{P})}{\epsilon}} =\int \int \eps^{-1} \cdot y \cdot \left[
\left( \frac{\tilde{p}_\eps(y , \bar{a}, \bar{x})}{\tilde{p}_\eps(\bar{a}, \bar{x})} - \frac{\tilde{p}(y , \bar{a}, \bar{x})}{\tilde{p}(\bar{a}, \bar{x})}\right)
\prod_{t=1}^T {\tilde{p}(x_t \mid \bar{a}_{t-1}, \bar{x}_{t-1})}  \tilde{p}(x_0) \right. \\
& \left.+ \sum_{k=1}^T \eps^{-1} \cdot {\tilde{p}(y \mid \bar{a}, \bar{x})}
\left(\frac{\tilde{p}_\eps(x_k , \bar{a}_{k-1}, \bar{x}_{k-1})}{\tilde{p}_\eps(\bar{a}_{k-1}, \bar{x}_{k-1}))}-\frac{\tilde{p}(x_k , \bar{a}_{k-1}, \bar{x}_{k-1})}{\tilde{p}(\bar{a}_{k-1}, \bar{x}_{k-1}))} \right)
 \prod_{t\in [T]\backslash k} {\tilde{p}(x_t \mid \bar{a}_{t-1}, \bar{x}_{t-1})} \tilde{p}(x_0) \right.\\
& \left.+ \eps^{-1} \cdot  {\tilde{p}(y \mid \bar{a}, \bar{x})}\prod_{t=1}^T {\tilde{p}(x_t \mid \bar{a}_{t-1}, \bar{x}_{t-1})}\left(\tilde{p}_\eps(x_0)- \tilde{p}(x_0) \right)\right]\dif \bar{x} \dif y
\end{align*}
}
{
\section{Influence Functions of Stochastic Optimization Programs}\label{sec-stochopt}  
}
In this section we apply the finite-differences approach for computing Gateaux derivatives to stochastic optimization functionals. We discuss how our framework can apply to linear optimization functionals with probability inputs in the coefficients. We illustrate for two estimands of interest in the causal inference literature that can be expressed as such: infinite-horizon off-policy optimization in reinforcement learning \citep{liu2018breaking}, and sensitivity analysis in causal inference under the marginal sensitivity model. 
% In this section, we will first recall these Gateaux derivative characterizations from the stochastic optimization literature (though they were derived for non-contextual settings). We then precisely derive the differences between     $\phi^{opt} (\delta_P({\tilde{o}})),   \phi (\delta_P({\tilde{o}})),     \hat\phi(\delta_P({\tilde{o}}))$ in the specific example of sensitivity analysis in causal inference. Finally, we generalize our derivations to a class of functionals that are the marginalizations of contextual linear programs (possibly with constraint matrices that vary in the conditioning). 

% In this subsection we consider a generic stochastic optimization problem under the distribution $P$, where the functional is defined as $
% \fnl(P) = \inf_{u\in M} \E[f(u,y)]$. The optimal solution $u^*(P) \in \arg\inf_{u\in M} \E_P[f(u,y)]$.

\subsection{Generalization to stochastic optimization programs}
% \az{swap order of these}
% Although so far we have discussed a number of different examples that have been studied directly in the causal inference literature, we now generalize our argumentation to a class of estimands that are {marginalizations of linear programs (with probability distribution inputs).}
\new{We first discuss our finite-differences approximation stability results for a general class of linear programs with uncertain coefficients and marginalized expectation objective and Lagrangian. In this section, we drop explicit discussion of $\lambda$-smoothing for brevity, but arguments of \Cref{sec-aipw} would give analogous additive approximation errors results. Under sufficient high-level regularity conditions (i.e., uniqueness and local feasibility), we analyze the analytical Gateaux derivative and its finite-difference approximation.
}
% We discuss how a perturbed matrix inverse point of view, sufficient regularity (nondegeneracy) ensures that the decomposition of \cref{eqn-sensitivityanalysis-productrulefd-optsolution} holds more generally.

% \citep[Thm. 17]{golstein1972}, restated in \citep[Thm. 9]{dupavcova1990stability}: 
% Let $M_0 \neq \varnothing$ be convex, closed and $P, Q \in \mathscr{P}$. Let $f(\cdot, P), g_i(\cdot, P), 1 \leqq i \leqq m$ be convex continuous on $M_0$. Let for programs (31), (33) the sets $X(P)$ and $V(P)$ be nonempty and bounded. Assume further that the functions $f(\cdot, Q), g_i(\cdot, Q), 1 \leqq i \leqq m$ are convex and finite for all $u$ belonging to a neighborhood of $X(P)$. Then the Gâteaux derivative of $\phi(P)$ in the direction of $Q-P$ exists and
% $$
% \begin{aligned}
% \phi^{\prime}(P ; Q-P) & =\min _{u \in X(P)} \max _{v \in V(P)}(L(u, v, Q)-L(u, v, P)) \\
% & =\max_{v \in V(P)} \min_{u \in X(P)}(L(u, v, Q)-L(u, v, P))
% \end{aligned}
% $$
% \az{how does this apply in our setting? re min/max the difference? } 
    The stochastic linear program could have uncertain objective, left-hand-side or right-hand-side constraint coefficients. Its objective and Lagrangian are effectively marginalizations over functions of a random variable $X$, which could be a subvector of the full observation random vector. 
    %We consider a stochastic linear program that is a marginalization  over a random variable $X$, a subset of the full joint observation. %The decision variable $u(x)$ is in a Banach space (it can be finite-dimensional, or a function, for example of $y$). 
    Suppose for simplicity of presentation $X$ is discretely distributed with cardinality $d_x$. \rev{Without loss of generality, we relabel its support to $\left\{1,2, \ldots, d_{x}\right\}$ } The decision vector is generically a function of $x$, i.e. $u \in \mathcal{U} \subseteq  \mathbb{R}^{d_x}$, and its entries describe the function value at a given $x$. 
    The objective coefficient $c(P)$ is a $d_x$-dimensional vector; its entries are functionals of components of the joint distribution, possibly evaluated at a particular value of $x$, so that $c \colon \mathcal{M}_{\mathrm{NP}} \to \mathbb{R}^{d_x}$.
    % The objective $c$ is a $d_x-$dimensional random vector, its entries are maps of components of the joint probability distribution evaluated at a particular value of $x$, so that $c \colon \mathcal{M}_{\mathrm{NP}} \to \mathbb{R}^{d_x}$.
    % $c \colon \mathbb{R}^{d_x} \to \mathbb{R}^{d_x}$. 
    %(In what follows, we denote the functional dependence on $x$ with $(x)$.)
When there are $j$ many constraints, then $M\in\mathcal{R}^{j}\times\mathcal{R}^{d_x}$. The coefficient matrix $M$ has entries that are maps of components of the joint probability density evaluated at $x$. The right-hand side vector is denoted $b \in \mathbb{R}^j$. %We generally assume that the entries of the coefficient matrix satisfy weak regularity conditions in their functional dependence on $V,$ e.g. when $V=X
    % =(\{A=1\},X)
    % $ then the coefficient entries are assumed to be smooth functions in $X.
    % (X).
    % $ 
    % We denote both the $\epsilon$-perturbed and $x$-conditional coefficients with 
   We use $(\cdot)^\epsilon$ to denote the perturbed coefficient vectors and matrices: that is, $c^\epsilon$ comprises of replacing its probability density estimates, components of $P$, with the corresponding component of $P_\epsilon,$ as defined previously. We analogously denote the perturbed left-hand side constraint coefficient matrix $M^{\epsilon},$ and right-hand-side coefficient vector $b^\epsilon$. The following definition summarizes this set-up. 
    % and $c^\epsilon(x)$. 
    % Define 
    % % $$ c(V) = \E[c(O)\mid V],\;\;b(V) = \E[b(O)\mid V], \text{ and } c^\eps(V) = \E_{P_\eps}[c(O)\mid V], \;\;b_\eps(V) = \E_{P_\eps}[b(O)\mid V]. $$
    %     $$ b(X) = \E[b(O)\mid X], \;\;b_\eps(X) = \E_{P_\eps}[b(O)\mid X]. $$
    % We assume that entries of the coefficient matrix are functionals of the distribution of the observation random variable. We generally assume that the entries of the coefficient matrix satisfy weak regularity conditions in their functional dependence on $V,$ e.g. when $V%=X
    % =(\{A=1\},X)
    % $ then the coefficient entries are assumed to be smooth functions in $%X.
    % (X).
    % $ We denote both the $\epsilon$-perturbed coefficients and $V=%X
    % (\{A=1\},X).
    % $ indexed conditional coefficients with the left-hand side constraint coefficient matrix $M^{\epsilon, x},$ right-hand-side coefficient vector $b^\epsilon(x)$.
    % % and $c^\epsilon(x)$. 
    % Define 
    % $$ c(V) = \E[c(O)\mid V],\;\;b(V) = \E[b(O)\mid V], \text{ and } c^\eps(V) = \E_{P_\eps}[c(O)\mid V], \;\;b_\eps(V) = \E_{P_\eps}[b(O)\mid V]. $$
    %     $$ b(X) = \E[b(O)\mid X], \;\;b_\eps(X) = \E_{P_\eps}[b(O)\mid X]. $$
    % \az{the argument we give only applies to $c$ a constant in $x$ currently}
    % $$ \fnl(P)& =  \E_{P}[f(u^*_P,y)] && \text{closed-form with first-stage primal solution $u^*_P \in\arg\inf \E_{P}[f(u,y)]$} $$
    \rev{\begin{definition}\label{eqn-stoch-ctxtlp}
         Define
\begin{align*}
& \textstyle \fnl(P) = c^\top  u^*  , 
 && \textstyle 
 \text{ where }  u^* \in \arg\max_{u } \left\{    c^\top  u \;
    \colon  M u  \;\leq b\right\} 
\\ & \textstyle \fnl(P_\eps)  = (c^\eps)^\top u^*_\eps \;, 
 &&   \textstyle 
 \text{ where } 
    u^*_\epsilon \in \arg\max_{u} \left\{ (c^\eps)^\top u 
    \colon M^{\epsilon} u \; \leq b^{\epsilon}\right\}.
 \end{align*}
 % \az{need $M$ to be indexed at least by every value of state (so it is marginalized) }
 % \az{this means we need different primal, dual subsets of observation random variable}
    \end{definition}
   }
   % We state this general formulation and then describe regularity conditions on the coefficients and on the optimization program. 
   \rev{How does this general formulation capture estimands of research interest? We introduce our first example,  infinite-horizon off-policy learning, before we introduce our regularity conditions and assumptions.
   \paragraph{Running example: linear programming representation for infinite-horizon offline reinforcement learning}}
   We show how our framework can derive generic bias corrections for infinite-horizon off-policy optimization in offline reinforcement learning, via the canonical \textit{linear programming} characterization  \citep{puterman2014markov,de2003linear}. Our application is relevant to a recent line of work on off-policy evaluation and learning (OPE/L) \citep{jl16,pb2016,bibaut2019more,kallus2022efficiently,tang2019doubly,kallus2020confounding,kallus2022stateful}. We recall the classical linear programming formulation of finding an optimal policy in a tabular (finite-state, finite-action) Markov decision process, due to \citet{puterman2014markov,de2003linear}. 

    We discuss the problem setup before introducing the linear programming formulation and its probability inputs. Consider an infinite-horizon Markov decision process with discount factor $\gamma$. The offline dataset is comprised of trajectories of (state, action, next state) observations: $\{(s^0_i,a^0_i, \dots, s_i^t, \dots, s_i^T)\}_{1:N}$. We derive our bias adjustment under a statistical model where the stationary distribution factorizes as ${p(s,a,s') = d(s,a) P(s'\mid s,a)}.$ From the joint state-action-state occupancy distribution $p(s,a,s')$, we estimate the marginalization $d(s,a)$, the stationary state-action occupancy distribution, which we use to estimate the transition probability $P(s'\mid a,s) = \frac{p(s,a,s')}{d(s,a)}.$ Let $\mu_0(s)$ denote the initial state distribution, estimated from offline data. Because we focus on the discrete case, we do not require smoothing.
% \paragraph{.}
\begin{example}[The optimal policy linear program]\label{ex-infhorizonope}
The \textit{optimal} value function solves the following linear program in a finite-state and finite-action setting; the objective value at optimality is the policy value. Here, $P_a \in \mathbb{R}^{\vert \mathcal{S} \vert \times \vert  \mathcal{S} \vert}$ is transition matrix at action $a$, with $P_a(s,s') = P(s'\mid s,a)$, and $\mu_0(s)$ is the marginal occupancy distribution of the initial state. We assume, as is fairly common in reinforcement learning, that the reward function is a known function of state and action, $r(s,a)$. Let ${r}_a=\{r(s,a)\}_{s\in \mathcal{S}}$ be the vector of reward values per state, for fixed action $a$.  We have:%\useshortskip
%and the state-averaged reward vector is $r_{a}(i)=\sum_{j \in \mathcal{S}} p_{i j}(a) r_{i j}(a)$. 
\begin{equation}\textstyle
\fnl_D(P) = \underset{V}{\operatorname{min}}
\{ (1-\gamma) {\mu_0}^{\top} {V} \colon \left(I-\gamma P_{a}\right) {V}-{r}_{a} \geq 0, \quad \forall a \in \mathcal{A} \}.
\label{eqn-primalmdplp}
\end{equation}
The dual formulation is well known to parametrize the stationary occupancy probabilities of the optimal policy:\useshortskip 
\begin{equation} \textstyle \fnl_P(P) = \underset{\mu}{\operatorname{max}} \left\{  \sum_{a \in \mathcal{A}} \mu_{a}^{\top} {r}_{{a}} \colon \sum_{a \in \mathcal{A}}(I-\gamma P_{a}^{\top}) \mu_{a}=(1-\gamma) {\mu_0}, \quad \mu_{a} \geq 0, \quad \forall a \in \mathcal{A}  \right\}. \label{eqn-dualmdplp}
\end{equation}
\end{example}
Recent work has revisited this formulation with interest in developing primal-dual algorithms \citep{wang2017randomized,lee2018stochastic,abbasi2019large,serrano2020faster}. 
   \begin{remark}[Infinite-horizon OPE \Cref{ex-infhorizonope} ($\fnl_D(P)$) in the notation of \Cref{eqn-stoch-ctxtlp}]
Without loss of generality, consider \Cref{eqn-stoch-ctxtlp} with minimization, i.e. let  $u^*(x) \in \min_{u(\cdot)} \left\{   \int   c(x) u(x) \intd x
    \colon \int M(x) u  \intd x\geq b\right\}$ We give $c,M,b$ corresponding to \Cref{ex-infhorizonope} and interpret.
       Let $\mathcal{S}$ be the state space and $x=s$. let the decision vector $u \in \mathbb{R}^{\vert \mathcal{S} \vert}$ correspond to $V(s)$. Then the objective coefficients correspond to $c(s) = (1-\gamma) \mu_0(s)$, the initial state distribution. Then $M \in \mathbb{R}^{\vert \mathcal{S} \vert \vert \mathcal A \vert \times \vert \mathcal{S} \vert}$ and $M = [\{ I - \gamma P_a\}_{a \in \mathcal{A}}]^\top, b = [\{r(\cdot,a)\}_{a \in \mathcal{A}}] $, where $[\{\cdot\}_{\circ}]$ indicates concatenation over $\circ$. 
       %$M = (I - \gamma P_a^\top).$ 
Rows of $M$ are transition probabilities conditional on $(\tilde{s},a);$ we let $M_{(\tilde{s},a),:}$ index the $(\tilde{s},a)$th row and we have that $M_{(\tilde{s},a),:} u = V(\tilde{s}) - \sum_{s'} P(s'\mid a,\tilde{s}) V(s').$ The RHS vector is $b = r_a \in \mathbb{R}^{\vert \mathcal S \vert \vert \mathcal A \vert}.$ 

Therefore the objective is the expected value over the initial state distribution, $\int c(s) u^*(s) \mathrm{d} s = \E_{\mu_0}[V^*(s_0)],$ and the constraints enforce that $V(s) \geq r(s,a) + \gamma \E[V(s')\mid s,a].$
   \end{remark}
\new{  \paragraph{Regularity conditions}  Having introduced the class of linear programs we consider, we now discuss regularity conditions required. A separate, but related, line of work in optimization studies the smoothness of the optimal value in the decision variable and provides other sufficient conditions  \citep{tercca2021envelope}.
Our analysis proceeds by applying an inverse function theorem to the optimal basis, so we require regularity conditions permitting invertibility. We maintain \Cref{asn-nondegeneracy,asn-locallysamebasis} which ensure a basic feasible solution. Let $\beta$ be the index vector of basic random variables (for the stochastic program indexed by $x$), so that $M^{\epsilon}_{\beta}(x), b_{\beta}^\epsilon(x) $ indexes the active basis of constraints. 
}
\new{
We assume strong duality holds in the optimization program and that the Lagrangian is pathwise differentiable. 
\rev{
Define the (primal) Lagrangian integrand
$
\mathcal{L}(P;u,v)
\coloneqq c^\top u + v^\top (M u - b), v \in \mathbb R_+^j,
$
where the entries of $c, M, b$ depend on $P$ as in \Cref{eqn-stoch-ctxtlp}. Define the Lagrangian value functional
\[
\Psi_{\mathcal L}(P)
:=
\sup_{u \in \mathcal U}
\inf_{v \in \mathbb R_+^j}
\mathcal{L}(P;u,v).\]
\begin{assumption}[Strong duality holds.]\label{asn-strong-duality}Under strong duality, the optimal value admits the saddle representation
\begin{equation*}
\Psi_{\mathcal{L}}(P)
=
\sup_{u \in \mathcal U}
\inf_{v \ge 0}
\mathcal{L}(P;u,v)
=
\inf_{v \ge 0}
\sup_{u \in \mathcal U}
\mathcal{L}(P;u,v).
\end{equation*}
\end{assumption}
}
\rev{\begin{assumption}[Pathwise differentiability]\label{asn-opt-pathwisedifferentiable}
Assume strong duality holds. The value functional
$
\Psi_{\mathcal L} : \mathcal M \to \mathbb R
$
is pathwise differentiable at $P$.
\end{assumption}
}
\Cref{asn-strong-duality} implies via Lagrangian duality with dual variable $v\in \mathbb{R}^{j}$ that the optimal  primal-dual function pair $u, v$ solves:
% \begin{align} 
% \coloneqq  
% \left[ 
$\inf_{u}\sup_{v} 
\{ c^\top u + \sum_k v_k (M_k u - b_k  )\}$.
% \right]
% \\
% \fnl_{ocl} &= \E_P[  c(X)^\top
% ( M_{\beta_X}(X)^{-1}b_{\beta_X}(X)) 
% ]
% \end{align} Lastly we make a high-level regularity assumption on the matrix coefficients that implies that the functional is pathwise differentiable. 
\Cref{asn-opt-pathwisedifferentiable} is a high-level regularity assumption on the objective and matrix coefficients, that their marginalizations $ c^\top u + \sum_l v_l (M_l u - b_l  )$ are pathwise differentiable. For example, in the infinite-horizon OPE example, the entries of $M$ are conditional densities evaluated at a value of $s$, which are \textit{not} pathwise differentiable, however the objective value and Lagrangian are. 
\Cref{asn-locallysamebasis,asn-opt-pathwisedifferentiable,asn-nondegeneracy} are high-level primitive conditions for asymptotic linearity of the stochastic optimization functional, so that it admits an influence function. 
% Note that non-uniqueness can lead to much more difficult challenges of non-regular inference. An interesting approach could be to introduce recent advances in smoothing for non-regular inference in other areas to this setting \citep{chen2023inference}. Otherwise, we proceed  with high-level assumptions of asymptotic linearity; similar assumptions are made in \citet{lam2021impossibility,duchi2021asymptotic}. Although generic arguments for stochastic optimization establish Hadamard differentiability \citep{shapiro2021lectures}, \citet{jeong2020assessing} also show that orthogonalized machine learning improvements are still retained under Hadamard differentiability. 
} 

\new{Similar high-level assumptions are made in generic studies of statistics of stochastic optimization \citet{lam2021impossibility,duchi2021asymptotic}. Indeed a challenge is verifying Gateaux differentiability, which can require structure of the functional, since stochastic optimization is generically Hadamard differentiable. See \Cref{apx-discussion-pathwisediff} for further discussions.
}
% \az{Stability condition here? where we move MDP example to after the general setting.}

\rev{
A common set of assumptions that ensure pathwise differentiability is to assume that the optimal solution is unique; in that case, we give primitive conditions that restrict $\epsilon$ to ensure that perturbations preserve the optimal solution.
\begin{assumption}\label{asn-nondegeneracy}
The optimal solution is unique; there is no degeneracy.
\end{assumption}
\rev{%{Role of the optimal basis.}
For a linear program in standard form, an optimal solution (under nondegeneracy) occurs at
a basic feasible solution determined by a set of active (binding) constraints \citep{bertsimas1997introduction}.
Let $\beta$ denote the index set of constraints corresponding to this optimal basis,
so that the optimal primal solution can be written in closed form as $u^* = M_\beta^{-1} b_\beta$, where $M_\beta$ and $b_\beta$ denote the submatrix and subvector indexed by $\beta$.}
\begin{assumption}\label{asn-locallysamebasis}
Assume that the optimal basis $\beta$ remains the same locally in a neighborhood near $\epsilon=0$, and finite differences are computed with $\eps$ within this neighborhood.
\end{assumption}
\begin{proposition}[Primitive conditions for same optimal basis (informal statement)]
\label{cor:eps-same-optimal-basis-informal}
Assume the optimal primal and dual solutions $u^*,v^*$ are bounded. There exist problem-dependent constants $K_M,K_b,K_c$ depending only on $ \|u^*\|_\infty, \|v^*\|_\infty, \|M\|_\infty, \|M^{-1}\|_\infty, \|c \|_\infty,$ and the margins of nondegeneracy (minimum value of nonzero basic entries of $u^*$ and reduced costs), so that for all $0\leq \varepsilon \leq \min \left\{K_M, K_b, K_c\right\}$, the same basis remains optimal. 
\end{proposition}
}
\rev{Here we give an informal statement of \Cref{cor:eps-same-optimal-basis} for brevity, whose full version is given in the appendix. \Cref{cor:eps-same-optimal-basis} gives primitive conditions for $\epsilon$-perturbations to be small enough to retain the same optimal basis. It leverages an elementary linear programming fact about the optimal basis: under \Cref{asn-nondegeneracy}, the optimal basic variables and non-basic reduced costs are nonzero, $u^*_B >0,r_N>0$ \citep{bertsekas1997nonlinear}. \Cref{cor:eps-same-optimal-basis} follows by deducing conditions on $\epsilon$ so that the perturbed basic variables $u^*_{B,\epsilon}>0$ and non-basic reduced costs $r_{N,\eps}^*>0$ maintain the same optimality conditions. These conditions on $\epsilon$ are then expressed in terms of known or observable problem quantities. }

\new{
%(Extension to continuous state spaces is immediate, but with more technical condition.) In this case, $c,u$ are vectors over the $X$ space. 
Next, we rewrite the optimal objective value via the inverse function theorem from the binding basis. %We call this the ``statistician's oracle" notion of the functional, since we study perturbations to the inverse function itself (rather than the original optimization program). This emulates the analysis of closed-form solutions: the ``statistician's oracle" solves the optimization in closed form, \textit{analytically} determining the optimal solution basis. 
}
\new{
\begin{definition}[Closed-form representation under basis stability]\label{def-oraclefnl}
    \begin{align}%\SwapAboveDisplaySkip
\fnl(P) &= c^\top
( M_{\beta}^{-1}b_{\beta}), \qquad 
\fnl(P_\eps) =  (c^\eps)^\top 
 ((M_{\beta}^\eps)^{-1}b_{\beta}^\eps ).  \label{eq-ocl-lpfnl}
% &= \E_P 
% \left[ \max_{u(X)\colon M(X)u\leq b(X) } \{   \E[c(X)^\top u(X) \mid X]\} 
% \right]
\end{align} 
\end{definition}
}
\new{
This represents the optimal solution as a function of the (potentially perturbed) probability distribution, so that the optimal solution too is perturbed. In general, if perturbations might enter the constraints, this requires evaluating a total derivative. 
}
% We focus on estimands that are the averages of linear programs with probability inputs in the constraint LHS matrix $M$ and/or RHS vector $b$:
\new{
\begin{remark}
    In the main text, we consider discrete state spaces for simpler statements. 
Many of the results can be extended to continuous state spaces, i.e. infinite or semi-infinite linear programs, under high-level regularity conditions such as strong duality. An analogue of the previous assumptions is the assumption that extreme points are basic feasible solutions (BFS).
\citet{ghate2009characterizing} gives a high-level sufficient condition: strictly positive support; i.e., where the infimum of the components of the decision variable is strictly positive. This is not \textit{necessary}, for example \citet{mohajerin2018infinite}, establishes other sufficient conditions for the infinite-horizon off policy evalution example. 
\end{remark}

Next we give arguments for an analytical influence function by evaluating the total Gateaux derivative at the optimal solution, i.e. the optimal solution is an inverse function at the binding basis. 
\begin{proposition}[Marginalization of linear programs]\label{prop-general-LP} 
Given \Cref{asn-locallysamebasis,asn-opt-pathwisedifferentiable,asn-nondegeneracy}, and assuming $\Psi(P)$ is as in \cref{eq-ocl-lpfnl}, we have:
% \begin{align*}
%   &   \frac{\intd}{\intd\eps} \fnl_{ocl}(P_\eps) \Big\vert_{\eps=0} 
%  = c(X_i)\left(M_\beta^{-1} b_\beta\right)(X_i) - \fnl(P) -(v^*)^\top  \left(\frac{dM^\eps_\beta}{d \eps}\Big\vert_{\eps=0}\right) u^*
%  +
%   c^\top 
% \left(    M_\beta^{-1}
% \frac{\intd}{\intd\eps}\{b_{\beta}^\eps\}\Big\vert_{\eps=0} \right). 
% \end{align*}
\begin{align*}
  \left.\frac{\intd}{\intd\eps}\fnl(P_\eps)\right|_{\eps=0}
  &=
  \left.\frac{\intd}{\intd \eps}(c^\eps)^\top\right|_{\eps=0}u^*
  -(v^*)^\top\left(\left.\frac{\intd M^\eps_\beta}{\intd \eps}\right|_{\eps=0}\right) u^*
  +c^\top\left(M_\beta^{-1}\left.\frac{\intd}{\intd\eps}b_{\beta}^\eps\right|_{\eps=0}\right),
\end{align*}
where $u^*:=M_\beta^{-1}b_\beta$.
\end{proposition}
}
\new{
Next we establish finite-difference approximation thereof. For this, we impose regularity assumptions that entries of $M$ are well-behaved mappings of the probability distributions; i.e., they are well-behaved under finite-difference approximation and incur linear approximation error. 
% for a generic linear program $\max \{ c^\top x\colon Mx = b\},$ with perturbations that are a linear additive perturbation in $\eps$, i.e. $M^\eps = M + \eps (G-M)$, we have the following relation between the analytical Gateaux derivative and the finite-difference: 
% \begin{corollary}
%   $$  \eps^{-1} (x^*(\eps) - x^*) = \frac{\intd }{\intd \eps} x^*(\eps) + O(\eps)
%   $$
% \end{corollary}
% \az{want to be careful about G and $G_\eps$ here}
\begin{assumption}[Invertibility of $M$ and its entries are well-behaved under finite-difference approximation]\label{asn-matrixentriesarewellapproxfnls}
Assume that there exists $\kappa > 0$ such that the smallest singular value $\sigma_{\min}(M_\beta)\geq \kappa$
 and 
% \begin{align*}
$M^{\eps}_\beta= M + \eps G_\eps= M + \eps G + O(\eps^2). 
$
% , 
% &= \epsilon\E_P [  G(X) ] + O(\eps^2) \\
% \E_P[b^{\eps}(X)-b^{\eps}(X)]
% &= \epsilon\E_P [  H(X) ] + O(\eps^2) 
% \end{align*} 
\end{assumption} 
}
\new{
\Cref{asn-matrixentriesarewellapproxfnls} is a high-level assumption that is satisfied if the coefficient matrix entries are compositions of transformations satisfying a finite-difference calculus; e.g., they are sums, products, quotients of component probability distributions. Our examples satisfy this assumption: the entries of the coefficient matrix $M$ are transition probabilities $p(s'\mid s,a)$, whose finite-difference approximation is given by \Cref{lemma-numerical-derivative-decomposition}. And the perturbed entries can be written as $M + \epsilon G_\eps,$ with entries of $\left(G_\epsilon^a\right)_{s, s^{\prime}}=\frac{\mathbb{I}\left[\tilde{s}, \tilde{a}, \tilde{s}^{\prime}\right]}{d_\epsilon(s,a)}-\frac{\mathbb{I}[\tilde{s}, \tilde{a}] P\left(s^{\prime} \mid s, a\right)}{d_\epsilon(s, a)}$.
Our next result establishes general finite-difference approximation stability under these regularity conditions. 
\begin{proposition}[Finite-difference approximation error, marginalizations of linear programs]\label{prop-fd-apx-error-lps}
% \vspace{-5pt}
% \begin{equation*}
Under \Cref{asn-matrixentriesarewellapproxfnls},
$
 \eps^{-1} (\Psi(P_\eps) - \Psi(P))   = 
 \frac{\intd}{\intd\eps} \fnl(P_\eps) \Big\vert_{\eps=0}  + O(\eps).
 $
% \end{equation*}
\end{proposition}
}
% \vspace{-5pt}

% \vspace{-5pt}
\new{
% \begin{remark}
Unlike the earlier case of the treatment mean in \Cref{cor-aipw-additivebias} and \Cref{prop-rate-double-robustness}, 
% even in a favorable discrete-state case without smoothing requirements, it does not appear that 
rate double robustness does not admit weaker numerical requirements on $\eps$ because of the perturbed matrix inversion in this example, rather than sum and product rules in previous examples. In the analysis, \Cref{eqn-valuefn-eps-nuisance} shows that $V_\eps(s)-V(s) = O(\epsilon)$, so that \Cref{prop-mdp-finitedifferenceapprox} holds with the same order of approximation error. 
% \end{remark}
}
% \az{put the remark here}

\rev{Importantly, our general analysis here extends to different linear programming estimators where the influence function is not currently known. Next, we discuss specific examples to highlight more specific term-by-term comparisons.}

% \az{What about perturbing $\alpha,\beta$? }

% \paragraph{Extension to perturbations in $c$}

% The following characterization is immediate from an application of Danskin's theorem for a fixed constraint set $M$, for linear $H$, and is due to \citep[Thm. 7]{dupavcova1990stability}, also discussed in \citep{shapiro2021lectures,shapiro1990differential}: 
% \begin{equation}
% \frac{d}{d\eps} \fnl^{opt}  (P_\eps)\mid_{\eps=0} 
%      = \min_{u \in X(P)}\E_{Q}[f(u,y)]- \fnl(P) 
%      \label{eqn-optdanskin} 
% \end{equation}
% Extensions hold also for the convex case and for local maximizers. On the other hand, these arguments are not specialized to the causal setting, as we do in the next subsections.  
% yw / can we write down Q here for P_\epsilon? 
% \az{what does evaluation at the measure $Q$ mean for delta measure?)}

\subsection{Stable finite-difference approximation for (constrained) infinite-horizon off-policy evaluation} 

% This is an example of the general stochastic linear program given in \Cref{eqn-stoch-ctxtlp}. 
% \begin{remark}[ ]
% We now restate \Cref{ex-infhorizonope} (infinite horizon off-policy optimization) in the notation of \Cref{eqn-stoch-ctxtlp}.
%   Let $O=(s,a,s'), X=s, u(s) = V(s), c(x) = (1-\gamma)\mu_0(s), M = [\{ I - \gamma P_a\}_{a \in \mathcal{A}}]^\top, b = [\{r(\cdot,a)\}_{a \in \mathcal{A}}] $, where $[\{\cdot\}_{\circ}]$ indicates concatenation over $\circ$. 
% \end{remark}
% We suppose nondegeneracy so that the optimal solution occurs at a basic feasible solution, and that perturbations $\eps$ are small enough to maintain the same basis $\beta$ locally.

We illustrate the prior general results in the context of the infinite-horizon OPE example of \Cref{ex-infhorizonope}. \rev{We let $V^*, \mu^*$ denote the optimal dual and primal optimization variables. }First we verify the analytical Gateaux derivative $\frac{\mathrm{d}}{\mathrm{d}\eps} \fnl_D(P_\eps) \Big\vert_{\eps=0}$, evaluated in the direction of $(\td{s}, \td{a},\td{s}').$ 
% We assume nondegeneracy (unique optimal solution); this can be relaxed by using a more sophisticated perturbation analysis. 
% \begin{assumption}
% The optimal solution is unique; there is no degeneracy.
% \end{assumption}

\begin{proposition}[Analytical Gateaux Derivative for \Cref{ex-infhorizonope}]\label{prop-infhorizonmdp} 
Suppose \Cref{asn-nondegeneracy}.\useshortskip %Perturb in the direction of a generic observation $o=(\td{s},\td{a},\td{s}'):$ 
\begin{equation}
    \frac{\mathrm{d}}{\mathrm{d}\eps} \fnl_D(P_\eps) \Big\vert_{\eps=0} =  (1-\gamma)V^*(\td{s})
    +\frac{\mu^*(\td{s},\td{a})}{d(\td{s},\td{a})}
    \left( r(\td{s},\td{a}) + \gamma V^*(\td{s}') - V^*(\td{s}) \right) -\fnl_D(P).
\end{equation}
\end{proposition}
% We do not claim great novelty of \Cref{prop-infhorizonmdp}; rather we recover known results with the alternative perturbation-based/Gateaux derivative argument. 
% \paragraph{Sketch of argument. }
% The statement of \Cref{prop-infhorizonmdp} is not surprising per se, since we can immediately verify the mean-zero sample version of the Bellman residual, and so one might plausibly derive this form from double robustness. Rather, we provide an alternative argument that applies perturbation analysis of optimization programs with the specific perturbations from the finite difference. We build on the sensitivity analysis characterization of \cite{freund1985postoptimal}, which studies perturbations of linear programs with respect to constraint matrix perturbations. For $\eps$ small enough, the \textit{active basis} remains the same. When the perturbation matrix is $P+\eps G$ for some matrix $G$, \cite{freund1985postoptimal} notes that the derivative of the optimal value is $\frac{\intd }{\intd\eps}(\fnl_D(P_\eps))=- \overline{\pi} G\overline{x}$ (where $(\overline{\pi}, \overline{x})$ are the dual and primal optimal solutions, respectively) can be obtained from the derivative of a matrix inverse, which yields higher-order expansions. We generalize this argument for the specific form of the perturbed transition probabilities.  

% yw / above in eq 8, can we remind people what V^* and u^* are? 

We next compute the approximation error that arises from evaluating finite differences, so long as $\eps$ is small enough to maintain the same active basis, which is empirically verifiable. Put simply, the analyst evaluates the functional by estimating the initial state distribution (objective coefficients) and transition probabilities (coefficient matrix coefficients) and solving the linear program. The analyst evaluates the perturbed distribution functional by perturbing the objective and matrix coefficients appropriately, and re-solving the linear program. 
\begin{proposition}[Error analysis from finite differencing]\label{prop-mdp-finitedifferenceapprox}
Suppose \Cref{asn-nondegeneracy}. Perturb in the direction of a generic observation $o=(\td{s},\td{a},\td{s}').$ Then: 
\begin{equation}
     \eps^{-1}(\fnl(P_\eps)-\fnl(P)) =  (1-\gamma)V^*_\eps(\td{s})
     +\frac{\mu^*(\td{s},\td{a})}{d_\eps(\td{s},\td{a})}
    \left( r(\td{s},\td{a}) + \gamma V^*(\td{s}') - V^*_\eps(\td{s}) \right) - \fnl_D(P) + O(\eps).
\end{equation}
\end{proposition}

\rev{\paragraph{Applying the framework:  infinite-horizon MDP LP.}
We overview the proof outline to highlight how we can finite-difference stability via composing our prior product/quotient rules, and a new matrix inverse rule. In \Cref{lemma-matrix-inverse-fd-stability}, we show finite-difference approximation stability for matrix inverse functions taking the form of \Cref{asn-matrixentriesarewellapproxfnls}, i.e. entrywise they are functionals that are stable under finite differences. \begin{equation} 
   \eps^{-1} ((M^\eps)^{-1}- M^{-1}) = 
     -  M^{-1}   G(M^{-1}) + O(\eps) %\tag{matrix inverse finite difference approximation}
     \label{eqn-matrix-inverse-fd-rule}
\end{equation} 
With this additional rule and our prior product/quotient rules and simplifications, certifying \Cref{prop-mdp-finitedifferenceapprox} follows by applying them as follows:
\begin{align*}
& \textstyle \eps^{-1}
(\fnl_D\left(P_\epsilon\right) -\fnl_D\left(P\right)  )
=(1-\gamma) \sum_s {
\eps^{-1}(\mu_0^\epsilon(s) V_\epsilon^*(s)-\mu_0(s) V^*(s))
} \\
& \textstyle=(1-\gamma) \sum_s\left(\mathbb{I}[\tilde{s}]-\mu_0(s)\right) V^*_\eps(s)+\sum_s \mu_0(s) 
\eps^{-1}\{V_\epsilon^*({s}) - V^*({s})
\} \tag{product rule }
\\ 
& \textstyle=(1-\gamma) \sum_s\left(\mathbb{I}[\tilde{s}]-\mu_0(s)\right) V^*_\eps(s)+\sum_s \mu_0(s) 
\{\left(I-\gamma P^\pi\right)^{-1}\left(G^\pi\right)\left(I-\gamma P^\pi\right)^{-1} r_\pi\}+O\left(\epsilon\right).  %\tag{\cref{eqn-matrix-inverse-fd-rule}}
\end{align*}
The last line follows by applying \cref{eqn-matrix-inverse-fd-rule} (the new matrix inverse finite difference stability rule) to $V_\epsilon^*=M_\epsilon^{-1} r_{a^*}$, where the matrix $M_\epsilon\coloneqq\left(I-\gamma P_{a^*}^\epsilon\right)$ satisfies the matrix perturbation representation of \Cref{asn-matrixentriesarewellapproxfnls} with perturbation matrix $\gamma G_\epsilon^a$ that has entries $\left(G_\epsilon^a\right)_{s, s^{\prime}}=\frac{\mathbb{I}\left[\tilde{s}, \tilde{a}, \tilde{s}^{\prime}\right]}{d_\epsilon(s,a)}-\frac{\mathbb{I}[\tilde{s}, \tilde{a}] P\left(s^{\prime} \mid s, a\right)}{d_\epsilon(s, a)}$
Denote $M_\epsilon\coloneqq\left(I-\gamma P_{a^*}^\epsilon\right)$, then $V_\epsilon^*=M_\epsilon^{-1} r_{a^*}$. Simplifying gives \Cref{prop-mdp-finitedifferenceapprox}.
}

\new{
\paragraph{Extension to continuous state space}
For notational clarity and ease of exposition, we focused on the case of a discrete state space. But the finite-difference analysis also extends directly to computational approximations of continuous state-space versions of the same formulation. We omit technical details re: finite approximations to infinite linear programs, for example \citet{mohajerin2018infinite} develop such approximations for the linear programming approach to dynamic programs.
}
% \citet{mohajerin2018infinite} study regularity properties and finite-dimensional approximations. They study 1) the approximation error from optimizing over value functions in  a specific basis and 2) high-probability bounds on approximation error from randomized constraint sampling (their Theorem 4.4). In short, 1) can be controlled via arguments for the approximation error of sieves (for example), while 2) can ensure the error from randomized approximation is $\leq \epsilon$ with probability $\geq 1-\beta,$ for a user's choice of $\beta,\epsilon$: the only thing that changes is perhaps the extent of computation required via the number of constraints being sampled. Therefore, judicious choices of parameters in the approach of \citet{mohajerin2018infinite} ensure these results also translate to continuous state spaces. (We omit details since they follow from \citet{mohajerin2018infinite}).}

%\added{This arises due to the finite-difference approximation of \textit{perturbed matrix inversion}, rather than repeated application of the product and quotient rules as in \Cref{prop-decomposition-gateaux-derivative-pdf-rep}.} 
\paragraph{Interpreting the finite-difference approach.} We take a moment to translate the benefits of a finite-difference approach to estimating a one-step adjustment in the context of reinforcement learning. In reinforcement learning, a common paradigm is that of ``model-based" reinforcement learning; i.e., learning a dynamics model (estimating the transition distribution) and plugging it into a planning problem or the linear programming formulation. A benefit of finite differences is that one can approximate doubly robust adjustments only with additional black-box functional evaluations. 

% One main benefit of this approach is that by computing these finite differences, we can obtain generic bias corrections via an argument that directly generalizes to other variants of the linear program. 
And, this argument directly generalizes to other variants of the linear program. For example, consider a relevant subclass of constrained Markov decision processes comprised of additional constraints on the policy variables: the linear programming formulation is particularly appealing because linear constraints can be directly added \citep{altman1999constrained}. Another benefit is that one-step estimation can be obtained by evaluating the functional at feasible points. %In our case, this corresponds to optimizing with respect to feasible probability distributions. 
% Although one can \textit{derive} doubly robust corrections from the Lagrangian, as \cite{tang2019doubly} does, and is implicit in perturbation-based approaches because the dual variables characterize the sensitivities, introducing importance sampling weights in the optimization problem (i.e in the constraints) easily leads to infeasibility in finite samples. 
% We emphasize that finite differences directly enables practical algorithmic evaluation of the bias correction. 
% \paragraph{Extensions of this characterization}
% So far we have presented the analytical derivative; again to build confidence in this procedure. But the numerical finite-difference approach provides a flexible way to obtain bias correction for additional arbitrary constraints; even if the solutions are not amenable to analysis in closed form. And, the finite-difference approach maintains feasibility because it reduces to evaluations of the (optimization) functional at feasible probability distributions. For example, we consider the important example of constrained Markov Decision Processes, for which the linear program representation is particularly appealing because additional constraints can simply be included in the optimization formulation. 
\begin{example}[Constrained policy optimization]\useshortskip
\begin{equation*} \textstyle \fnl_P(P) = \operatorname{max} \left\{  \sum_{a \in \mathcal{A}} \mu_{a}^{\top} {r}_{{a}} \colon \sum_{a \in \mathcal{A}}(I-\gamma P_{a}^{\top}) \mu_{a}=(1-\gamma) {\mu_0}, \;\; \mu_{a} \geq 0, \;\; \forall a \in \mathcal{A}, \; \mu \in \mathcal{P}  \right\}. %\label{eqn-dualmdplp}
\end{equation*}
\end{example}
The additional constraint $\mu \in \mathcal{P}$ includes any linearly representable constraints on state occupancy; this can be used to ensure safety, e.g.
\cite{paternain2019constrained} develops a framework with a ``caution function," a convex function on the state-action occupancy measures. We note that doubly-robust policy evaluation has been studied in \cite{tang2019doubly}, but our arguments are more general. 
% while this formulation has been built upon for policy learning in \cite{nachum2019algaedice}. The first work focuses on policy evaluation. Unsurprisingly, 
% Our derivation via perturbation analysis is similar to \citet[Thm. 4.1]{tang2019doubly}'s illustration via the Lagrangian, but our arguments generalize to similarly-structured optimization programs. %This is expected since the Lagrangian provides the dual characterization that also underlies perturbation analysis. %Doubly robust evaluation has been studied in \cite{kallus2022efficiently} (although efficiency is subtle due to regularity issues). 
% Relative to the doubly robust estimators derived elsewhere, this influence function may be higher variance. %Note this approach attains doubly robust estimation of the optimal policy's \textit{value}.

\new{
\subsection{Orthogonalized sensitivity analysis for the Marginal Sensitivity Model} 
We previously described general arguments for obtaining influence functions for stochastic linear program functionals. We now describe a more complex example, sensitivity analysis in causal inference, which is a marginalization of a pointwise separable stochastic linear program. We obtain the nonparametric influence function by taking the total derivative of the optimization program, and derive a finite-difference approximation.%For simplicity, we discuss approximation error with discrete delta-function perturbations, noting that approximation error with the smoothed perturbations follows in the analogous way as in the previous sections. 
Ultimately, this more complex example shows how our analytical framework could be used to study questions of contemporary research interest, beyond the ATE itself.
}
\new{
% \paragraph{Setup.}
Sensitivity analysis for causal inference obtains bounds under violations of assumptions. It can be generally written as a stochastic optimization problem. We develop a case study based on the well-studied marginal sensitivity model \citep{tan2006distributional}, a variant of Rosenbaum's sensitivity model \citep{rosenbaum2002sensitivity}, which we describe briefly. 
}
\new{
One viewpoint on sensitivity analysis in causal inference under the marginal sensitivity model \citep{tan2006distributional} is a robust or adversarial version of inverse propensity weighting. The inverse propensity weighting estimator obtains identification under unconfoundedness since the propensity score can debias the data via $e(x)^{-1}$. In the absence of unconfoundedness, the full-information inverse propensity score, $w(x,y) = e(x,y)^{-1}$, would debias the data, but is unknown. Instead, bounds on functionals can be obtained by optimizing over unknown perturbations to the inverse propensity score $\tilde{W}(x) = e(x)^{-1}$, which can always be estimated but is not sufficient to debias the data. A natural assumption is that the extent of unobserved confounding is not ``too large" and this is imposed by  analyst-chosen restrictions on how far the unknown inverse propensity score $w(x,y)$ can deviate from its estimated nominal/marginal value. In the marginal sensitivity model, an analyst specifies $\Lambda$ which uniformly bounds the odds ratio:\useshortskip $$ \Lambda^{-1} \leq \frac{e_1(x,y) / (1-e_1(x,y))}{e_1(x)/(1-e_1(x))} \leq \Lambda.$$ Note by Bayes' theorem this also implies a bound on the likelihood ratio to the counterfactual outcome density. The adversarial weight is still required to be a valid likelihood ratio where $\E[\indic{A=1}w(X,Y)\mid X]=1.$ 
}
\new{
These restrictions correspond to constraints on the unknown weight $w$. The odds ratio restriction corresponds to linear interval bounds 
% $( \Lambda+\pi_{b}(a \mid s)^{-1}(1-\Lambda)) \leq W'(x,y) \leq (\pi_{b}(a \mid s)^{-1}+\Lambda^{-1}\left(1-\pi_{b}(a \mid s)^{-1}\right) )$. A bit agnostically, we will refer to these upper and lower bounds as $a_\Lambda(x) \leq W'(x,y) \leq b_\Lambda(x).$ 
$ %a_{\Lambda}(x,a)
%\coloneqq 
% \mathcal{U}_a = \left\{ W'_a(x,y) \colon
( 1+\Lambda^{-1} \cdot\left(e_a(x)^{-1}-1\right))\leq 
%W'
w_a(x,y) \leq
1+\Lambda \cdot\left(e_a(x)^{-1}-1\right).%\right\}
%\coloneqq b_\Lambda(x).
$
}
\new{
Combining these constraints yields a linear optimization that optimizes
the objective function, for $A=a,$ $\E[Y  w_a (X,Y)\indic{A=a} \mid X=x ]$ 
over a weight function $w_a(x,y)$. Finally, we reparametrize with $w_a(x,y) = W_a(x,y)/e_a(x)$,
 where $W_a(x,y)$ satisfies:%\useshortskip
% ; we denote the pointwise optimal solution $g(x)$:
% \begin{equation}
%   g(x) =   \sup \left\{ 
%   \E[Y  W'_a \indic{A=a} \mid X=x ] \colon \E[ W'_a \indic{A=a} \mid X=x] = 1; \;\; W'_a(x,y) \in \mathcal{U}_a 
%   %a(x,a) \leq W'_a(x,y)\leq  b(x,a), \forall y
%     \right\}
% \end{equation}
% \az{ this should be $W'(x,y) = W(x,y)/P(A=1\mid X=x)$ }
\begin{equation*}
\alpha_a(x)
% \coloneqq (e_1(x) +\Lambda^{-1} \left(1-e_a(x)\right))
\coloneqq e_a(x)(1-\Lambda^{-1}) +\Lambda^{-1} 
\leq W_a(x,y) \leq
% e_1(x)+\Lambda \left(1-e_1(x)\right)
e_a(x)(1-\Lambda)+\Lambda 
\coloneqq \beta_a(x),
\end{equation*}
 % We reparametrize with $W'(x,y) = W(x,y)/P(A=1)$:
%  so the optimal integrand is:%\useshortskip 
% \begin{equation*}
%    g_a(x) =      \sup \left\{ 
%   \\E[Y  W_a \mid A=a,X=x ] \colon \E[ W_a \mid A=a,X=x] = 1; \;\; \alpha_a(x) \leq W_a(x,y) \leq \beta_a(x)
%   % e_a(x)
%   % a(x,a)  \leq W_a(x,y)\leq  e_a(x) b(x,a), \forall  y
%     \right\}.
% \end{equation*}
% \vspace{-5pt}
We summarize the bounds problem in the following example.
\begin{example}\label{ex-sensitivityanalysis}
We consider the case of $A=1$ and drop subscripts on optimization variables for brevity. %Assume that $\pi_b(a\mid x)$ is known.
Given the marginal sensitivity model with sensitivity parameter $\Lambda$, the bounds on the treated mean are:%\useshortskip 
$$
\Psi(P)= \sup_{W(x,y)} \left\{ 
  \E[Y  W(X,Y) \mid A=1] \colon \E[ W (X,Y)\mid A=1,X] = 1; \;\; \alpha(x) \leq W(x,y)\leq  \beta(x), \forall  y,x
    \right\}.
    $$
The optimization is pointwise separable so that $\fnl(P) = \E[g(X)\mid A=1],$ where $g(x)$ solves:
\begin{align*}%\SwapAboveDisplaySkip
    g(x)= 
    & 
    \sup_{W} \left\{ 
  \E[Y  W \mid A=1,X=x] \colon \E[ W \mid A=1,X=x] = 1, \forall x; \;\; \alpha(x) \leq W(x,y)\leq  \beta(x), \forall  y
    \right\}. %&& \text{(Primal)}
% \\
%     \text{dual: } & \min_{\lambda \in \mathbb{R}} \left\{ 
%  P(A=1)\left(a(x)\E[(\lambda-Y)_- \mid A=1, X=x])+ b(x) \E[(Y -\lambda)_+\mid A=1,X=x]\right) + \lambda 
%     \right\}.
    % \\
%         = &  \left\{ 
% \left(\alpha(x)\E[ (\lambda^*-Y)_-\mid A=1, X=x])+ \beta(x) \E[(Y -\lambda^*)_+ \mid A=1,X=x]\right) + \lambda^*
%     \right\} &&
\end{align*}
% We next will show that 
% $$ \frac{\intd}{\intd \eps}  
% \{\fnl(P_\eps) \}\vert_{\eps = 0}
% = \textstyle \E_P\left[\E_P\left[     Y   \frac{\mathrm{d}}{\mathrm{d}\eps}\left\{W_\eps^*\right\}\Big\vert_{\eps = 0}\mid A=1,X \right]\right]
% +\frac{\intd}{\intd \eps}   \{  \fnl^{opt}(P_\eps) \}\vert_{\eps = 0}
% $$
% Previous work has studied the functional form of the $W^*(X,Y)$ optimal primal solution. The optimal value is a marginalization of the optimal primal solution $g(x),$ which we define in general for possibly $\epsilon$-perturbed distributions below. 
% \begin{align*} 
% g_\eps(x) &=  \max_{W \in \mathcal{U}(x) 
% } \left\{  \E_{P_\eps} [Y  W \mid A=1,X=x] \colon \E_{P_\eps}[ W \mid A=1,X=x] = 1, \forall x; \;\; \alpha(x) \leq W(x,y)\leq  \beta(x), \forall  y     \right\} %\\
%  % \mathcal{U}(x) &= \{ W(x,y) \colon   \E_{P_\eps}[ W \mid A=1,X=x] = 1; \;\; \alpha(x) \leq W(x,y)\leq  \beta(x), \forall  y \} 
% \end{align*} 
% We call this the closed-form representation of the optimal value with the ``oracle" optimal primal function, $\fnl(P),$ which for example can be obtained by analysis. 
\end{example}
}
% \vspace{-5pt}
% \new{
 Let $(W^*(x,y), \lambda^*(x))$ denote the primal-dual optimal pair conditionally on $X=x$. Define $\tau = \Lambda /(1+\Lambda)$. By leveraging recent closed-form characterizations of the conditionally sharp MSM solution \citep{dorn2021doubly}, we can identify that the minimizing $\lambda$ dual variable is the conditional quantile, and the adversarial weight function is a threshold: 
\begin{align*} %\SwapAboveDisplaySkip
\lambda^*_a(x) &=   Q^{\tau}(y\mid x,a), \text{ where } Q^{\tau}(y\mid x,a) \coloneqq  \inf_{z} \{ z \colon 
        P^{b}( Y \leq z \mid x,a) 
        \geq \tau \}.\\
        % W^* &= \alpha(x) \indic{Y \leq  Q^{\tau}_t(x,a)} + \beta(x) \indic{Y \geq Q^{\tau}_t(x,a)} 
                W^*_a(x,y)&= \alpha(x) \indic{y \leq  \lambda^*_a} + \beta(x) \indic{y \geq \lambda^*_a}. 
\end{align*} 
This corresponds to the closed-form notion defined in \Cref{def-oraclefnl}. 
% since the ``statistician's oracle" determines the functional form of the solution, which is also perturbed:%\useshortskip
% \begin{align*}
% \fnl(P) = \E[ Y   W^*(X,Y) \mid A=1 ], \qquad 
% \fnl(P_\eps) =\E_{P_\eps}[ Y   W^*_\eps(X,Y) \mid A=1 ]
% %= \E[  \alpha(X) (Y - \lambda^*_a(X) )_{-}  + \beta(X) (Y -\lambda^*_a(X))_+ \mid A=1]
% \end{align*}
% }
% and the dual
% \begin{align*}
%     = & \min_{\lambda \in \mathbb{R}} \left\{ 
% \left(\alpha(x)\E[ (Y-\lambda )_-\mid  A=1,X=x])+ \beta(x) \E[(Y -\lambda)_+ \mid A=1,X=x]\right) + \lambda 
%     \right\} && \text{(Dual)}\end{align*} 
% The robust estimate, under the conditionally sharp restriction, can be written as the expectation of the conditional optimal solution: 
% $$ \sup \{ \E[Y W\indic{A=1}]\colon W \in \mathcal{U}_{\op{MSM}}] \} = \E[g(X)] $$
% \vspace{-5pt}

\new{
% \paragraph{Potential benefits of automating the MSM}
We discuss some potential benefits of ``automating" sensitivity models. Recent work studies custom constraints or reformulations that might be more appropriate for one setting or another. \cite{kallus2021minimax,jin2022sensitivity,huang2022variance} provide such generalizations, such as replacing the $L_\infty$ constraint of the MSM with $L_1,L_2,$ f-divergences, etc. Recent work has obtained state-of-the-art performance for these and related estimators by deriving influence functions \citep{yadlowsky2018bounds,dorn2021doubly,scharfstein2021semiparametric,bonvini2022sensitivity,zhang2022sharp}. 
}

\new{
So far, to the best of our knowledge, most of these arguments study closed-form solutions for the optimization problem. Indeed, specializing to the specific choice of constraints is first-best when it matches the analyst's desiderata. But if it does not match, general arguments for debiased estimation for general optimization problems can improve expressivity. %Further, interactive procedures that query analysts sequentially for revealed preferences also provide use cases for automatic debiasing. 
\rev{For example, an analyst might impose a global budget on deviations from the nominal weights,
$\mathbb{E}\!\left[|W_1(X,Y)-1| \mid A=1\right] \le B$,
which limits the total amount of unobserved confounding across the population and couples the optimization across values of $X$, or smoothness constraints $\mathbb{E}\!\left[|W_1(X,Y)-W_1(X',Y)|\right] \le L \|X-X'\|$.}
Computational ways of deriving estimators for minor variations of a class of functionals could also enable estimation in interactive settings.
}
\paragraph{Influence functions via generic arguments.}
In the next results, we obtain a nonparametric influence function via generic optimization-based arguments that could extend to other variants, then we show finite-difference approximation. We start with the analytical influence function. % We illustrate how our more general argument recovers the statistically sharper influence function obtained previously via specialized arguments. 
% \vspace{-5pt}
% \paragraph{Analytical Gateaux derivative of the closed-form solution}
%  We briefly outline the argument of deriving an analytical derivative of $   \fnl(P),$ additionally with respect to dependence of the functional $W^*(X,Y)$, which will highlight how these two perspectives and derivations on the same functional are different. 
 % After an interchange of differentiation and integration this essentially follows from the product rule. 
 % When we also seek orthogonality with respect to the nuisance function $W^*(X,Y),$ i.e. the optimal basis, we compute the product rule with an additional term (perturbing $W^*_\eps(X,Y)$). Note that derivations for the influence function in the stochastic optimization literature omit this last term. 
\new{
\begin{theorem}[Nonparametric influence function for sensitivity analysis]\label{proposition-if-closedform} Dropping treatment subscripts for brevity, perturb towards ${o=(X_i,A_i,Y_i)}$:
\begin{align*}%\SwapAboveDisplaySkip
&  \frac{\intd}{\intd \eps}  
\{\fnl(P_\eps) \}\vert_{\eps = 0} =  \frac{\indic{A_i=1}({Y_i}W^*({X_i},{Y_i}) - \E_P[YW^* \mid A=1,{X_i}])}{p(A=1\mid {X_i})} + g(X_i)\\
&\qquad \qquad \qquad \qquad   + \lambda(X_i)  \cdot \frac{1-\alpha}{1-\tau}\indic{A_i=1} \cdot { (\indic{{Y_i} \leq \lambda(X_i)} - \tau)} - \fnl(P)
\end{align*}
\end{theorem}
%  We work with the oracle formulation of the dual, analogously defined with respect to $\lambda^*(x)$: 
% (Note that \citep{jeong2020assessing}
We also obtain the influence function of the perturbed solution in addition to an AIPW-like term. 
}
% \flag{double check tildes everywhere or nah?}
% \new{
% % \textit{The finite difference.}
% % Let the perturbed optimal solution, denoted as $W^*_\eps(X,Y)$, achieve optimality for the perturbed optimization problem $g_\eps(x)$. 
% % , but it only requires \textit{functional evaluations}, where we obtain a \textit{numerical equivalence} in value of the functional under optimization and closed-form representations $\fnl^{opt}(P)=\fnl(P),$ for all distributions $P$. Therefore, because of equivalence of functional evaluations, 
%  % analyze error relative to the analytical IF of the closed-form solution. 
% }
% \new{
% The perturbed functional is 
% \begin{align*}%\SwapAboveDisplaySkip
% &\Psi(P_\eps)= 
%   \E_{P_\eps} [Y  W_\eps \mid A=1], \text{ where } \\
%   &W_\eps(X,Y) \in \underset{W(X,Y)}{\arg\sup} \left\{ 
%   \E_{P_\eps}[Y  W \mid A=1,X] \colon \E_{P_\eps}[ W \mid A=1,X] = 1, \forall x; \;\; \alpha(x) \leq W(x,y)\leq  \beta(x), \forall  y
%     \right\} 
% \end{align*}
%      }
\rev{Next we study finite-difference approximation. Finite differences evaluates:
\begin{align*}%\SwapAboveDisplaySkip
    &\eps^{-1}(\fnl(P_\eps) - \fnl(P)) 
    \\
    &
   \qquad  =  
\E_{P_\eps}[ \E_{P_\eps}[YW^*_\eps(X,Y)\mid A=1,X]\mid A=1] - \E_P[ \E_P[YW^*(X,Y)\mid A=1,X]\mid A=1], \\
&\text{where }W_\eps^*(x,y) \in \underset{\alpha(x) \leq W(x,y)\leq  \beta(x), \forall  y}{\arg\sup} \left\{ 
  \E_{P_\eps}[Y  W \mid A=1,X=x] \colon \E_{P_\eps}[ W \mid A=1,X=x] = 1
    \right\} 
\end{align*}
One way to quickly verify stability of the approximation error is via the product rule from our analytical framework.}
\new{Our next result characterizes the finite-difference approximation.
% The finite-difference analogue of the product rule in \cref{eqn-sensitivity-productruledecomp-2} is: 
% \begin{equation}\label{eqn-sensitivityanalysis-productrulefd-optsolution} \Delta \left\{ (y W(x,y))  p(y\mid A=1,x)   p (x)\right\}  =\Delta \left\{ (y W(x,y))\right\} \cdot  p(y\mid A=1,x) p (x)  + (y W_\eps(x,y)) \Delta \left\{  p(y\mid A=1,x)  p (x) \right\} 
% \end{equation} 
\begin{theorem}[Finite-difference approximation for sensitivity analysis.]\label{prop-sensanalysis-finitediff}
Perturb in the direction of $o=(X_i,A_i,Y_i).$
The finite-difference approximation is: 
\begin{align*}%\SwapAboveDisplaySkip
\eps^{-1}(\fnl(P_\eps) - \fnl(P)) 
% \\
% &
&=\frac{\indic{A_i=1}(Y_iW^*  - \E_P[YW^* \mid A=1,X_i])}{p_\eps(A=1\mid X_i)} +  g_\eps(X_i) \\
&\qquad+  \lambda(X_i)  \cdot \frac{1-\alpha}{1-\tau}\indic{A_i=1} \cdot { (\indic{{Y_i} \leq \lambda(X_i)} - \tau)} - \fnl(P) + O(\eps) %+ O(\hsmoothdelta^\beta)
\\
& =   \frac{\intd}{\intd \eps}  
\{\fnl(P_\eps) \}\vert_{\eps = 0} + O(\eps)
\end{align*} 
\end{theorem}
}
\rev{
We can use the product rule from \Cref{prop-prodquotchainrule} to deduce where the finite-difference approximation stability arises. 
\paragraph{Applying the framework: sensitivity analysis.}
$$
\begin{aligned}
&\eps^{-1}(\fnl(P_\eps)-\fnl(P))=
% \\&
\textstyle\eps^{-1}
\left(\int y W_\epsilon^* p_\epsilon(y \mid A=1, x) p_\epsilon(x) \mathrm{d} x-\int y W^* p(y \mid A=1, x) p(x) \mathrm{d} x\right)
\\
& =\textstyle
\eps^{-1}\int y W_\epsilon^*( p_\epsilon(y \mid A=1, x) p_\epsilon(x)
- p(y \mid A=1, x) p(x)) \mathrm{d} x \\
& \qquad \textstyle+ \eps^{-1} \int y (W_\epsilon^*-W^*)p(y \mid A=1, x) p(x)\mathrm{d} x 
\end{aligned}
$$
The first term of the above generates the AIPW-like term with integrand $y W^*(x,y)$. The second term arises from perturbing the optimization solution, and generates the quantile perturbation term $\lambda\left(X_i\right) \cdot \tau^{-1}(1-\alpha) \mathbb{I}\left[A_i=1\right] \cdot\left(\mathbb{I}\left[Y_i \geq \lambda\left(X_i\right)\right]-\tau\right)$, which we show is stable via an implicit function argument. The approximation stability result of \Cref{prop-sensanalysis-finitediff} follows by showing that the remainder term $\int y (W_\epsilon^*-W^*)(p_\epsilon(y \mid A=1, x) p_\epsilon(x)
-p(y \mid A=1, x) p(x))\mathrm{d} x $ is ultimately higher order.
}
\begin{remark}[Comparison to other frameworks for debiasing optimization or other influence function derivations.]
    Prior works have investigated influence functions for optimization as a reliability assessment \citep{shapiro2021lectures,dupavcova1990stability}. \citet{gupta2021debiasing} debiases non-causal stochastic optimization, but with uncertain objective coefficients only such that Danskin's theorem applies. \cite{olma2021nonparametric} obtains a Neyman-orthogonal adjustment for ``truncated conditional expectation" of the form $m(\tau, x)=\frac{1}{\tau} \mathbb{E}[Y \mathbb{I}\{Y \leq q_\tau( X)\} \mid X=x]$, not necessarily with causal structure. %The structure of their argument could be applied to the coherent-risk measure functional as well.  
They derive a Neyman-orthogonal adjustment with respect to the quantile nuisance function $q_\tau( X)$. \cite{jeong2020assessing} study a worst-case treatment effect functional that reduces to a coherent measure of risk. Their argument studies an influence function with Danskin's theorem. \rev{However, in their influence function analysis, they assume the quantile function is effectively known, i.e. their influence function analysis excludes the perturbation effects on the quantile function. }

\end{remark}

\section{Empirical Validation}\label{sec-application} 
% \az{to edit} 
\subsection{$(\eps,\lambda)$ plot (AIPW)}As an empirical illustration, we first present an $(\eps,\lambda)$ plot, ranging over the finite-difference approximation error $\eps$ and the smoothing kernel bandwidth $\lambda$, of the final numerical approximation error. This type of plot was proposed in \citet{carone2018toward} as a tool for practitioners to choose the approximation parameters. We conduct a validation of our decomposition and theoretical characterization. Our data-generating process includes a piecewise linear outcome model (such that kernel regression is misspecified). We conduct plug-in evaluation of the mean potential outcome with kernel density estimates. In \Cref{fig:1a} we include an $(\epsilon,\lambda)$-plot for the simple case of AIPW (see \cite{carone2018toward} for more discussion and examples). We consider a one-dimensional case with uniformly distributed $X$, piecewise-linear $Y$, and smooth propensity scores that are logistic in $\sin(X)$. We use $n=500$ and fix the bandwidth $h=0.05.$ We use colors to denote magnitude of the mean absolute error, included in text on the heatmap. We study the estimation of a mean under missingness, $\E[Y(1)]$; results extend analogously to the ATE, $\E[Y(1)-Y(0)]$ by symmetry of treatment and control. By contrast, \Cref{fig:12} illustrates the estimation error of various strategies with the comparable kernel-based estimates (DM is regression adjustment).  Note that even in this simple setting AIPW offers benefits and sample efficiency relative to DM. \rev{\Cref{fig:1a} shows that the finite-difference approximation error can be quite finely controlled, to the extent that differences in approximation would not show up in our error comparison in \Cref{fig:12}. Instead, in our next subsection we conduct a more thorough empirical comparison for the sensitivity analysis examples. }
%We then apply the tools that we derived in the previous section to analyze more complicated settings. We study longitudinal data (dynamic treatment regime) and the policy linear program for an infinite-horizon Markov decision with linear constraints, with attention to how different functionals have different interactions between approximation error and statistical properties.  
\begin{figure}
     \centering
     \begin{subfigure}[b]{0.47\textwidth}
         \centering
         \includegraphics[height=0.65\textwidth]{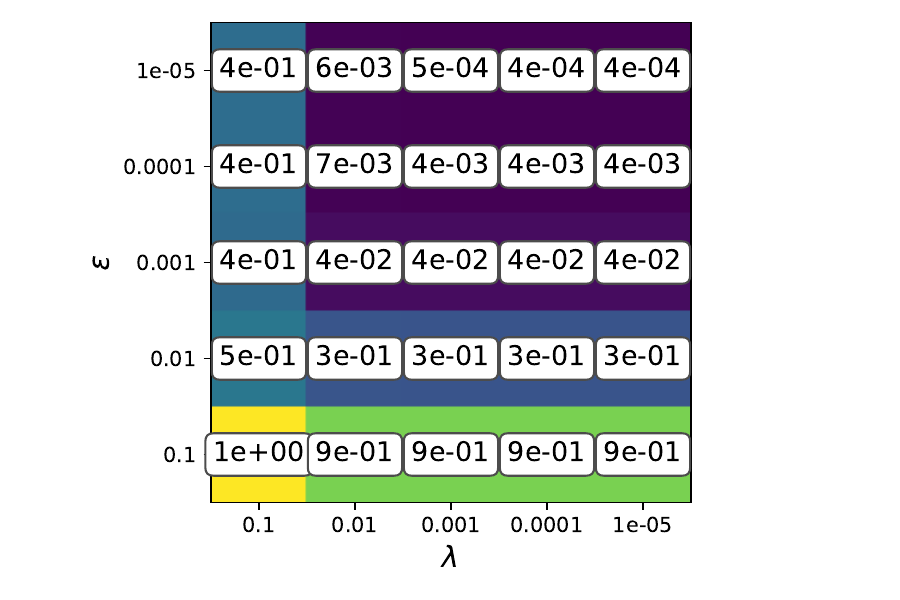}
         \caption{\label{fig:1a}}
     \end{subfigure}
     \hfill
     \begin{subfigure}[b]{0.47\textwidth}
         \centering
         \includegraphics[height=0.65\textwidth]{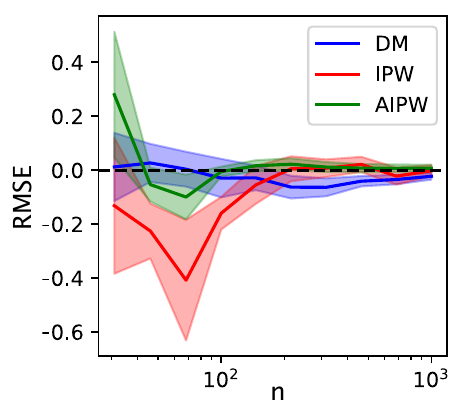}
         \caption{\label{fig:12}}
     \end{subfigure}
     \caption{(a) Epsilon-lambda plot for mean potential outcome: MAE in approximation of one-step adjustment. (b) Comparison in error in estimation of $\E[Y(1)]$ over $n$ \rev{for AIPW, IPW, DM methods.}}
\end{figure}
\rev{
\subsection{Marginal sensitivity analysis}
Next we include an empirical study implementing finite differences for the sensitivity analysis functional of \Cref{ex-sensitivityanalysis}.
}
\rev{
\paragraph{Simplified $\lambda$-smoothing for the marginalized functional} We make a few simplifications that reduce dependence of the approach on $\lambda$-smoothing: first, given what we know about the functional as a marginalization, in the one-step estimator, we evaluate \textit{un-smoothed} perturbations with respect to $p(X)$, i.e. $p_\eps(x)=(1-\eps)p(x)+\delta(x)$. (This could still admit custom constraints in the $X$-conditional uncertainty set). Second, we use a uniform histogram kernel with bandwidth $\lambda$ for the remaining perturbation in $Y$. %We fix $\lambda=2\epsilon$ throughout. 
\paragraph{Local stability in $\epsilon$}
Note that finite-difference approximations of optimization-based functionals are piecewise smooth. Under local stability of the active basis, linear optimization functionals have locally linear derivatives so that the finite-difference error from secant approximation is linear in $\eps.$ Given this, we omit an $\epsilon,\lambda$ plot for the optimization functional since the error is well-behaved in $\eps$ and we have simplified the $\lambda$-approximation as above. 
\paragraph{Implementation details}
Next, in order to implement the stochastic optimization of \Cref{ex-sensitivityanalysis} on finite samples, we use numerical integration for plug-in evaluation of the functional (integration in the objective and constraint), and introduce a feasibility relaxation to ensure feasibility at all sample sizes. 
% Although the sensitivity analysis functional of \Cref{ex-sensitivityanalysis} is stated with expectations, and is a doubly-infinite linear program in the population, we calculate this in finite samples using numerical integration with respect to plug-in density estimates from $\tilde{P}$. 
% We computationally approximated the sensitivity analysis functional of \Cref{ex-sensitivityanalysis} using numerical integration of plug-in density estimates from $\tilde{P}$. 
We expect that numerical integration is ultimately more compatible with $\lambda$-smoothed delta function perturbations instead of typical sample-average approximation \citep{shapiro2021lectures}. %We use numerical integration over $M$ grid points: the conditional density $f(y \mid x, A=1)$ is estimated by a product-kernel KDE and converted to quadrature weights $q_j = \hat{f}(y_j \mid x) \Delta y$ on a uniform $Y$-grid. 
We use a grid of size $M$ for numerical integration. Let $y_1,\ldots,y_M$ be the gridded outcome values, $\Delta y$ the grid width, $\tilde{p}_{jx}^\Delta = \tilde{p}(y_j \mid x) \Delta y$ the quadrature weights for numerical integration, and recall that $\alpha(x)$, $\beta(x)$ are the sensitivity bounds on the importance weight $w$ (derived from the marginal sensitivity model). We solve
\begin{align*}
g(x)=\max_{w_1,\ldots,w_M \geq 0,\,t \geq 0} \left\{ \sum_{j=1}^{M} w_j \, y_j\;\; 
\text{s.t. } \;\; \sum_{j=1}^{M} w_j = 1;\;\alpha(x) \, \tilde{p}_{jx}^\Delta \, t \le w_j \le \beta(x) \, \tilde{p}_{jx}^\Delta \, t, \;\; j = 1,\ldots,M \right\}
 % w_j \ge 0, \quad t \ge 0.
% \max_{w_1,\ldots,w_M,\,t} \quad & \sum_{j=1}^{M} w_j \, y_j
% \label{eq:lp-obj}
% \\
% \text{s.t.} \quad & \sum_{j=1}^{M} w_j = 1,
% \label{eq:lp-sum}
% \\
% & \alpha(x) \, \tilde{p}_j \, t \le w_j \le \beta(x) \, \tilde{p}_j \, t, \qquad j = 1,\ldots,M,
% \label{eq:lp-bounds}
% \\
% & w_j \ge 0, \quad t \ge 0.
% \label{eq:lp-nonneg}
\end{align*}
The scaling variable $t$ yields a homogeneous formulation which ensures finite-sample feasibility and asymptotically-vanishing bias. 
}

\begin{table}[ht]
\centering

\caption{\rev{Sensitivity analysis model, bias (estimate $-$ $\Psi^*$) with standard errors (SE $= \mathrm{std}/\sqrt{n_{\mathrm{mc}}}$). $n_{\mathrm{mc}}=100$, $\Psi^*=2.4782$, less-smooth outcome.}}
\label{tab:one-step-bias}
\rev{
\begin{tabular}{lrrr}
\toprule
$n$ & Baseline & Binning (Perturb.) & Analytical (KDE-total) 
% & Analytical (KDE-env.)
\\
\midrule
75 & $+0.0610 \pm 0.0227$ & $\mathbf{+0.0156 \pm 0.0190}$& $-0.0750 \pm 0.0234$% & $-0.1155 \pm 0.0234$
\\
100 & $+0.0684 \pm 0.0192$ & $\mathbf{+0.0206 \pm 0.0166}$& $-0.0496 \pm 0.0185$ %& $-0.0857 \pm 0.0184$ 
\\
150 & $+0.0469 \pm 0.0135$ & $\mathbf{+0.0095 \pm 0.0120}$& $-0.0575 \pm 0.0144$ %& $-0.0873 \pm 0.0145$ 
\\
200 & $+0.0328 \pm 0.0135$ & $\mathbf{-0.0068 \pm 0.0118}$& $-0.0627 \pm 0.0127$% & $-0.0965 \pm 0.0125$ 
\\
\bottomrule
\end{tabular}
}
\end{table}

\begin{table}[ht]
\centering
\caption{\rev{Sensitivity analysis model, RMSE vs.\ ground truth with standard errors. Same setup as Table~\ref{tab:one-step-bias}.}}
\label{tab:one-step-rmse}
\rev{
\begin{tabular}{lrrr}
\toprule
$n$ & Baseline & Binning (Perturb.) & Analytical (KDE-total) %& Analytical (KDE-env.) 
\\
\midrule
75 & $0.2354 \pm 0.0181$ & $\mathbf{0.1905 \pm 0.0131}$ & $0.2457 \pm 0.0161$ 
% & $0.2612 \pm 0.0170$ 
\\
100 & $0.2041 \pm 0.0164$ & $\mathbf{0.1671 \pm 0.0121}$ & $0.1916 \pm 0.0132$ 
% & $0.2026 \pm 0.0136$ 
\\
150 & $0.1427 \pm 0.0108$ & $\mathbf{0.1200 \pm 0.0099}$& $0.1548 \pm 0.0095$ %& $0.1689 \pm 0.0101$ 
\\
200 & $0.1392 \pm 0.0097$ & $\mathbf{0.1183 \pm 0.0074}$ & $0.1418 \pm 0.0093$ %& $0.1580 \pm 0.0100$ 
\\
\bottomrule
\end{tabular}
}
\end{table}

% One-step comparison: bias and RMSE (mean $\pm$ SE)
% $n_{mc}=50$, $\Psi^*=2.2486$, delta='uniform_binning', $\lambda/\varepsilon=0.2$, outcome\_smoothness='linear', bandwidth\_scale=1.0. Use \usepackage{booktabs} for \toprule/\midrule/\bottomrule.

\begin{table}[ht]
\centering
\caption{\rev{Sensitivity analysis example, $d=10$. Bias (estimate $-$ $\Psi^*$) and RMSE vs ground truth (mean $\pm$ SE, SE $= \mathrm{std}/\sqrt{n_{\mathrm{mc}}}$). Gaussian Mixture Models for $\tilde{P}$ throughout.}}
\label{tab:one-step-bias-rmse-10d}
\rev{
\begin{tabular}{lrrrr}
\toprule
$n$ & \multicolumn{2}{c}{Bias} & \multicolumn{2}{c}{RMSE} \\
\cmidrule(lr){2-3} \cmidrule(lr){4-5}
 & Baseline & Binning~(Perturb.) & Baseline & Binning~(Perturb.) \\
\midrule
100 & $-0.1320 \pm 0.0215$ & $\mathbf{-0.1206 \pm 0.0199}$ & $0.2013 \pm 0.0176$ & $\mathbf{0.1852 \pm 0.0188}$ \\
200 & $-0.1319 \pm 0.0131$ & $\mathbf{-0.0813 \pm 0.0130}$ & $0.1610 \pm 0.0121$ & $\mathbf{0.1227 \pm 0.0098}$ \\
300 & $-0.1153 \pm 0.0103$ & $\mathbf{-0.0631 \pm 0.0112}$ & $0.1364 \pm 0.0088$ & $\mathbf{0.1011 \pm 0.0083}$ \\
400 & $-0.1062 \pm 0.0106$ & $\mathbf{-0.0525 \pm 0.0100}$ & $0.1298 \pm 0.0116$ & $\mathbf{0.0880 \pm 0.0102}$ \\
\bottomrule
\end{tabular}
}
\end{table}
\rev{
\paragraph{Results - small experiments.} In \Cref{tab:one-step-bias,tab:one-step-rmse} we include tables describing the results by bias and root-mean-squared error (RMSE). We first set-up a simple one-dimensional example, see \Cref{apx-experimental-details} for more details on the data-generating process. We choose $\Lambda = 1.65$. We fix $\epsilon = 0.01$.
}
%The target parameter is the marginal mean potential outcome under treatment, $\Psi = \E[Y(1)]$,
% evaluated under a marginal sensitivity model with $\Lambda = 1.65$.} 
\rev{We compare our approach (Binning (Perturb.)) to plug-in estimation (Baseline) and an implementation of the analytical influence function (Analytical (KDE-total)).}
\rev{We find finite-sample improvements where our perturbation method achieves consistent bias improvements upon the baseline approach. Even in our small example, we also see consistent improvements upon the analytical IF: we expect this is due to the variance from the inverse propensity scores in finite samples. One benefit of the perturbations approach is that it maintains linear-programming feasibility of all evaluations, whereas evaluating the analytical IF can violate such constraints. This is analogous to potential benefits from TMLE when outcomes are in a bounded range.}

\rev{
For probability density estimation, we use product Gaussian kernels with bandwidths set by Silverman's rule of thumb. We also compare to the analytical influence function, the total derivative in \Cref{proposition-if-closedform} (called $\textrm{Analytical (KDE-total)}$). %As an ablation and comparison to prior literature, we also compare to the influence function derived by using the envelope theorem and not perturbing the quantile function, \Cref{thm-analytical-optif} (called $\textrm{Analytical (KDE-env.)}$). These use the same KDE-based plug-in estimation for the outcome model, but 
We strengthen the comparison by using the true logistic regression parametrization for the propensity score. The optimal $W^*(x,y)$ is evaluated at $Y_i$ via the linear programming solution's value on the nearest numerical integration grid point. We set $\lambda(X_i)$ to the $\tau$-quantile of the estimated conditional distribution. %The optimal importance weights $W^*(x,y)$ from the linear program are evaluated at the observed $Y_i$ either by interpolation of the LP solution on the $Y$-grid or, in an oracle variant, by the closed-form two-level solution $W^* = \alpha$ if $Y \le \lambda(x)$ and $W^* = \beta$ otherwise, with $\lambda(x)$ the $(1-\tau)$-quantile of the (estimated or true) conditional distribution. Term~3 of the influence function involves the quantile $\lambda(X_i)$; we set it to the $(1-\tau)$-quantile of the KDE quadrature weights $\tilde{p}$ on the grid (CDF-based), or optionally from the LP solution (LP-CDF). 
% No separate quantile perturbation is used for the main results; the perturbation is in the direction of the empirical delta $\delta_{o_i}$ for the one-step (binning) estimator, and the analytical IF uses the same KDE and LP outputs without an extra quantile tilt.
% \paragraph{Optimization problem solved in practice.}
% The robust conditional expectation $g(x) = \sup_{w \in \mathcal{W}(x)} \mathbb{E}[w(Y) Y \mid X=x, A=1]$ is implemented by solving a finite-dimensional linear program over the $M$ grid points. 
% The bounds $\alpha(x)$, $\beta(x)$ are computed from the estimated propensity $e(x)$ and sensitivity parameter $\Gamma = e^{\texttt{log\_gamma}}$ via
% \begin{equation}
% \alpha(x) = \frac{1}{p_{\mathrm{hi}}(x)}, \qquad \beta(x) = \frac{1}{p_{\mathrm{lo}}(x)},
% \qquad
% p_{\mathrm{hi}} = \frac{\Gamma e}{1-e+\Gamma e}, \quad p_{\mathrm{lo}} = \frac{e/\Gamma}{1-e+e/\Gamma}.
% \end{equation}
% The marginal estimator is $\hat{\Psi} = (1/n_{\mathrm{tr}}) \sum_{j: A_j=1} \hat{g}(X_j)$, and the one-step (binning) estimator is $\hat{\Psi}_{\mathrm{corrected}} = \hat{\Psi} + \bar{\phi}$ with $\bar{\phi} = (1/n) \sum_i \hat{\phi}_i$ and $\hat{\phi}_i = (1/\varepsilon)(\hat{\Psi}_\varepsilon^i - \hat{\Psi})$, where $\hat{\Psi}_\varepsilon^i$ is the marginal computed under the $\varepsilon$-perturbed distribution with the $\lambda$-bin delta at observation $i$.
}
 \rev{\paragraph{Scaling up} Next we scale up our experiments to a more moderate dimension $d=10$ with a nonlinear single-index model (see \Cref{apx-experimental-details} for details). Instead of kernel density estimates, we use Gaussian mixture models (GMMs) for plug-in density estimates. 
 \Cref{tab:one-step-bias-rmse-10d} includes the results. Note that under the parametrization of GMMs, the plug-in density could be misspecified or sensitive to hyperparameter tuning - as a result some asymptotic bias remains in \Cref{tab:one-step-bias-rmse-10d}. We also investigated other black-box ML conditional density estimation approaches \citep{izbicki2017converting,sugiyama2010conditional}, but find that these do worse than the simpler GMM. Therefore, performance of our approach is also highly dependent on the reliability of conditional density estimation, which remains an open problem in general. }
% -----------------------------------------------------------------------------
% Tables: one-step comparison, bias and RMSE (lambda_bins, n_mc=100 example)
% -----------------------------------------------------------------------------
% Caption note: $n_{\mathrm{mc}}=100$, $\Psi^*=2.4782$, delta='lambda_bins',
% $\lambda/\varepsilon=0.2$, outcome\_smoothness='less\_smooth', bandwidth\_scale=1.0.

% \vspace{-5pt}

% \subsection{Mean potential outcome}

% \vspace{-5pt}

\section{Conclusions}

We have presented a constructive algorithm that approximates Gateaux derivatives by finite differencing, with a focus on the statistical functionals used in causal inference and on stochastic optimization. 
    \rev{While the AIPW functional admits weaker requirements for computational approximation (\Cref{prop-rate-double-robustness}), this depends on the structure of the functional --- \Cref{prop-mdp-finitedifferenceapprox} does not since the functional includes matrix inversion.}
% Our analysis of different functionals surfaced fine-grained differences in opportunities for weakening requirements for computational approximation: \Cref{prop-rate-double-robustness} and \Cref{prop-mdp-finitedifferenceapprox} have different, estimand-dependent implications (re: Q2). 
We also studied examples of causal estimands of recent interest in the literature that can be written as linear stochastic optimization functionals, and provided generic arguments for the nonparametric influence function, inspired by perturbation analysis of linear programs (re: Q3). %Obtaining orthogonality adjustments also requires perturbing the solution with respect to probability distribution primitives, unlike typical envelope theorems. 
Follow-up directions to this work include reducing unfavorable approximation rates with regards to dimension dependence due to the smoothed perturbation, exploring adaptive methods that do not require identification of the plug-in functional, and extensive further empirical comparisons. 

\if\forjournal 0
\paragraph{Acknowledgements.} Angela Zhou gratefully acknowledges support from the Foundations of Data Science Institute and the Simons Institute's Program on Causality. Part of this work was done while the author was visiting the Simons Institute for the Theory of Computing. Yixin Wang acknowledges National Science Foundation Grant NSF-CHE-2231174.  Michael Jordan acknowledges support from the Vannevar Bush Faculty Fellowship program
under grant number N00014-21-1-2941 and a European Research Council Synergy Grant.
\fi 
% \clearpage 

% \bibliographystyle{abbrvnat} 
\bibliography{auto-semipara}

% Acknowledgements and Disclosure of Funding should go at the end, before appendices and references

% \acks{}

\appendix

\section{Overview and background for influence functions}\label{apx-if}
For a broad overview of influence functions and their role in causal inference, see \cite{tsiatis2006semiparametric,hines2022demystifying,kennedy2022semiparametric}. 
% \begin{equation*}
%     \Psi_n^{os} - \Psi= \Psi(\hat{\mathbb{P}}) + \Pn (\phi(\obs; \Pn)) - \fnl(\bP)= (\Pn-\bP)(\phi(O; \mathbb P)) + (\Pn - \bP) (\phi(\obs; \Pn) - \phi(\obs; \bP)) + R_2(\Pn,\bP)
% \end{equation*}
% Typical analyses establish statistical properties via the first term (sample average term) while the second term is the \textit{empirical process term} (argued to be small via Donsker conditions and sample-splitting) and the last term is the second-order remainder term, typically requiring ad-hoc arguments.

% Computing Gateaux derivatives (analytical, or empirical) is useful for statistical estimation because they can be used to derive estimators, i.e. by solving the estimating equation or via the one-step estimator. 
We focus on functionals that admit an asymptotically linear representation:
$$\begin{aligned} \Psi(\tP)-\Psi(P)
% &=\int \phi_{P_{1}}(u) d\left(P_{1}-P\right)(u)+R_2\left(P_{1}, P\right)
% \\ 
&=-\int \phi(u;\tP) d P(u)+R_2\left(\tP, P\right), \end{aligned}$$
where the influence function, $\phi(u;\tP)$, captures the first-order bias in the functional.  That is, we focus on functionals that are smooth enough to admit a \textit{von Mises expansion} with a mean-zero, finite-variance influence function and a second-order remainder term $R_2\left(P_{1}, P\right)$. 
\section{Proofs for \Cref{sec-aipw}}

 \subsubsection{Supplementary lemmas}

 \begin{lemma}[Intermediate decomposition]\label{lemma-numerical-derivative-decomposition}
\begin{align*}&\frac{1}{\epsilon} 
      \int  \int y  \left( 
  \frac{\tepspnloweryax{j}    }{
        \tepspnalowerx{j} }
         -   \frac{ \tpnloweryax{j} }{ \tpnalowerx{j}  }\right) \tpnlowerx \; \mathrm{d}y \; \mathrm{d}x \\
                  &\qquad =  \int
    \frac{\tpnlowerx{j} }{   \tepspnalowerx{j}} \indic{A_i = 1}\smoothdel{(x_i)}(x)\left\{
        \left( \int y \; \smoothdel{y_i}(y) \mathrm{d}y \right) - 
  \E_{\td{p}}[Y\mid A=1,x]
        \right\}\mathrm{d}x.
\end{align*} 
\end{lemma}

\begin{proposition}[Gateaux derivative of probability density representation]\label{prop-decomposition-gateaux-derivative-pdf-rep}
  Consider perturbations in the direction of $o_i=(x_i,a_i,y_i)$. Let $\td{p}(x), \td{p}(A=1,x), \td{p}(y,A=1,x)$ denote kernel density estimates. Consider the mean potential outcome functional $\fnl(P)=\E[Y(1)]$, then: 
\begin{align*} \frac{\Psi(\td{P}_{\eps_\ptbdir} ) - \Psi(\td{P} )}{\epsilon} 
                   &=    %(1-\eps) 
                   \int 
 \frac{\tilde{p}(x) }{   \tilde{p}_\eps (A=1,x) } \indic{a_i = 1}\smoothdel{x_i}(x)\left\{
        \left( \int y \; \smoothdel{y_i}(y) \mathrm{d}y \right) - 
  \E_{\tP}[Y\mid A=1,x]
        \right\} \;\mathrm{d}x
        \\&\qquad 
+\left( \td{\mathbb{E}}_{\tP}[Y\mid A=1,X=x_{\ptbdir}] - \Psi(\td{P}) \right). \\
& \qquad + \eps 
 \int (\eps)^{-1}
 \frac{ \indic{a_i = 1} }{   \tilde{p}_\eps (A=1,x) }\smoothdel{x_i}(x)\left\{
        \left( \int y \; \smoothdel{y_i}(y) \mathrm{d}y \right) - 
  \E_{\tP}[Y\mid A=1,x]
        \right\} \{ \smoothdel{x_{\ptbdir}} - \td{p}(x)  \}\;\mathrm{d}x
\end{align*} 
\end{proposition}

\subsubsection{Proofs} 

\begin{proof}[Proof of \Cref{prop-prodquotchainrule}]
\rev{Product rule: 
    \begin{align*}
        &\epsilon^{-1}\left(f_\epsilon(u) g_\epsilon(v)-f(u) g(v)\right)\\
        &= \epsilon^{-1}\left(f_\epsilon(u) g_\epsilon(v)- f(u)g_\epsilon(v) + f(u)g_\epsilon(v) -f(u) g(v)\right)
             \tag{$\pm f(u)g_\epsilon(v)$} \\
       &  = f(u) \cdot \epsilon^{-1}\left(g_\epsilon(v)-g(v)\right)+g_\epsilon(v) \cdot \epsilon^{-1}\left(f_\epsilon(u)-f(u)\right)
    \end{align*}
Quotient rule: 
\begin{align*}
    & \epsilon^{-1} \left( \frac{f_\epsilon(u)}{g_\epsilon(v)} - \frac{f(u)}{g(v)} \right) \\
    &= \epsilon^{-1} \left( \frac{f_\epsilon(u)g(v) - f(u)g_\epsilon(v)}{g(v)g_\epsilon(v)} \right) \\    % & \text{Factor by grouping terms in the numerator:} \\
    &= \epsilon^{-1} \left( \frac{g(v)(f_\epsilon(u) - f(u)) - f(u)(g_\epsilon(v) - g(v))}{g(v)g_\epsilon(v)} \right) \tag{$\pm f(u)g(v)$}
\end{align*}
    }
%     \rev{Chain rule:
% %Let $g: \mathbb{R}^{d_v} \to \mathbb{R}$ and $f: \mathbb{R} \to \mathbb{R}$. We analyze the finite difference of the composition $(f \circ g)(v) = f(g(v))$.
% \begin{align*}
%     & \epsilon^{-1}\left(f(g_\epsilon(v)) - f(g(v))\right) \\
%     &= \epsilon^{-1} \left( \frac{f(g_\epsilon(v)) - f(g(v))}{g_\epsilon(v) - g(v)} \cdot (g_\epsilon(v) - g(v)) \right) %\tag{Multiply and divide by $g_\epsilon(v) - g(v)$}
%     \\
%     &= \left( \frac{f(g_\epsilon(v)) - f(g(v))}{g_\epsilon(v) - g(v)} \right) \cdot \left( \epsilon^{-1} (g_\epsilon(v) - g(v)) \right)
% \end{align*}
% The second term is the finite-difference approximation of the derivative of $g$ at $v$. The first term is the finite-difference approximation of the derivative of $f$ evaluated over the interval $[g(v), g_\epsilon(v)]$. As $\epsilon \to 0$, this mirrors the structure of the standard chain rule, $f'(g(v)) \cdot g'(v)$.
% }
\end{proof}
\rev{
\begin{proof}[Proof of \Cref{eq-cndlexp-fd} ]
By definition of $p_\eps$,
\[
\E_{P_\epsilon}[h(X)]
= \int h(x)p_\epsilon(x)\,dx
= (1-\epsilon)\E_P[h(X)] + \epsilon \int h(x)\smoothdel{(x_i)}(x)\,dx.
\]
Rearranging gives
\[
\frac{1}{\epsilon}\big(\E_{P_\epsilon}[h(X)]-\E_P[h(X)]\big)
=
\int h(x)\smoothdel{(x_i)}(x)\,dx - \E_P[h(X)].
\]
Since 
$\E_{P_\epsilon}[Y\mid X=x]
=
\frac{\int y\,p_\epsilon(y,x)\,dy}{p_\epsilon(x)}$, 
we obtain
\begin{align*}
\E_{P_\epsilon}[Y\mid X=x]
&=
\frac{(1-\epsilon)\int y\,p(y,x)\,dy + \epsilon \smoothdel{(x_i)}(x)\int y\,\smoothdel{(y_i)}(y)\,dy}{p_\epsilon(x)} \\
&=
\frac{(1-\epsilon)p(x)\E_P[Y\mid X=x] + \epsilon \smoothdel{(x_i)}(x)\int y\,\smoothdel{(y_i)}(y)\,dy}{p_\epsilon(x)}.
\end{align*}
Subtracting \(\E_P[Y\mid X=x]\) and simplifying,
\begin{align*}
\E_{P_\epsilon}[Y\mid X=x]-\E_P[Y\mid X=x]
&=
\frac{\epsilon \smoothdel{(x_i)}(x)}{p_\epsilon(x)}
\Big(
\int y\,\smoothdel{(y_i)}(y)\,dy - \E_P[Y\mid X=x]
\Big).
\end{align*}
Dividing by \(\epsilon\) yields the result.
\end{proof}
}
% \if\forjournal 0
% \begin{proof}[Proof of \Cref{prop-decomposition-gateaux-derivative-pdf-rep}]
% \else
\rev{\begin{proof}[Proof of \Cref{thm-one-step-sufficient-condition}]
Adding and subtracting $\Psi_{\mathrm{OS}}$:
\[
\Psi_n-\Psi(P)
=
\underbrace{(\Psi_{\mathrm{OS}}-\Psi(P))}_{= \frac{1}{n}\sum_{i=1}^n \phi(O_i)+o_p(n^{-1/2})}
+
(\Psi_n-\Psi_{\mathrm{OS}}) 
\]
% By assumption,
% \[
% \Psi_{\mathrm{OS}}-\Psi(P)
% =
% \frac{1}{n}\sum_{i=1}^n \phi(O_i)+o_p(n^{-1/2}).
% \]
% Also, since $\Psi_n$ and $\Psi_{\mathrm{OS}}$ share the same plug-in term $\Psi(\tilde P)$,
By the rate requirement \eqref{eqn-one-step-decomposition-empiricalappx-vanishes-fast},
\[
\Psi_n-\Psi_{\mathrm{OS}}
=
\frac{1}{n}\sum_{i=1}^n (\tilde{\phi}(O_i)-\phi_{\tilde{\eta}}(O_i))
=
o_p(n^{-1/2}).
\]
 Combining the two displays yields the result.
\end{proof}}
\begin{proof}{Proof of \Cref{prop-decomposition-gateaux-derivative-pdf-rep}.}
We have:
\begin{align*}
\frac{\Psi(\td{P}_{\eps_\ptbdir} ) - \Psi(\td{P} )}{\epsilon}&= \frac 1\epsilon
\int  \bces{
\int y
\left[ 
\frac{ \tepspnloweryax{j}  }{\tepspax{x} }  
  \left( (1-\eps)\td{p}({x})+ \eps \smoothdel{x_{\ptbdir}} \right) - 
\frac{ \tpnloweryax{j}  }{\tpnalowerx{j} } \td{p}(x)\right]
\mathrm{d}y
  }\mathrm{d}x\\
  & = 
  %(1-\eps)  
  \left[ 
  \int \td{p}(x)  \int y
\left(
   \frac{\tepspnloweryax{j}    }{
        \td{p}_\eps(A=1,{x}) }
         -   \frac{ \tpnloweryax{j} }{ \tpnalowerx{j}  }\right)\;\mathrm{d}y\;\mathrm{d}x
         \right] 
        \\
        &\qquad + %\eps 
       \cancel{\frac{1}{\epsilon}  } \cancel{\epsilon} 
    \int \int     y 
 \frac{ \tpnloweryax{j}  }{ \td{p}(A=1,{x})  } 
    \{ \smoothdel{x_{\ptbdir}} - \td{p}(x)  \}\mathrm{d}y\;\mathrm{d}x. \\
   &  \qquad + \cancel{\frac{1}{\epsilon}  } \cancel{\epsilon} 
    \int \int     y  
    \left\{ \frac{ \tepspnloweryax{j}  }{\tepspnalowerx{j} } 
    - \frac{ \tpnloweryax{j}  }{\tpnalowerx{j} } 
    \right\}
    \{ \smoothdel{x_{\ptbdir}} - \td{p}(x)  \}\mathrm{d}y\;\mathrm{d}x.
  \end{align*} 
We expand $\left( (1-\eps)\td{p}({x})+ \eps \smoothdel{x_{\ptbdir}} \right)$ and add/subtract  $    {\int \int     y 
 \frac{ \tpnloweryax{j}  }{ \td{p}(A=1,{x})  } 
    \{ \smoothdel{x_{\ptbdir}} - \td{p}(x)  \}\mathrm{d}y\;\mathrm{d}x}$. The first term can be expanded using the decomposition of \Cref{lemma-numerical-derivative-decomposition}. We recognize the second term as $
\left( \td{\mathbb{E}}_{\tP}[Y\mid A=1,X=x_{\ptbdir}] - \Psi(\td{P}) \right)$. 
We obtain:
\begin{align*} \frac{\Psi(\td{P}_{\eps_\ptbdir} ) - \Psi(\td{P} )}{\epsilon} 
                   &=    %(1-\eps) 
                   \int 
 \frac{\tilde{p}(x) }{   \tilde{p}_\eps (A=1,x) } \indic{a_i = 1}\smoothdel{x_i}(x)\left\{
        \left( \int y \; \smoothdel{y_i}(y) \mathrm{d}y \right) - 
  \E_{\tP}[Y\mid A=1,x]
        \right\} \;\mathrm{d}x
        \\&\qquad 
+\left( \td{\mathbb{E}}_{\tP}[Y\mid A=1,X=x_{\ptbdir}] - \Psi(\td{P}) \right). \\
& \qquad + 
 \int 
 \frac{ \indic{a_i = 1} }{   \tilde{p}_\eps (A=1,x) }\smoothdel{x_i}(x)\left\{
        \left( \int y \; \smoothdel{y_i}(y) \mathrm{d}y \right) - 
  \E_{\tP}[Y\mid A=1,x]
        \right\} \{ \smoothdel{x_{\ptbdir}} - \td{p}(x)  \}\;\mathrm{d}x
\end{align*} 
\end{proof}

\begin{proof}{Proof of \Cref{lemma-numerical-derivative-decomposition}.} 
%          \begin{align*}&\frac{1}{\epsilon} 
%       \int  \int y  \left( 
%   \frac{\tepspnloweryax{j}    }{
%         \tepspnalowerx{j} }
%          -   \frac{ \tpnloweryax{j} }{ \tpnalowerx{j}  }\right) \tpnlowerx \; \mathrm{d}y \; \mathrm{d}x \\
%                   &\qquad =  \int
%     \frac{\tpnlowerx{j} }{   \tepspnalowerx{j}} \indic{a_i = 1}\smoothdel{(x_i)}(x)\left\{
%         \left( \int y \; \smoothdel{y_i}(y) \mathrm{d}y \right) - 
%   \E_{\td{p}}[Y\mid A=1,x]
%         \right\}\mathrm{d}x.
% \end{align*} 
\rev{First consider the inner term $\frac{1}{\epsilon} 
       \int y  \left( 
  \frac{\tepspnloweryax{j}    }{
        \tepspnalowerx{j} }
         -   \frac{ \tpnloweryax{j} }{ \tpnalowerx{j}  }\right) \tpnlowerx \; \mathrm{d}y. $ 
Note that $\mu_1(x)=\int y\,\tfrac{\tpnloweryax{j}}{\tpnalowerx{j}}dy$, and $\mu_{1,\eps}(x)=\int y\,\tfrac{\tepspnloweryax{j}}{\tepspnalowerx{j}}dy.$ By applying Lemma~\ref{eq-cndlexp-fd} with the conditioning set $\{ A=1,X=x\},$ and the product kernel at $(y_i,1,x_i)$, we obtain that:
%$$\epsilon^{-1}(\mu_\eps(x)-\mu(x))=\tilde p_\eps(A=1,x)^{-1}\int (y-\mu(x))\{\tilde\delta(y,x)-p(y\mid A=1,x)\tilde\delta(x)\}dy,$$ 
% which under the product kernel at $(y_i,1,x_i)$ becomes 
$$\frac{1}{\epsilon} 
       \int y  \left( 
  \frac{\tepspnloweryax{j}    }{
        \tepspnalowerx{j} }
         -   \frac{ \tpnloweryax{j} }{ \tpnalowerx{j}  }\right) \tpnlowerx \; \mathrm{d}y = \frac{\indic{a_i=1}\smoothdel{x_i}(x)}{\tilde p_\eps(A=1,x)}
         \left(\int y\smoothdel{y_i}(y)dy-\mu(x)\right).$$ Multiplying by $\tilde p(x)$ and integrating yields the statement. 
         }
         One can also verify by direct computation. 
    %For simplicity we consider a single summand in the sum (the claim follows by applying the same argument to every summand). %Recall that ${\tilde P_\epsilon(A=1,X_j)=         (1-\eps)\tpnax{j}
         % + \epsilon \tindic{(o_{\backslash y})_{\tilde{\ptbdir}}}}$
         %and expanding the latter yields:
\begin{align*}
    % &\frac{\Psi(\tilde{P}_\epsilon) - \Psi(\tilde P)}{\epsilon} \\
   & =\int \frac{\tpnlowerx{j}}{\epsilon}  \int y  
    \left[     
    % \sum_{j}^n
   \frac{ (1-\eps) \tpnloweryax{j}  + \epsilon \smoothdel{o_\ptbdir}     }{
   \tepspnalowerx{j}
        %  (1-\eps)\tpnax{j}
        %  + \epsilon \tindic{(o_{\backslash y})_{\tilde{\ptbdir}}}
         }  
         -   \frac{ \tpnloweryax{j} }{ \tpnalowerx{j}  }
         \right] \mathrm{d}y \; \mathrm{d}x 
         \\
        &=\int  \frac{ \tpnlowerx{j}}{\epsilon} \int y
       %\sum_{j}^n
  \left[ \frac{    
   \left\{ \cancel{(1-\eps)} \tpnloweryax{j}  + \epsilon \smoothdel{o_\ptbdir} \right\}  
   \tpnalowerx{j}    }{    \tepspnalowerx{j}
    %  \left\{    (1-\eps)\tpnax{j}
        %  + \epsilon \tindic{(o_{\backslash y})_{\tilde{\ptbdir}}} \right\} 
         \tpnalowerx{j}
         }  \right.\\
         &\qquad \qquad \qquad\qquad\left.
         -   \frac{ \tpnloweryax{j}    \left\{\cancel{(1-\eps)} \tpnalowerx{j}  + \epsilon \smoothdel{(a_{\ptbdir},x_{\ptbdir})} \right\}   }{ \tpnalowerx{j}   
         \tepspnalowerx{j}
        %  \left\{(1-\eps) \tpnloweryax{j}  + \epsilon \tindic{(o_{\backslash y})_{\tilde{\ptbdir}}}\right\} 
        }
        \right] \mathrm{d}y \;\mathrm{d}x  \\
             & =\int  \frac{ \tpnlowerx{j}}{   \cancel{\epsilon } } \int y
        \left[ 
       %\sum_{j}^n
   \cancel{\epsilon } \frac{    
\smoothdel{o_\ptbdir}
     }{    \tepspnalowerx{j}
         }  
         -  
         \cancel{\epsilon } \frac{ \tpnloweryax{j}  \smoothdel{(a_{\ptbdir},x_{\ptbdir})}   }{ \tpnalowerx{j}   
         \tepspnalowerx{j}
        %  \left\{(1-\eps) \tpnloweryax{j}  + \epsilon \tindic{(o_{\backslash y})_{\tilde{\ptbdir}}}\right\} 
        }
        \right] \mathrm{d}y \; \mathrm{d}x \\
                     & = \int \frac{ \tpnlowerx{j}}{ \tepspnalowerx{j} } \smoothdel{(a_{\ptbdir},x_{\ptbdir})} \; 
                     \left\{ \int y {    
\smoothdel{y_\ptbdir}
     }\mathrm{d}y
         -  
    \E_{\tilde{p}}[Y \mid A=1, x]\right\} \mathrm{d}x 
    % \\
%                      & = \int \frac{ \tpnlowerx{j}}{ \tepspnalowerx{j} }     \indic{a_i = 1}\smoothdel{(x_i)}(x)
%  \; 
%                      \left\{ \int y {    
% \smoothdel{y_\ptbdir}
%      }\mathrm{d}y  
%          -  
%     \E_{\tilde{p}}[Y \mid A=1, x]\right\}  \mathrm{d}x.
    \end{align*} 
\end{proof}

% \if \forjournal 0 
% \begin{proof}[Proof of \Cref{lemma-rates}.]
% \else
\begin{proof}{Proof of \Cref{lemma-rates}.}
% \fi 
To bound $\E[(\td{e}_\eps(X) - e(X))^2],$ argue similarly and observe that since $\td{e}_\eps=\td{p}_\eps(A=1 \mid x)=\td{p}_\eps( A=1, {x})/\td{p}(x),$ we have:
\begin{align*}  
\E[(\td{e}_\eps(X)-\td{e}(X))^2]&=
\int (((1-\eps) \td{p}(A=1,x) + \eps \smoothdel{a,X_i}(x)) - \td{p}(A=1,x))^2 /\td{p}(x)^2 \dif x  \cdot \td{p}(x)\\&=  \eps^2 \int(\smoothdel{a,X_i}(x) - \td{p}(A=1,x) )^2 /\td{p}(x)\dif x \\
&=  \eps^2 \int((\smoothdel{a,X_i}(x)^2 - 2\smoothdel{a,X_i}(x)\td{p}(A=1,x) + \td{p}(A=1,x)^2)/\td{p}(x)  \dif x \\
& \leq \eps^2\int \nu^{-1} (\smoothdel{a,X_i}(x)^2 +1 )  \dif x \\
&=  \eps^2 (\nu^{-1} \hsmoothdelta^{-d}R(\smoothdel{})+\nu^{-1})  = O(\eps^2  \hsmoothdelta^{-d}).
% &= \textstyle \eps^2 \int\left( {\smoothdel{a,X_i}(x)}/{ \td{p}(A=1,x)} - 1 \right )^2  \td{e}^2(x)  \cdot \td{p}(x)\dif x \\
% &\textstyle \leq \eps^2 \Big\|{ \frac{\smoothdel{a,X_i}(x)}{ \td{p}(A=1,x)} - 1 }\Big\|_\infty^2 \\
% &\textstyle \leq \eps^2\lambda^{-d}
\end{align*} 
In the second-to-last inequality we lower bound $\tilde{p}(x) \geq \nu$ and upper bound $\td{p}(A=1,x) \leq 1.$ In the last inequality we apply a change of variables to identify the term that arises in analysis of variance of kernel density estimates. We incur the $\lambda^{-d}$ dependence since we assumed that multivariate kernels are product kernels. 
The roughness term $R(\smoothdel{}) = \int \tilde{\delta}^2{}(u) du $ depends on the particular $\tilde{\delta}$ kernel function and is a kernel-dependent constant. Hence the term scales overall as $\hsmoothdelta^{-d}$ as in standard analyses of variance of kernel density estimates \citep[see, e.g.,][pg. 23-26]{ullah1999nonparametric}.

% $r_{k}(\lambda):=\left\|\frac{d H_{x, \lambda}}{d P}\right\|_{k, P}$
% Uniform kernel in $d$ dimensions: $r_{k}(\lambda)=\lambda^{-d_{1}(1-1 / k)}$
Next observe that: 
\begin{align*}
\MoveEqLeft{\norm{
   \td{\mu}_\eps(X) - \mu(X)
   }}
   \\
   &\leq     
%   \E[(
\norm{
   \td{\mu}_\eps(X) -\td{\mu}(X) 
   }
%   )^2] 
   + 
   \norm{
%   \E[(
   \td{\mu}(X) - \mu(X)
   }
%   )^2] 
   \\
    & = \textstyle \left(
    \int \left\{ \int y \left(\frac{(1-\eps)\td{p}(y,A=1,x) + \eps \smoothdel{o}}{(1-\eps) \td{p}(A=1,x)+\eps \smoothdel{a,x} } - \frac{\td{p}(y,A=1,x)}{\td{p}(A=1,x)} \right) \mathrm{d}y\right\}^2 p(x) \mathrm{d}x
      \right)^{\frac 12}
    + \norm{
    % \E[(
    \td{\mu}(X) - \mu(X)
    }
    % )^2]  
    \\
                  & = \textstyle 
                  \left( 
                  \int \eps^2 \left\{
    \frac{ \indic{a_i = 1}\smoothdel{(x_i)}(x)}{   \tepspnalowerx{j}} \left\{
        y_i - 
  \E_{\td{p}}[Y\mid A=1,x]
        \right\} \right\}^2 p(x) \mathrm{d}x 
        \right)^{\frac 12}+ 
        \norm{
        % \E[(
        \td{\mu}(X) - \mu(X)
        }
        %)^2]
        \\
    & =  O(\eps \lambda^{-d/2})+
    % \E[(
    \norm{
    \td{\mu}(X) - \mu(X)
    },
    %)^2] 
\end{align*}
where the first inequality comes from the triangle inequality, and the second equality from arguments similar to those in the previous bound. 
\end{proof}

% \if\forjournal 0 
% \begin{proof}[Proof of \Cref{prop-rate-double-robustness}.]
% \else
\begin{proof}{Proof of \Cref{prop-rate-double-robustness}.}
% \fi 
% We will decompose the claim of interest into the error of the (unsmoothed) finite-difference from AIPW, and the error of the  smoothed vs. unsmoothed finite difference. 
  Let $$\td{e}(X)=\td{p}(A=1,X)/\td{p}(X)=\td{p}(A=1\mid X), \td{\mu}(X)$$ denote the propensity and conditional outcome nuisances arising from plug-in estimation of $\tP$. 
  
  Let $${\td{e}_\epsilon(X)= \td{p}_\eps(A=1\mid X) = \tepspnax{}}/{\tpnx{}}$$
  denote the nuisance arising from plug-in estimation at the perturbed distribution.

% Define 
% \begin{equation} \ipwindicphi_\ptbdir(\obs_i) = 
%  \bces{
%  (1-\epsilon) %\tpnx{\ptbdir}
% \left(\frac{Y_\ptbdir\indic{A_\ptbdir=1} - \td{\E}[ Y\mid A=1, X_{\ptbdir} ]}{ \td{e}_\eps(X_i) } \right)+ \tilde{\mathbb{E}}[Y\mid A=1, X_{\ptbdir} ]}
% - \Phi(\tP).
% \end{equation} 
%  That is, $\ipwindicphi_\ptbdir(x)$ defines the Gateaux derivative adjustment, similar to the output of the empirical numerical approximation of \Cref{prop-decomposition-gateaux-derivative-pdf-rep}, but with evaluation of ``IPW''-type terms with the \textit{unsmoothed} indicator rather than the smoothed indicator. (The conditional expectation nuisance remains the \textit{h-smoothed} conditional expectation of \Cref{defn-implicit-condexp-nuisance}.) 

% The influence function adjustment is $\sum_k \frac{1}{\epsilon} (\Phi(P_{\epsilon \tindic{o_{k}}}) - \Phi(P))$
First observe that by \Cref{cor-aipw-additivebias}:
\begin{align*}
&\frac{\Psi(\tilde{P}_\epsilon^{\ptbdir} ) - \Psi(\tilde P)}{\epsilon}\textstyle =\\
& \frac{\indic{A_i=1} }{   \tepspnamidx{i}} 
   \left( Y_i - 
  \tilde{\mu}(X_i)\right)
        % \\&\textstyle\qquad 
+\left(\tilde{\mu}(X_i)- \Psi(\td{P}) \right) +  \int \{\td{\mu}_\eps(x) - \td{\mu}(x)\}\{\smoothdel{(x_i)}(x)-
    {p}(x)\}\,\mathrm{d}x+O(\hsmoothdelta^\beta).\\
    % & \qquad +   \int \{\td{\mu}_\eps(x) - \td{\mu}(x)\}\{\smoothdel{(x_i)}(x)-
    % \tilde{p}(x)\}\,\mathrm{d}x
%  \frac{\Psi(\tilde{P}_\epsilon^{\ptbdir} ) - \Psi(\tilde P)}{\epsilon}&\textstyle =
%   (1-\eps)\frac{\indic{A_i=1} }{   \tepspnamidx{i}} 
%    \left( Y_i - 
%   {\E_{\tP}[Y\mid A=1,X_i]} \right)
%         % \\&\textstyle\qquad 
% +\left( {\mathbb{E}}_{\tP_\eps}[Y\mid A=1,X_{\ptbdir}] - \Psi(\td{P}) \right) + O(\hsmoothdelta^J).\\
\end{align*}

% yw / is it by chance possible to remind the reader what J means here?

This decomposition characterizes convergence of an AIPW estimator with possibly biased but asymptotically consistent nuisances $\td{\mu}, \td{e}_\epsilon$ in terms of the AIPW estimator with oracle nuisances---a standard analysis---and an additional term that vanishes on the order of $\epsilon$ (assuming strong overlap and boundedness). 
We denote the higher-order error term $\Delta_i$: 
\begin{align*}
 \Delta_i\coloneqq    \int \{\td{\mu}_\eps(x) - \td{\mu}(x)\}\{\smoothdel{(x_i)}(x)-
    {p}(x)\}\,\mathrm{d}x
   % = \{\td{\mu}_\eps^\lambda(X_i) - \td{\mu}^\lambda(X_i)\}
\end{align*}

Define $\Gamma^*(O)$ as the doubly robust score for AIPW with \textit{oracle nuisances} $e^*, \mu^*$, and let $\Gamma^*_j$ denote its empirical counterpart at $O_j=(X_k, A_k, Y_k)$: 
$$\Gamma^*(O) = { \indic{A=1}\frac{Y-\mu^*(X)}{e^*(X)} + \mu^*(X)  }.$$
We will study the decomposition relative to the AIPW estimand with oracle nuisance estimates, with $\Gamma(O; e,\mu)$ denoting the above score but with nuisance functions $e,\mu$: 
\begin{align*} \textstyle 
\MoveEqLeft{(\fnl(\tP) + \frac 1n \sumn{\ptbdir} \htdel_{\ptbdir}) -\E[ \Gamma^*(O; e^*,\mu^*) ]
% \fnl_n(\tP) - \E[ \Gamma^*(O; e^*,\mu^*) ]
} \\
& = \E_n[\Gamma(O; \tilde{e}_\eps,\td{\mu})] -\E[ \Gamma^*(O; e^*,\mu^*) ] 
+\E_n[\Delta_i]
%\E_n[ \td{\mu}_\eps(X)- \td{\mu}(X) ]    
+ O(\lambda^\beta) \\
& = (\E_n-\E)[\Gamma(O; \td{e}_\eps,\td{\mu})] + \E[ \Gamma(O; \td{e}_\eps,\td{\mu}) - \Gamma^*(O; e^*,\mu^*) ] +
+\E_n[\Delta_i]
% \E_n[ \td{\mu}_\eps(X)- \td{\mu}(X) ] 
+ O(\lambda^\beta),
\end{align*}
where the second equality follows by \Cref{prop-decomposition-gateaux-derivative-pdf-rep}, and the third by adding/subtracting $\E[\Gamma(O; \tilde{e}_\eps,\tilde{\mu})].$

The first term, $(\E_n-\E)[\Gamma(O; \td{e}_\eps,\td{\mu})]$ is the standard empirical process term of the decomposition, obtained by the Donsker assumption and classical stability results for Donsker classes~\citep{van2000asymptotic}:
$$(\E_n-\E)[\Gamma(O; \td{e}_\eps,\td{\mu})]={(\E_n-\E)[\Gamma(O; {e}^*,\mu^*)]+ o_p(n^{-\frac 12}).}$$
Hence, we have
\begin{equation}
\fnl_n(\tP) - \E[ \Gamma^*(O; e^*,\mu^*) ] 
= \E[ \Gamma(O; \td{e}_\eps,\td{\mu}) - \Gamma^*(O; e^*,\mu^*) ] +
\E_n[\Delta_i]
%\E_n[ \td{\mu}_\eps(X)- \td{\mu}(X) ] 
+ O(\lambda^\beta) + o_p(n^{-\frac 12}). \label{eqn-aipw-decomposition}
\end{equation}
We obtain the product error conditions by standard decomposition of the remaining first term:
% $\hat{\psi}-\psi_{0}=\left(\mathbb{P}_{n}-\mathbb{P}\right) m(Z ; \hat{\eta})+\mathbb{P}\{m(Z ; \hat{\eta})-m(Z ; \bar{\eta})\}$
% (\fnl(\tP) + \frac 1n \sumn{\ptbdir} \htdel_{\ptbdir}) -\frac 1n \sumn{\ptbdir} \Gamma_\ptbdir^* 
% %= \underbrace{\frac 1n \sumn{\ptbdir} (\htdel_{\ptbdir}
% % - \ipwindicphi_\ptbdir) }_{T_1}+ \underbrace{\frac 1n \sumn{\ptbdir}(\fnl(\tP)+\ipwindicphi_\ptbdir - \Gamma_\ptbdir^*)}_{T_2}
% $$
% Define the induced nuisance function (without smoothing) as $\Eindic [Y\mid A=1,X=X_{\ptbdir}]$ :
% \begin{align*} 
% \Eindic[Y\mid A=1,X=X_{\ptbdir}]&= \sum_{j=1}^n \left( \frac{Y_j \sum_{i=1}^n K_h(X_j-X_i) K_h(Y_j-Y_i)\mathbb{I}[A_i=1] }{\sum_{i=1}^n K_h(X_j-X_i)\mathbb{I}[A_i=1]} \right) \indic{x_{\ptbdir}}
% \end{align*} 
% \underline{Bounding $T_1$}:  
% $$\frac 1n \sumn{\ptbdir} (\htdel_{\ptbdir} 
% - \ipwindicphi_\ptbdir) = (1-\epsilon) \sumn{\ptbdir} \left[ 
% \frac{Y_\ptbdir(\indic{A_\ptbdir=1} - \tindic{Y_i} )}{\td{e}_\eps}(X_i)  \right]  =  (1-\epsilon)\E_n[ Y/\td{e}_\eps (\indic{O'=O}-\tindic{O'=O} ]=  (1-\epsilon)O(h^J)$$
% \underline{Bounding $T_2$}: 
\begin{align*}
  \MoveEqLeft{\lefteqn{ \E[ \Gamma(O; \td{e}_\eps,\td{\mu}) - \Gamma^*(O; e^*,\mu^*) ]}}\\
%   \left\{ \sum_{ j}  \frac{1}{\epsilon} (\Phi(P_{\epsilon \indic{o_{{j} }}}) - \Phi(P))
%   \right\} - \E_n \bkts{ \indic{A=1}\frac{Y-\mu^*(X)}{e^*(X)} + \mu^*(X)  } 
&= 
   \E\bkts{
   \paren{\indic{A=1}\frac{Y-\td{\mu}(X)}{\td{e}_\epsilon(X) } + \td{\mu}(X)} }  - 
  \E\bkts{\paren{\indic{A=1}\frac{Y-\mu^*(X)}{e^*(X)} + \mu^*(X)}}    %-  \E_n\bkts{\eps \paren{\indic{A=1}\frac{Y-\td{\mu}}{\td{e}_\epsilon }} }
 \\
&= 
  \E \bkts{\indic{A=1} \paren{ \frac{1}{\td{e}_\epsilon(X)} - \frac{1}{e^*(X)} }(Y - \mu^*(X))} 
  % -  \E \bkts{ \paren{\frac{1}{e^*(X)}-1} (\td{\mu}(X) -\mu^*(X)) } 
  \\&\qquad 
  + \E \bkts{\indic{A=1} \paren{\frac{1}{\td{e}_\epsilon(X)} - \frac{1}{e^*(X)} } (\td{\mu}(X)-\mu^*(X)) } \\
  &= 
  \E \bkts{\indic{A=1} \paren{\frac{1}{\td{e}_\epsilon(X)} - \frac{1}{e^*(X)} } (\td{\mu}(X)-\mu^*(X)) } \\
  &\leq \nu^{-1} \|\td{e}_\epsilon(X) - {e^*(X)}\|_{2,P} \times  \|\td{\mu}(X)-\mu^*(X)\|_{2,P}.
\end{align*}
% \begin{align*}
%   & (\fnl(\tilde P) +\frac 1n \sumn{\ptbdir} \htdel_{\ptbdir}) - \frac 1n \sumn{\ptbdir}\Gamma^*_\ptbdir\\
% %   \left\{ \sum_{ j}  \frac{1}{\epsilon} (\Phi(P_{\epsilon \indic{o_{{j} }}}) - \Phi(P))
% %   \right\} - \E_n \bkts{ \indic{A=1}\frac{Y-\mu^*(X)}{e^*(X)} + \mu^*(X)  } 
% &= 
%   \E_n\bkts{
%   \paren{\indic{A=1}\frac{Y-\td{\mu}}{\td{e}_\epsilon } + \td{\mu}}  - 
% \paren{\indic{A=1}\frac{Y-\mu^*}{e^*} + \mu^*} - \eps \paren{\indic{A=1}\frac{Y-\td{\mu}}{\td{e}_\epsilon }}
% } \\
% &= 
%   \E_n \bkts{\indic{A=1}( 1/\td{e}_\epsilon - 1/e^* )(Y - \mu^*)} -  \E_n \bkts{(1/e^*-1) (\td{\mu} -\mu^*) } 
%   + \E_n \bkts{\indic{A=1} (1/\td{e}_\epsilon - 1/e^*) (\td{\mu}-\mu^*) }\\
%   &\qquad -\eps \E_n[ {\indic{A=1}\paren{Y-\td{\mu}}/{\td{e}_\epsilon }}] 
% % + \E_n \bkts{\td\mu (1 - \frac{ \tpax{X} }{\tepspax{X} })}
% % +  \Phi(P)\E_n\bkts{ \frac{ \tpax{X} }{\tepspax{X} }}
% \end{align*}
The second equality follows by adding/subtracting $\indic{A=1}/\td{e}_\epsilon$). The third equality follows from iterated expectations. The last inequality follows from the Cauchy-Schwarz inequality and overlap. Therefore, $\epsilon$ must satisfy a rate so that $\td{e}_\epsilon$ converges at typical double robustness fast rates, such as $n^{-1/4}$, under a suitable sample splitting scheme. 
% \begin{align*}
%      &= 
%   \E_n \bkts{\indic{A=1}( 1/\td{e}_\epsilon - 1/e^* )(Y - \mu^*)} -  \E_n \bkts{(1/e^*-1) (\td{\mu} -\mu^*) } 
%   + \E_n \bkts{\indic{A=1} (1/\td{e}_\epsilon - 1/e^*) (\td{\mu}-\mu^*) }\\
%   &\qquad 
% + \E_n \bkts{\td\mu (1 - \frac{ \tpax{X} }{\tepspax{X} })}
% +  \Phi(P)\E_n\bkts{ \frac{ \tpax{X} }{\tepspax{X} }}
% \end{align*}
% But the additional bias terms from numerical approximation, include
% $$\E_n \bkts{\td\mu (1 - \frac{ \tpax{X} }{\tepspax{X} })}
% +  \Phi(P)\E_n\bkts{ \frac{ \tpax{X} }{\tepspax{X} }}+ \left\{ \sum_{ j}  \frac{1}{\epsilon} (
% \Phi(P_{\epsilon \indic{o_{{j} }}})-
% \Phi(P_{\epsilon \tindic{o_{{j} }}}) )
%   \right\}
% $$

% The second to last term is the highest-order dependence on $\eps$: \Cref{lemma-numerical-derivative-decomposition} shows that this is $O(\eps)$, leading to the $o(n^{-\frac 12})$ requirement for $\eps$ to preserve rate double robustness achieved by the other terms. 

% Now suppose $\mu, e$ converge in $L_2$-norm at rates $r_1, r_2$ respectively satisfying rate-double robustness so that $r_1+r_2\geq n^{-\frac 12}$. Then $\epsilon$ must converge on the order of $O(n^{-\frac 12})$. 
% \az{new}
For the last remaining error term of \Cref{eqn-aipw-decomposition}, we decompose $\Delta_i$ as
\begin{align}
\Delta_i
&=
\int \{\td{\mu}_\eps(x) - \td{\mu}(x)\}\{\tilde\delta_{X_i}^\lambda(x)-{p}(x)\}\,dx
+
\int \{\td{\mu}_\eps(x) - \td{\mu}(x)\}\{p(x)-\tilde p(x)\}\,dx
\label{eqn-apx-secondorderproductdecomposition}\end{align}

The first term satisfies
\begin{align*} \textstyle & \E_n[
\int \{\td{\mu}_\eps(x) - \td{\mu}(x)\}\{\tilde\delta_{X_i}^\lambda(x)-p(x)\}\,dx
] \\
& = (\E_n - \E)[ \td{\mu}_\eps^\lambda(x) - \td{\mu}^\lambda(x) ]
+ \E[ \{\td{\mu}_\eps^\lambda(x) - \td{\mu}^\lambda(x)\}- \{\td{\mu}_\eps(x) - \td{\mu}(x)\} ]
% & =(\E_n - \E)[ {\mu}_\eps^\lambda(x) - {\mu}^\lambda(x) ] + o_p(n^{-\frac 12})
% + O(\lambda^\beta)
\end{align*} 
We apply Chebyshev's inequality to the remaining empirical process term on the small $\eps$ perturbation. We can bound the variance as follows: 
$$\operatorname{Var}\left((\E_n - \E)[ \td{\mu}_\eps^\lambda(x) - \td{\mu}^\lambda(x) ]\right)=\frac{\operatorname{Var}\left(\td{\mu}_\eps^\lambda(x) - \td{\mu}^\lambda(x) \right)}{n} \leq \frac{\mathbb{E}\left[(\td{\mu}_\eps^\lambda(x) - \td{\mu}^\lambda(x) )^2\right]}{n}$$
Applying Chebyshev's inequality and \Cref{lemma-rates}, we obtain: 
\begin{align*} \textstyle 
    \E_n[
\int \{\td{\mu}_\eps(x) - \td{\mu}(x)\}\{\tilde\delta_{X_i}^\lambda(x)-p(x)\}\,dx
] 
& = O_p\left(\frac{\left\|\td{\mu}_\eps(x) - \td{\mu}(x)\right\|}{\sqrt{n}}\right) + O(\lambda^\beta) \\
& = O_p\left(\frac{\epsilon \lambda^{-d / 2}}{\sqrt{n}}\right)+ O(\lambda^\beta)\\
& = o_p(n^{-\frac 12}).\tag{by the rate assumptions}
\end{align*}

For the second term of \cref{eqn-apx-secondorderproductdecomposition}, note that by 
\Cref{lemma-numerical-derivative-decomposition}, we have the finite-difference representation
\[
\td{\mu}_\epsilon(x)-\td{\mu}(x)
=
\epsilon\,
\frac{\smoothdel{(x_i)}(x)}{\td{p}_\epsilon(A=1\mid x)}
\left\{ Y_i^\lambda - \td{\mu}(x)\right\}
+ O(\epsilon \lambda^\beta),
\]
where $Y_i^\lambda$ denotes the kernel-smoothed response.
Under strong overlap $\td{p}_\epsilon(A=1\mid x)\ge \nu>0$,% taking $L_2(P)$ norms yields
\[
\|\td{\mu}_\epsilon-\td{\mu}\|_{2,P}
\lesssim
\epsilon\,
\left\|
\frac{\indic{A=1}}{\td{p}_\epsilon(A=1\mid X)}
\{Y-\td{\mu}(X)\}
\right\|_{2,P}
+
O(\epsilon\lambda^\beta).
\]
By iterated expectations,
\[
\E\!\left[
\frac{\indic{A=1}}{\td{p}_\epsilon(A=1\mid X)}
\{Y-\td{\mu}(X)\}
\,\middle|\,X
\right]
=
\frac{e^*(X)}{\td{e}_\epsilon(X)}
\{\mu^*(X)-\td{\mu}(X)\},
\]
so that the only vanishing component is the regression error.
Hence, by rate assumptions, this is higher order:
\[
\|\td{\mu}_\epsilon-\td{\mu}\|_{2,P}
= O\left(
\epsilon
(\|\td{\mu}-\mu^*\|_{2,P}
+
O(\lambda^\beta))
\right)=o_p(n^{-\frac 12}).
\]

\end{proof}

\section{Proofs for \Cref{sec-application}}
\subsection{Empirical Gateaux derivative for multi-stage dynamic treatment regimes}

We study the estimation properties of the empirical Gateaux derivative for multi-stage DTR from the probability density representation. In $T$-stage dynamic treatment regimes, the causal quantity of interest is the mean potential outcome
$\EE{}{Y^{\bar{a}}}$, where $\bar{a} = (a_0, \ldots, a_{T})$ is the (deterministic) treatment strategy. Assuming $Y^{\bar{a}}$ is sequentially ignorable given treatment and covariate history $(\bar{A}_t, \bar{X}_t)$ at each time $t$, this causal quantity can be identified by the $g$-formula
\begin{equation*} \textstyle \EE{}{Y^{\bar{a}}} = \int \EE{}{Y \,|\, \bar{A}=\bar{a}, \bar{X}=\bar{x}}\prod_{t=1}^T \tilde{p}(x_t \,|\, \bar{a}_{t-1}, \bar{x}_{t-1})\dif \bar{x},\end{equation*}
where $\bar{A} = (A_0, \ldots, A_{t}), \bar{X} = (X_0, \ldots, X_{t})$.

To derive its influence function, we take the empirical Gateaux derivative by considering the perturbation in the direction of $o_i=(\bar{x}_i, \bar{a}_i, y_i)$, where $\bar{x}_i = (x_{i0}, \ldots, x_{it})$ and $\bar{a}_i = (a_{i0}, \ldots, a_{it})$:
\begin{align}
    \Psi(\tilde{P}^i_\epsilon) &=\int \left(\int y \cdot \frac{\tilde{p}_\epsilon(y , \bar{A}=\bar{a}, \bar{X}=\bar{x})}{\tilde{p}_\epsilon(\bar{A}=\bar{a}, \bar{X}=\bar{x})} \dif y\right)\prod_{t=1}^T \frac{\tilde{p}_\epsilon(x_t , \bar{a}_{t-1}, \bar{x}_{t-1})}{\tilde{p}_\epsilon(\bar{a}_{t-1}, \bar{x}_{t-1}))} \cdot \tilde{p}_\epsilon(x_0)\dif \bar{x}.
\end{align}

Below we characterize how this empirical Gateaux derivative at the smoothed distribution differs from the one at the (unsmoothed) estimated distribution. The following result is analogous to \Cref{prop-decomposition-gateaux-derivative-pdf-rep} but is extended to the dynamic treatment regimes.

% Specifically, in \Cref{eq:term1,eq:term2}, smoothed perturbation leads to a kernel smoothed evaluation of a conditional expectation.  \Cref{eq:term3,eq:term4,eq:term5} involves $T$ terms of inverse probability weighted conditional expecation functional each. Under the same conditions as in Lemma 1, the biases due to smoothed conditional expectation are of the order $O(\lambda^J)$ where $\lambda$ is the bandwidth of the kernel and the nuisance function is $J$-times continuously differentiable. Thus the empirical Gateaux derivative preserves the rate double robustness in dynamic treatment regimes under similar conditions as in Theorem 1.

% \pagebreak
% \if\forjournal 0 
% \begin{proof}[Proof of \Cref{prop-fd-dtr}]
% \else
\begin{proof}{Proof of \Cref{prop-fd-dtr}.}
% \fi 
We first provide a more explicit statement of the result in \Cref{prop-fd-dtr}. 

\begin{align}
    &\frac{\Psi(\tilde{P}^i_\epsilon)-\Psi(\tilde{P})}{\epsilon} \nonumber\\
    &= \left[\int \EE{\tilde{p}_\epsilon}{Y \,|\, \bar{A}=\bar{a}, \bar{X}_{1:T}=\bar{x}_{1:T}, X_0=x_{i0}} \prod_{t=1}^{T} \tilde{p}(x_t \,|\, \bar{a}_{t-1}, \bar{x}_{1:t-1}, x_{i0})\dif \bar{x}  - \Psi(\tilde{P})\right] \nonumber\\
    &\quad+  \int\tilde{p}(x_0)\left\{\left[\frac{\tilde{\delta}^\lambda_{(\bar{a}_{i},\bar{x}_{i})}}{\tilde{p}_\epsilon(\bar{A}=\bar{a}, \bar{X}=\bar{x})}{}\left[\int y \cdot \tilde{\delta}^\lambda_{y_i} \dif y - \EE{\tilde{p}}{Y\,|\, \bar{A}=\bar{a}, \bar{X}=\bar{x} }\right] \right]\prod_{t=1}^T \frac{\tilde{p}(x_t , \bar{a}_{t-1}, \bar{x}_{t-1})}{\tilde{p}(\bar{a}_{t-1}, \bar{x}_{t-1}))}\right\} \dif \bar{x}\nonumber\\
    &\quad+ \sum_{s=1}^{T}\int\tilde{p}(x_0)\left\{\left(\int y \cdot \frac{\tilde{p}_\epsilon(y , \bar{A}=\bar{a}, \bar{X}=\bar{x})}{\tilde{p}_\epsilon(\bar{A}=\bar{a}, \bar{X}=\bar{x})} \dif y\right)\left[\left\{\prod_{t=s+1}^{T}\frac{\tilde{p}(x_t , \bar{a}_{t-1}, \bar{x}_{t-1})}{\tilde{p}(\bar{a}_{t-1}, \bar{x}_{t-1}))} \prod_{t=1}^{s-1}\frac{\tilde{p}_\epsilon(x_t , \bar{a}_{t-1}, \bar{x}_{t-1})}{\tilde{p}_\epsilon(\bar{a}_{t-1}, \bar{x}_{t-1}))} \right.\right.\right.\nonumber\\
    &\qquad\qquad\times\left.\left.\left.\frac{1}{\tilde{p}_\epsilon(\bar{a}_{s-1}, \bar{x}_{s-1}))}\left[\tilde{\delta}^\lambda_{(\bar{a}_{i,s-1}, \bar{x}_{i,s})}) - \frac{\tilde{p}(x_s , \bar{a}_{s-1}, \bar{x}_{s-1})}{\tilde{p}(\bar{a}_{s-1}, \bar{x}_{s-1}))}\tilde{\delta}^\lambda_{(\bar{a}_{i,s-1}, \bar{x}_{i,s-1})}\right]\right\}\right]\right\} \dif \bar{x}\nonumber\\
    &\quad + \epsilon\sum_{s=1}^{T}\int \EE{\tilde{p}_\epsilon}{Y \,|\, \bar{A}=\bar{a}, \bar{X}_{1:T}=\bar{x}_{1:T}, X_0=x_{i0}} \nonumber\\
    &\qquad\qquad\times\left[\left\{\prod_{t=s+1}^{T}\frac{\tilde{p}(x_t , \bar{a}_{t-1}, \bar{x}_{1:t-1}, x_{i0})}{\tilde{p}(\bar{a}_{t-1}, \bar{x}_{1:t-1}, x_{i0}))} \prod_{t=1}^{s-1}\frac{\tilde{p}_\epsilon(x_t , \bar{a}_{t-1}, \bar{x}_{1:t-1}, x_{i0})}{\tilde{p}_\epsilon(\bar{a}_{t-1}, \bar{x}_{1:t-1}, x_{i0}))} \frac{1}{\tilde{p}_\epsilon(\bar{a}_{s-1}, \bar{x}_{1:s-1}, x_{i0}))}\right.\right.\nonumber\\
    &\qquad\qquad\qquad\qquad\times\left.\left.\left[\tilde{\delta}^\lambda_{(\bar{a}_{i,s-1}, \bar{x}_{i,s})}) - \frac{\tilde{p}(x_s , \bar{a}_{s-1}, \bar{x}_{1:s-1}, x_{i0})}{\tilde{p}(\bar{a}_{s-1}, \bar{x}_{1:s-1}, x_{i0}))}\tilde{\delta}^\lambda_{(\bar{a}_{i,s-1}, \bar{x}_{i,s-1})}\right]\right\}\right]\dif \bar{x}.
\end{align}
For comparison, the exact efficient influence function under the non-Markovian dynamic treatment regime is
\begin{align}
&\left[\EE{\tilde{p}(\bar{x}_{1:T})}{ \EE{\tilde{p}}{Y \,|\, \bar{A}=\bar{a}, \bar{X}_{1:T}=\bar{x}_{1:T}, X_0=x_{i0}} }  - \Psi(\tilde{P})\right]\nonumber\\
    &\quad+\frac{\mathbb{I}{(\bar{a}_{i})}}{\tilde{p}(\bar{A}=\bar{a}\,|\,\bar{X}=\bar{x})}{}\left[ Y_i - \EE{\tilde{p}(\bar{x})}{\EE{\tilde{p}}{Y\,|\, \bar{A}=\bar{a}, \bar{X}=\bar{x} }}\right]  \nonumber\\
    &\quad+\sum_{s=1}^T\frac{\mathbb{I}{(\bar{a}_{i,T-s}, \bar{x}_{i,T-s+1})}}{\tilde{p}_\epsilon(\bar{a}_{T-s}\,|\, \bar{x}_{T-s})}\left[\int \EE{\tilde{p}}{Y\,|\, \bar{A}=\bar{a}, \bar{X}=\bar{x}}{\prod_{t=T-s+1}^{T-1} \tilde{p}(x_t \,|\, \bar{a}_{t-1}, \bar{x}_{t-1})\dif \bar{x}_{(T-s+1):(T-1)}}\right]\nonumber\\
    &\quad- \sum_{s=1}^T\frac{\mathbb{I}{(\bar{a}_{i,T-s}, \bar{x}_{i,T-s})}}{\tilde{p}(\bar{a}_{T-s}\,|\, \bar{x}_{T-s})}\left[\int\EE{\tilde{p}}{Y\,|\, \bar{A}=\bar{a}, \bar{X}=\bar{x}}\prod_{t=T-s}^{T-1} \tilde{p}(x_t \,|\,\bar{a}_{t-1}, \bar{x}_{t-1})  \dif\bar{x}_{(T-s):(T-1)}\right].\nonumber
\end{align}

This proposition performs a decomposition of the empirical derivative. It quantifies the additive bias due to the smoothed perturbation compared to the (raw) perturbation at the observation $o_i$.

We first decompose the empirical Gateaux derivative:
\begin{align}
    &\frac{\Psi(\tilde{P}^i_\epsilon)-\Psi(\tilde{P})}{\epsilon} \nonumber\\
    &= \frac{1}{\epsilon}  \left[\int \left(\int y \cdot \frac{\tilde{p}_\epsilon(y , \bar{A}=\bar{a}, \bar{X}=\bar{x})}{\tilde{p}_\epsilon(\bar{A}=\bar{a}, \bar{X}=\bar{x})} \dif y\right)\prod_{t=1}^T \frac{\tilde{p}_\epsilon(x_t , \bar{a}_{t-1}, \bar{x}_{t-1})}{\tilde{p}_\epsilon(\bar{a}_{t-1}, \bar{x}_{t-1}))} \cdot [(1-\epsilon)\tilde{p}(x_0) + \epsilon \tilde{\delta}^\lambda_{x_{i0}}]\dif \bar{x} \right.\nonumber\\
    &\quad\left.- \int \left(\int y \cdot \frac{\tilde{p}(y , \bar{A}=\bar{a}, \bar{X}=\bar{x})}{\tilde{p}(\bar{A}=\bar{a}, \bar{X}=\bar{x})} \dif y\right)\prod_{t=1}^T \frac{\tilde{p}(x_t , \bar{a}_{t-1}, \bar{x}_{t-1})}{\tilde{p}(\bar{a}_{t-1}, \bar{x}_{t-1}))} \cdot \tilde{p}(x_0)\dif \bar{x} \right]\label{eq:dec-1}\\
    &
    \textstyle =  \int \left[\left(\int y \cdot \frac{\tilde{p}_\epsilon(y , \bar{A}=\bar{a}, \bar{X}=\bar{x})}{\tilde{p}_\epsilon(\bar{A}=\bar{a}, \bar{X}=\bar{x})} \dif y\right)\prod_{t=1}^T \frac{\tilde{p}_\epsilon(x_t , \bar{a}_{t-1}, \bar{x}_{t-1})}{\tilde{p}_\epsilon(\bar{a}_{t-1}, \bar{x}_{t-1}))} \cdot \tilde{\delta}^\lambda_{x_{i0}}\right.\nonumber\\
    &\textstyle \quad-\left.\left(\int y \cdot \frac{\tilde{p}(y , \bar{A}=\bar{a}, \bar{X}=\bar{x})}{\tilde{p}(\bar{A}=\bar{a}, \bar{X}=\bar{x})} \dif y\right)\prod_{t=1}^T \frac{\tilde{p}(x_t , \bar{a}_{t-1}, \bar{x}_{t-1})}{\tilde{p}(\bar{a}_{t-1}, \bar{x}_{t-1}))} \cdot \tilde{p}(x_0) \right]\dif \bar{x} \nonumber\\
    &\textstyle \quad+ \frac{1}{\epsilon} \left[\int\tilde{p}(x_0)\left[ \left(\int y \cdot \frac{\tilde{p}_\epsilon(y , \bar{A}=\bar{a}, \bar{X}=\bar{x})}{\tilde{p}_\epsilon(\bar{A}=\bar{a}, \bar{X}=\bar{x})} \dif y\right)\prod_{t=1}^T \frac{\tilde{p}_\epsilon(x_t , \bar{a}_{t-1}, \bar{x}_{t-1})}{\tilde{p}_\epsilon(\bar{a}_{t-1}, \bar{x}_{t-1}))}\right. \right.\nonumber\\
    &\textstyle \quad\left.\left.-  \left(\int y \cdot \frac{\tilde{p}(y , \bar{A}=\bar{a}, \bar{X}=\bar{x})}{\tilde{p}(\bar{A}=\bar{a}, \bar{X}=\bar{x})} \dif y\right)\prod_{t=1}^T\frac{\tilde{p}(x_t , \bar{a}_{t-1}, \bar{x}_{t-1})}{\tilde{p}( \bar{a}_{t-1}, \bar{x}_{t-1})} \right] \dif \bar{x} \right]\label{eq:dec-2} \\
    \end{align}
    \begin{align}
    & \textstyle = \left[\int \EE{\tilde{p}_\epsilon}{Y \,|\, \bar{A}=\bar{a}, \bar{X}_{T}=\bar{x}_{T}}\prod_{t=1}^T \tilde{p}_\epsilon(x_t \,|\, \bar{a}_{t-1}, \bar{x}_{t-1})\tilde{p}(x_0) \tilde{\delta}^\lambda_{x_{i0}}\dif \bar{x} - \Psi(\tilde{P})\right] \nonumber\\
    &\textstyle \quad+ \frac{1}{\epsilon} \left[\int\tilde{p}(x_0)\left[ \left(\int y \cdot \frac{\tilde{p}_\epsilon(y , \bar{A}=\bar{a}, \bar{X}=\bar{x})}{\tilde{p}_\epsilon(\bar{A}=\bar{a}, \bar{X}=\bar{x})} \dif y\right)\prod_{t=1}^T \frac{\tilde{p}_\epsilon(x_t , \bar{a}_{t-1}, \bar{x}_{t-1})}{\tilde{p}_\epsilon(\bar{a}_{t-1}, \bar{x}_{t-1}))}\right. \right.\nonumber\\
    &\textstyle \quad\left.\left.-  \left(\int y \cdot \frac{\tilde{p}(y , \bar{A}=\bar{a}, \bar{X}=\bar{x})}{\tilde{p}(\bar{A}=\bar{a}, \bar{X}=\bar{x})} \dif y\right)\prod_{t=1}^T \frac{\tilde{p}(x_t , \bar{a}_{t-1}, \bar{x}_{t-1})}{\tilde{p}(\bar{a}_{t-1}, \bar{x}_{t-1}))} \right] \dif \bar{x} \right]\nonumber\\
    &\textstyle = \left[\int \EE{\tilde{p}_\epsilon}{Y \,|\, \bar{A}=\bar{a}, \bar{X}_{1:T}=\bar{x}_{1:T}, X_0=x_{i0}} \prod_{t=1}^{T} \tilde{p}_\epsilon(x_t \,|\, \bar{a}_{t-1}, \bar{x}_{1:t-1}, x_{i0})\dif \bar{x}  - \Psi(\tilde{P})\right] \nonumber\\
    &\textstyle \quad+ \frac{1}{\epsilon} \left[\int\tilde{p}(x_0)\left[ \left(\int y \cdot \frac{\tilde{p}_\epsilon(y , \bar{A}=\bar{a}, \bar{X}=\bar{x})}{\tilde{p}_\epsilon(\bar{A}=\bar{a}, \bar{X}=\bar{x})} \dif y\right)\prod_{t=1}^T \frac{\tilde{p}_\epsilon(x_t , \bar{a}_{t-1}, \bar{x}_{t-1})}{\tilde{p}_\epsilon(\bar{a}_{t-1}, \bar{x}_{t-1}))}\right. \right.\nonumber\\
    &\textstyle \quad\left.\left.-  \left(\int y \cdot \frac{\tilde{p}(y , \bar{A}=\bar{a}, \bar{X}=\bar{x})}{\tilde{p}(\bar{A}=\bar{a}, \bar{X}=\bar{x})} \dif y\right)\prod_{t=1}^T \frac{\tilde{p}(x_t , \bar{a}_{t-1}, \bar{x}_{t-1})}{\tilde{p}(\bar{a}_{t-1}, \bar{x}_{t-1}))} \right] \dif \bar{x} \right].\label{eq:dec-3}
    \end{align}

\Cref{eq:dec-1} is due to the definition of the smoothed perturbation. \Cref{eq:dec-2} separates the effect of smoothed perturbation into two pieces: the evaluation of $y$ and the evaluation of the density ratios. \Cref{eq:dec-3} is due to the definition of $\EE{\tilde{p}_\epsilon}{\cdot}$.

\paragraph{Calculating the second term of \Cref{eq:dec-3}.}
We first decompose this term into two difference terms:
\begin{align}
    &\left(\int y \cdot \frac{\tilde{p}_\epsilon(y , \bar{A}=\bar{a}, \bar{X}=\bar{x})}{\tilde{p}_\epsilon(\bar{A}=\bar{a}, \bar{X}=\bar{x})} \dif y\right)\prod_{t=1}^T \frac{\tilde{p}_\epsilon(x_t , \bar{a}_{t-1}, \bar{x}_{t-1})}{\tilde{p}_\epsilon(\bar{a}_{t-1}, \bar{x}_{t-1}))}\\
    &\qquad \qquad-  \left(\int y \cdot \frac{\tilde{p}(y , \bar{A}=\bar{a}, \bar{X}=\bar{x})}{\tilde{p}(\bar{A}=\bar{a}, \bar{X}=\bar{x})} \dif y\right)\prod_{t=1}^T \frac{\tilde{p}(x_t , \bar{a}_{t-1}, \bar{x}_{t-1})}{\tilde{p}(\bar{a}_{t-1}, \bar{x}_{t-1}))}\nonumber\\
    &=\left(\int y \cdot \frac{\tilde{p}_\epsilon(y , \bar{A}=\bar{a}, \bar{X}=\bar{x})}{\tilde{p}_\epsilon(\bar{A}=\bar{a}, \bar{X}=\bar{x})} \dif y\right)\left[\prod_{t=1}^T \frac{\tilde{p}_\epsilon(x_t , \bar{a}_{t-1}, \bar{x}_{t-1})}{\tilde{p}_\epsilon(\bar{a}_{t-1}, \bar{x}_{t-1}))}-  \prod_{t=1}^T \frac{\tilde{p}(x_t , \bar{a}_{t-1}, \bar{x}_{t-1})}{\tilde{p}(\bar{a}_{t-1}, \bar{x}_{t-1}))}\right]\nonumber\\
    &\quad+ \left[\left(\int y \cdot \frac{\tilde{p}_\epsilon(y , \bar{A}=\bar{a}, \bar{X}=\bar{x})}{\tilde{p}_\epsilon(\bar{A}=\bar{a}, \bar{X}=\bar{x})} \dif y\right)- \left(\int y \cdot \frac{\tilde{p}(y , \bar{A}=\bar{a}, \bar{X}=\bar{x})}{\tilde{p}(\bar{A}=\bar{a}, \bar{X}=\bar{x})} \dif y\right)\right]\prod_{t=1}^T \frac{\tilde{p}(x_t , \bar{a}_{t-1}, \bar{x}_{t-1})}{\tilde{p}(\bar{a}_{t-1}, \bar{x}_{t-1}))}.
\end{align}

The second difference term can be decomposed in a similar way as in Lemma 2:
\begin{align}
    \MoveEqLeft{ \left(\int y \cdot \frac{\tilde{p}_\epsilon(y , \bar{A}=\bar{a}, \bar{X}=\bar{x})}{\tilde{p}_\epsilon(\bar{A}=\bar{a}, \bar{X}=\bar{x})} \dif y\right)- \left(\int y \cdot \frac{\tilde{p}(y , \bar{A}=\bar{a}, \bar{X}=\bar{x})}{\tilde{p}(\bar{A}=\bar{a}, \bar{X}=\bar{x})} \dif y\right) }\nonumber\\
    &\textstyle =\int y \cdot \left[\frac{\tilde{p}_\epsilon(y , \bar{A}=\bar{a}, \bar{X}=\bar{x})}{\tilde{p}_\epsilon(\bar{A}=\bar{a}, \bar{X}=\bar{x})} - \frac{\tilde{p}(y , \bar{A}=\bar{a}, \bar{X}=\bar{x})}{\tilde{p}(\bar{A}=\bar{a}, \bar{X}=\bar{x})}\right] \dif y\nonumber\\
    &\textstyle =\int y \cdot \left[\frac{(1-\epsilon)\tilde{p}(y , \bar{A}=\bar{a}, \bar{X}=\bar{x}) + \epsilon \tilde{\delta}^\lambda_{o_i}}{\tilde{p}_\epsilon(\bar{A}=\bar{a}, \bar{X}=\bar{x})} - \frac{\tilde{p}(y , \bar{A}=\bar{a}, \bar{X}=\bar{x})}{\tilde{p}(\bar{A}=\bar{a}, \bar{X}=\bar{x})}\right] \dif y\nonumber\\
    &\textstyle =\int y \cdot \left[\frac{((1-\epsilon)\tilde{p}(y , \bar{A}=\bar{a}, \bar{X}=\bar{x}) + \epsilon \tilde{\delta}^\lambda_{o_i}) \tilde{p}(\bar{A}=\bar{a}, \bar{X}=\bar{x})}{\tilde{p}_\epsilon(\bar{A}=\bar{a}, \bar{X}=\bar{x})\tilde{p}(\bar{A}=\bar{a}, \bar{X}=\bar{x})} \right.\\
    &\textstyle \quad\quad\left.- \frac{\tilde{p}(y , \bar{A}=\bar{a}, \bar{X}=\bar{x}) ((1-\epsilon)\tilde{p}(\bar{A}=\bar{a}, \bar{X}=\bar{x})+\epsilon \tilde{\delta}^\lambda_{(\bar{a}_{i},\bar{x}_{i})})}{\tilde{p}_\epsilon(\bar{A}=\bar{a}, \bar{X}=\bar{x})\tilde{p}(\bar{A}=\bar{a}, \bar{X}=\bar{x})}\right] \dif y\nonumber\\
    &\textstyle =\int y \cdot \left[\frac{(\epsilon \tilde{\delta}^\lambda_{o_i}) \tilde{p}(\bar{A}=\bar{a}, \bar{X}=\bar{x})}{\tilde{p}_\epsilon(\bar{A}=\bar{a}, \bar{X}=\bar{x})\tilde{p}(\bar{A}=\bar{a}, \bar{X}=\bar{x})} - \frac{\tilde{p}(y , \bar{A}=\bar{a}, \bar{X}=\bar{x}) (\epsilon \tilde{\delta}^\lambda_{(\bar{a}_{i},\bar{x}_{i})})}{\tilde{p}_\epsilon(\bar{A}=\bar{a}, \bar{X}=\bar{x})\tilde{p}(\bar{A}=\bar{a}, \bar{X}=\bar{x})}\right] \dif y\nonumber\\
    & \textstyle =\frac{\epsilon\tilde{\delta}^\lambda_{(\bar{a}_{i},\bar{x}_{i})}}{\tilde{p}_\epsilon(\bar{A}=\bar{a}, \bar{X}=\bar{x})}{}\int y \cdot \left[( \tilde{\delta}^\lambda_{y_i})  - \frac{\tilde{p}(y , \bar{A}=\bar{a}, \bar{X}=\bar{x}) }{\tilde{p}(\bar{A}=\bar{a}, \bar{X}=\bar{x})}\right] \dif y\nonumber\\
    &=\frac{\epsilon\tilde{\delta}^\lambda_{(\bar{a}_{i},\bar{x}_{i})}}{\tilde{p}_\epsilon(\bar{A}=\bar{a}, \bar{X}=\bar{x})}{}\left[\int y \cdot \tilde{\delta}^\lambda_{y_i} \dif y - \EE{\tilde{p}}{Y\,|\, \bar{A}=\bar{a}, \bar{X}=\bar{x} }\right]. \nonumber
\end{align}

We next analyze the first difference term by induction:
\begin{align}
    \MoveEqLeft{\prod_{t=1}^T \frac{\tilde{p}_\epsilon(x_t , \bar{a}_{t-1}, \bar{x}_{t-1})}{\tilde{p}_\epsilon(\bar{a}_{t-1}, \bar{x}_{t-1}))}-  \prod_{t=1}^T \frac{\tilde{p}(x_t , \bar{a}_{t-1}, \bar{x}_{t-1})}{\tilde{p}(\bar{a}_{t-1}, \bar{x}_{t-1}))}}\nonumber\\
    &=\prod_{t=1}^{T-1}\frac{\tilde{p}_\epsilon(x_t , \bar{a}_{t-1}, \bar{x}_{t-1})}{\tilde{p}_\epsilon(\bar{a}_{t-1}, \bar{x}_{t-1}))} \left[  \frac{\tilde{p}_\epsilon(x_T , \bar{a}_{T-1}, \bar{x}_{T-1})}{\tilde{p}_\epsilon(\bar{a}_{T-1}, \bar{x}_{T-1}))} - \frac{\tilde{p}(x_T , \bar{a}_{T-1}, \bar{x}_{T-1})}{\tilde{p}(\bar{a}_{T-1}, \bar{x}_{T-1}))}\right]\nonumber\\
    &\quad + \frac{\tilde{p}(x_T , \bar{a}_{T-1}, \bar{x}_{T-1})}{\tilde{p}(\bar{a}_{T-1}, \bar{x}_{T-1}))} \left[ \prod_{t=1}^{T-1} \frac{\tilde{p}_\epsilon(x_t , \bar{a}_{t-1}, \bar{x}_{t-1})}{\tilde{p}_\epsilon(\bar{a}_{t-1}, \bar{x}_{t-1}))}-  \prod_{t=1}^{T-1} \frac{\tilde{p}(x_t , \bar{a}_{t-1}, \bar{x}_{t-1})}{\tilde{p}(\bar{a}_{t-1}, \bar{x}_{t-1}))}\right].
\end{align}

Next we evaluate the key term in this induction $\frac{\tilde{p}_\epsilon(x_s , \bar{a}_{s-1}, \bar{x}_{s-1})}{\tilde{p}_\epsilon(\bar{a}_{s-1}, \bar{x}_{s-1}))} - \frac{\tilde{p}(x_s , \bar{a}_{s-1}, \bar{x}_{s-1})}{\tilde{p}(\bar{a}_{s-1}, \bar{x}_{s-1}))}$:
\begin{align}
     &\frac{\tilde{p}_\epsilon(x_s , \bar{a}_{s-1}, \bar{x}_{s-1})}{\tilde{p}_\epsilon(\bar{a}_{s-1}, \bar{x}_{s-1}))} - \frac{\tilde{p}(x_s , \bar{a}_{s-1}, \bar{x}_{s-1})}{\tilde{p}(\bar{a}_{s-1}, \bar{x}_{s-1}))} \nonumber\\
     &= \frac{((1-\epsilon)\tilde{p}(x_s , \bar{a}_{s-1}, \bar{x}_{s-1}) + \epsilon \tilde{\delta}^\lambda_{o_i})\tilde{p}(\bar{a}_{s-1}, \bar{x}_{s-1}))}{\tilde{p}_\epsilon(\bar{a}_{s-1}, \bar{x}_{s-1}))\tilde{p}(\bar{a}_{s-1}, \bar{x}_{s-1}))} - \frac{\tilde{p}(x_s , \bar{a}_{s-1}, \bar{x}_{s-1})((1-\epsilon)\tilde{p}(\bar{a}_{s-1}, \bar{x}_{s-1}))+\epsilon \tilde{\delta}^\lambda_{o_i})}{\tilde{p}(\bar{a}_{s-1}, \bar{x}_{s-1}))\tilde{p}_\epsilon(\bar{a}_{s-1}, \bar{x}_{s-1}))}\nonumber\\
     &= \frac{(\epsilon \tilde{\delta}^\lambda_{(\bar{a}_{i,s-1}, \bar{x}_{i,s})})}{\tilde{p}_\epsilon(\bar{a}_{s-1}, \bar{x}_{s-1}))} - \frac{\tilde{p}(x_s , \bar{a}_{s-1}, \bar{x}_{s-1})(\epsilon \tilde{\delta}^\lambda_{(\bar{a}_{i,s-1}, \bar{x}_{i,s-1})})}{\tilde{p}(\bar{a}_{s-1}, \bar{x}_{s-1}))\tilde{p}_\epsilon(\bar{a}_{s-1}, \bar{x}_{s-1}))}\nonumber\\
     &= \frac{\epsilon}{\tilde{p}_\epsilon(\bar{a}_{s-1}, \bar{x}_{s-1}))}\left[\tilde{\delta}^\lambda_{(\bar{a}_{i,s-1}, \bar{x}_{i,s})}) - \frac{\tilde{p}(x_s , \bar{a}_{s-1}, \bar{x}_{s-1})}{\tilde{p}(\bar{a}_{s-1}, \bar{x}_{s-1}))}\tilde{\delta}^\lambda_{(\bar{a}_{i,s-1}, \bar{x}_{i,s-1})}\right].
\end{align}
Thus we have the following inductive relationship in $T$:
\begin{align}
    \MoveEqLeft{ \prod_{t=1}^T \frac{\tilde{p}_\epsilon(x_t , \bar{a}_{t-1}, \bar{x}_{t-1})}{\tilde{p}_\epsilon(\bar{a}_{t-1}, \bar{x}_{t-1}))}-  \prod_{t=1}^T \frac{\tilde{p}(x_t , \bar{a}_{t-1}, \bar{x}_{t-1})}{\tilde{p}(\bar{a}_{t-1}, \bar{x}_{t-1}))}}\nonumber\\
    &=\textstyle\prod_{t=1}^{T-1}\frac{\tilde{p}_\epsilon(x_t , \bar{a}_{t-1}, \bar{x}_{t-1})}{\tilde{p}_\epsilon(\bar{a}_{t-1}, \bar{x}_{t-1}))} \frac{\epsilon}{\tilde{p}_\epsilon(\bar{a}_{T-1}, \bar{x}_{T-1}))}\left[\tilde{\delta}^\lambda_{(\bar{a}_{i,T-1}, \bar{x}_{i,T})}) - \frac{\tilde{p}(x_T , \bar{a}_{T-1}, \bar{x}_{T-1})}{\tilde{p}(\bar{a}_{T-1}, \bar{x}_{T-1}))}\tilde{\delta}^\lambda_{(\bar{a}_{i,T-1}, \bar{x}_{i,T-1})}\right]\nonumber\\
    &\textstyle\quad + \frac{\tilde{p}(x_T , \bar{a}_{T-1}, \bar{x}_{T-1})}{\tilde{p}(\bar{a}_{T-1}, \bar{x}_{T-1}))} \left[ \prod_{t=1}^{T-1} \frac{\tilde{p}_\epsilon(x_t , \bar{a}_{t-1}, \bar{x}_{t-1})}{\tilde{p}_\epsilon(\bar{a}_{t-1}, \bar{x}_{t-1}))}-  \prod_{t=1}^{T-1} \frac{\tilde{p}(x_t , \bar{a}_{t-1}, \bar{x}_{t-1})}{\tilde{p}(\bar{a}_{t-1}, \bar{x}_{t-1}))}\right]\\
    &\textstyle=\sum_{s=T-1}^{T}\left\{\prod_{t=s+1}^{T}\frac{\tilde{p}(x_t , \bar{a}_{t-1}, \bar{x}_{t-1})}{\tilde{p}(\bar{a}_{t-1}, \bar{x}_{t-1}))} \prod_{t=1}^{s-1}\frac{\tilde{p}_\epsilon(x_t , \bar{a}_{t-1}, \bar{x}_{t-1})}{\tilde{p}_\epsilon(\bar{a}_{t-1}, \bar{x}_{t-1}))} \frac{\epsilon}{\tilde{p}_\epsilon(\bar{a}_{s-1}, \bar{x}_{s-1}))}\right.\nonumber\\
    &\textstyle\qquad\left.\qquad\times\left[\tilde{\delta}^\lambda_{(\bar{a}_{i,s-1}, \bar{x}_{i,s})}) - \frac{\tilde{p}(x_s , \bar{a}_{s-1}, \bar{x}_{s-1})}{\tilde{p}(\bar{a}_{s-1}, \bar{x}_{s-1}))}\tilde{\delta}^\lambda_{(\bar{a}_{i,s-1}, \bar{x}_{i,s-1})}\right]\right\}\nonumber\\
    &\textstyle\quad + \prod_{t=T-1}^{T}\frac{\tilde{p}(x_t , \bar{a}_{t-1}, \bar{x}_{t-1})}{\tilde{p}(\bar{a}_{t-1}, \bar{x}_{t-1}))} \left[ \prod_{t=1}^{T-2} \frac{\tilde{p}_\epsilon(x_t , \bar{a}_{t-1}, \bar{x}_{t-1})}{\tilde{p}_\epsilon(\bar{a}_{t-1}, \bar{x}_{t-1}))}-  \prod_{t=1}^{T-2} \frac{\tilde{p}(x_t , \bar{a}_{t-1}, \bar{x}_{t-1})}{\tilde{p}(\bar{a}_{t-1}, \bar{x}_{t-1}))}\right]\\
    &=\sum_{s=1}^{T}\left\{\prod_{t=s+1}^{T}\frac{\tilde{p}(x_t , \bar{a}_{t-1}, \bar{x}_{t-1})}{\tilde{p}(\bar{a}_{t-1}, \bar{x}_{t-1}))} \prod_{t=1}^{s-1}\frac{\tilde{p}_\epsilon(x_t , \bar{a}_{t-1}, \bar{x}_{t-1})}{\tilde{p}_\epsilon(\bar{a}_{t-1}, \bar{x}_{t-1}))} \frac{\epsilon}{\tilde{p}_\epsilon(\bar{a}_{s-1}, \bar{x}_{s-1}))}\right.\nonumber\\
    &\qquad\qquad\times\left.\left[\tilde{\delta}^\lambda_{(\bar{a}_{i,s-1}, \bar{x}_{i,s})}) - \frac{\tilde{p}(x_s , \bar{a}_{s-1}, \bar{x}_{s-1})}{\tilde{p}(\bar{a}_{s-1}, \bar{x}_{s-1}))}\tilde{\delta}^\lambda_{(\bar{a}_{i,s-1}, \bar{x}_{i,s-1})}\right]\right\}.\label{eq:induct-diff}
\end{align}
%az{$T-1\to t-1$ }
These calculations imply that
\begin{align}
\MoveEqLeft{\textstyle\frac{\Psi(\tilde{P}^i_\epsilon)-\Psi(\tilde{P})}{\epsilon}} \nonumber\\
    &\textstyle= \left[\int \EE{\tilde{p}_\epsilon}{Y \,|\, \bar{A}=\bar{a}, \bar{X}_{1:T}=\bar{x}_{1:T}, X_0=x_{i0}} \prod_{t=1}^{T} \tilde{p}_\epsilon(x_t \,|\, \bar{a}_{t-1}, \bar{x}_{1:t-1}, x_{i0})\dif \bar{x}  - \Psi(\tilde{P})\right] \nonumber\\
    &\textstyle\quad+ \frac{1}{\epsilon} \left[\int\tilde{p}(x_0)\left[ \left(\int y \cdot \frac{\tilde{p}_\epsilon(y , \bar{A}=\bar{a}, \bar{X}=\bar{x})}{\tilde{p}_\epsilon(\bar{A}=\bar{a}, \bar{X}=\bar{x})} \dif y\right)\prod_{t=1}^T \frac{\tilde{p}_\epsilon(x_t , \bar{a}_{t-1}, \bar{x}_{t-1})}{\tilde{p}_\epsilon(\bar{a}_{t-1}, \bar{x}_{t-1}))}\right. \right.\nonumber\\
    &\textstyle\quad\left.\left.-  \left(\int y \cdot \frac{\tilde{p}(y , \bar{A}=\bar{a}, \bar{X}=\bar{x})}{\tilde{p}(\bar{A}=\bar{a}, \bar{X}=\bar{x})} \dif y\right)\prod_{t=1}^T \frac{\tilde{p}(x_t , \bar{a}_{t-1}, \bar{x}_{t-1})}{\tilde{p}(\bar{a}_{t-1}, \bar{x}_{t-1}))} \right] \dif \bar{x} \right]\\
    % \end{align}
    % \begin{align}
    &\textstyle= \left[\int \EE{\tilde{p}_\epsilon}{Y \,|\, \bar{A}=\bar{a}, \bar{X}_{1:T}=\bar{x}_{1:T}, X_0=x_{i0}} \prod_{t=1}^{T} \tilde{p}_\epsilon(x_t \,|\, \bar{a}_{t-1}, \bar{x}_{1:t-1}, x_{i0})\dif \bar{x}  - \Psi(\tilde{P})\right] \nonumber\\
    &\textstyle\quad+ \frac{1}{\epsilon} \int\tilde{p}(x_0)\left\{\left(\int y \cdot \frac{\tilde{p}_\epsilon(y , \bar{A}=\bar{a}, \bar{X}=\bar{x})}{\tilde{p}_\epsilon(\bar{A}=\bar{a}, \bar{X}=\bar{x})} \dif y\right)\left[\prod_{t=1}^T \frac{\tilde{p}_\epsilon(x_t , \bar{a}_{t-1}, \bar{x}_{t-1})}{\tilde{p}_\epsilon(\bar{a}_{t-1}, \bar{x}_{t-1}))}-  \prod_{t=1}^T \frac{\tilde{p}(x_t , \bar{a}_{t-1}, \bar{x}_{t-1})}{\tilde{p}(\bar{a}_{t-1}, \bar{x}_{t-1}))}\right]\right.\nonumber\\
    &\textstyle\quad+ \left.\left[\left(\int y \cdot \frac{\tilde{p}_\epsilon(y , \bar{A}=\bar{a}, \bar{X}=\bar{x})}{\tilde{p}_\epsilon(\bar{A}=\bar{a}, \bar{X}=\bar{x})} \dif y\right)- \left(\int y \cdot \frac{\tilde{p}(y , \bar{A}=\bar{a}, \bar{X}=\bar{x})}{\tilde{p}(\bar{A}=\bar{a}, \bar{X}=\bar{x})} \dif y\right)\right]\prod_{t=1}^T \frac{\tilde{p}(x_t , \bar{a}_{t-1}, \bar{x}_{t-1})}{\tilde{p}(\bar{a}_{t-1}, \bar{x}_{t-1}))}\right\} \dif \bar{x}  
    \end{align}
    \begin{align}
    &\textstyle= \left[\int \EE{\tilde{p}_\epsilon}{Y \,|\, \bar{A}=\bar{a}, \bar{X}_{1:T}=\bar{x}_{1:T}, X_0=x_{i0}} \prod_{t=1}^{T} \tilde{p}_\epsilon(x_t \,|\, \bar{a}_{t-1}, \bar{x}_{1:t-1}, x_{i0})\dif \bar{x}  - \Psi(\tilde{P})\right] \nonumber\\
    &\textstyle\quad+ \frac{1}{\epsilon} \int\tilde{p}(x_0)\left\{\left(\int y \cdot \frac{\tilde{p}_\epsilon(y , \bar{A}=\bar{a}, \bar{X}=\bar{x})}{\tilde{p}_\epsilon(\bar{A}=\bar{a}, \bar{X}=\bar{x})} \dif y\right)\left[\prod_{t=1}^T \frac{\tilde{p}_\epsilon(x_t , \bar{a}_{t-1}, \bar{x}_{t-1})}{\tilde{p}_\epsilon(\bar{a}_{t-1}, \bar{x}_{t-1}))}-  \prod_{t=1}^T \frac{\tilde{p}(x_t , \bar{a}_{t-1}, \bar{x}_{t-1})}{\tilde{p}(\bar{a}_{t-1}, \bar{x}_{t-1}))}\right]\right\} \dif \bar{x}\nonumber\\
    &\textstyle\quad+ \frac{1}{\epsilon} \int\tilde{p}(x_0)\left\{\left[\frac{\epsilon\tilde{\delta}^\lambda_{(\bar{a}_{i},\bar{x}_{i})}}{\tilde{p}_\epsilon(\bar{A}=\bar{a}, \bar{X}=\bar{x})}{}\left[\int y \cdot \tilde{\delta}^\lambda_{y_i} \dif y - \EE{\tilde{p}}{Y\,|\, \bar{A}=\bar{a}, \bar{X}=\bar{x} }\right] \right]\prod_{t=1}^T \frac{\tilde{p}(x_t , \bar{a}_{t-1}, \bar{x}_{t-1})}{\tilde{p}(\bar{a}_{t-1}, \bar{x}_{t-1}))}\right\} \dif \bar{x}   %\\
    \end{align}
    \begin{align}
    &\textstyle= \left[\int \EE{\tilde{p}_\epsilon}{Y \,|\, \bar{A}=\bar{a}, \bar{X}_{1:T}=\bar{x}_{1:T}, X_0=x_{i0}} \prod_{t=1}^{T} \tilde{p}_\epsilon(x_t \,|\, \bar{a}_{t-1}, \bar{x}_{1:t-1}, x_{i0})\dif \bar{x}  - \Psi(\tilde{P})\right] \nonumber\\
    &\textstyle\quad+ \frac{1}{\epsilon} \int\tilde{p}(x_0)\left\{\left(\int y \cdot \frac{\tilde{p}_\epsilon(y , \bar{A}=\bar{a}, \bar{X}=\bar{x})}{\tilde{p}_\epsilon(\bar{A}=\bar{a}, \bar{X}=\bar{x})} \dif y\right)\left[\sum_{s=1}^{T}\left\{\prod_{t=s+1}^{T}\frac{\tilde{p}(x_t , \bar{a}_{t-1}, \bar{x}_{t-1})}{\tilde{p}(\bar{a}_{t-1}, \bar{x}_{t-1}))} \prod_{t=1}^{s-1}\frac{\tilde{p}_\epsilon(x_t , \bar{a}_{t-1}, \bar{x}_{t-1})}{\tilde{p}_\epsilon(\bar{a}_{t-1}, \bar{x}_{t-1}))} \right.\right.\right.\nonumber\\
    &\textstyle\qquad\qquad\times\left.\left.\left.\frac{\epsilon}{\tilde{p}_\epsilon(\bar{a}_{s-1}, \bar{x}_{s-1}))}\left[\tilde{\delta}^\lambda_{(\bar{a}_{i,s-1}, \bar{x}_{i,s})}) - \frac{\tilde{p}(x_s , \bar{a}_{s-1}, \bar{x}_{s-1})}{\tilde{p}(\bar{a}_{s-1}, \bar{x}_{s-1}))}\tilde{\delta}^\lambda_{(\bar{a}_{i,s-1}, \bar{x}_{i,s-1})}\right]\right\}\right]\right\} \dif \bar{x}\nonumber\\
    &\textstyle\quad+ \frac{1}{\epsilon} \int\tilde{p}(x_0)\left\{\left[\frac{\epsilon\tilde{\delta}^\lambda_{(\bar{a}_{i},\bar{x}_{i})}}{\tilde{p}_\epsilon(\bar{A}=\bar{a}, \bar{X}=\bar{x})}{}\left[\int y \cdot \tilde{\delta}^\lambda_{y_i} \dif y - \EE{\tilde{p}}{Y\,|\, \bar{A}=\bar{a}, \bar{X}=\bar{x} }\right] \right]\prod_{t=1}^T \frac{\tilde{p}(x_t , \bar{a}_{t-1}, \bar{x}_{t-1})}{\tilde{p}(\bar{a}_{t-1}, \bar{x}_{t-1}))}\right\} \dif \bar{x}  
\\    &= \left[\int \EE{\tilde{p}_\epsilon}{Y \,|\, \bar{A}=\bar{a}, \bar{X}_{1:T}=\bar{x}_{1:T}, X_0=x_{i0}} \prod_{t=1}^{T} \tilde{p}_\epsilon(x_t \,|\, \bar{a}_{t-1}, \bar{x}_{1:t-1}, x_{i0})\dif \bar{x}  - \Psi(\tilde{P})\right] \nonumber\\
    &\quad+ \int\tilde{p}(x_0)\left\{\left(\int y \cdot \frac{\tilde{p}_\epsilon(y , \bar{A}=\bar{a}, \bar{X}=\bar{x})}{\tilde{p}_\epsilon(\bar{A}=\bar{a}, \bar{X}=\bar{x})} \dif y\right)\left[\sum_{s=1}^{T}\left\{\prod_{t=s+1}^{T}\frac{\tilde{p}(x_t , \bar{a}_{t-1}, \bar{x}_{t-1})}{\tilde{p}(\bar{a}_{t-1}, \bar{x}_{t-1}))} \prod_{t=1}^{s-1}\frac{\tilde{p}_\epsilon(x_t , \bar{a}_{t-1}, \bar{x}_{t-1})}{\tilde{p}_\epsilon(\bar{a}_{t-1}, \bar{x}_{t-1}))} \right.\right.\right.\nonumber\\
    &\qquad\qquad\times\left.\left.\left.\frac{1}{\tilde{p}_\epsilon(\bar{a}_{s-1}, \bar{x}_{s-1}))}\left[\tilde{\delta}^\lambda_{(\bar{a}_{i,s-1}, \bar{x}_{i,s})}) - \frac{\tilde{p}(x_s , \bar{a}_{s-1}, \bar{x}_{s-1})}{\tilde{p}(\bar{a}_{s-1}, \bar{x}_{s-1}))}\tilde{\delta}^\lambda_{(\bar{a}_{i,s-1}, \bar{x}_{i,s-1})}\right]\right\}\right]\right\} \dif \bar{x}\nonumber\\
    &\quad+  \int\tilde{p}(x_0)\left\{\left[\frac{\tilde{\delta}^\lambda_{(\bar{a}_{i},\bar{x}_{i})}}{\tilde{p}_\epsilon(\bar{A}=\bar{a}, \bar{X}=\bar{x})}{}\left[\int y \cdot \tilde{\delta}^\lambda_{y_i} \dif y - \EE{\tilde{p}}{Y\,|\, \bar{A}=\bar{a}, \bar{X}=\bar{x} }\right] \right]\prod_{t=1}^T \frac{\tilde{p}(x_t , \bar{a}_{t-1}, \bar{x}_{t-1})}{\tilde{p}(\bar{a}_{t-1}, \bar{x}_{t-1}))}\right\} \dif \bar{x}. \label{eq:dec-3-0} 
    \end{align}

\paragraph{Calculating the first term of \Cref{eq:dec-3-0}.}
The first term of \Cref{eq:dec-3-0} can be decomposed as follows:
\begin{align}
    \MoveEqLeft{\int \EE{\tilde{p}_\epsilon}{Y \,|\, \bar{A}=\bar{a}, \bar{X}_{1:T}=\bar{x}_{1:T}, X_0=x_{i0}} \prod_{t=1}^{T} \tilde{p}_\epsilon(x_t \,|\, \bar{a}_{t-1}, \bar{x}_{1:t-1}, x_{i0})\dif \bar{x}}\nonumber\\
    & = \int \EE{\tilde{p}_\epsilon}{Y \,|\, \bar{A}=\bar{a}, \bar{X}_{1:T}=\bar{x}_{1:T}, X_0=x_{i0}} \prod_{t=1}^{T} \tilde{p}(x_t \,|\, \bar{a}_{t-1}, \bar{x}_{1:t-1}, x_{i0})\dif \bar{x}\\
    &\qquad + \int \EE{\tilde{p}_\epsilon}{Y \,|\, \bar{A}=\bar{a}, \bar{X}_{1:T}=\bar{x}_{1:T}, X_0=x_{i0}} \nonumber\\
    &\qquad\qquad\times\left[\prod_{t=1}^T \frac{\tilde{p}_\epsilon(x_t , \bar{a}_{t-1}, \bar{x}_{1:t-1}, x_{i0})}{\tilde{p}_\epsilon(\bar{a}_{t-1}, \bar{x}_{1:t-1}, x_{i0}))}-  \prod_{t=1}^T \frac{\tilde{p}(x_t , \bar{a}_{t-1}, \bar{x}_{1:t-1}, x_{i0})}{\tilde{p}(\bar{a}_{t-1}, \bar{x}_{1:t-1}, x_{i0}))}\right]\dif \bar{x}.
\end{align}

Following a similar calculation as in \Cref{eq:induct-diff}, we have that
\begin{align}
    &\prod_{t=1}^T \frac{\tilde{p}_\epsilon(x_t , \bar{a}_{t-1}, \bar{x}_{1:t-1}, x_{i0})}{\tilde{p}_\epsilon(\bar{a}_{t-1}, \bar{x}_{1:t-1}, x_{i0}))}-  \prod_{t=1}^T \frac{\tilde{p}(x_t , \bar{a}_{t-1}, \bar{x}_{1:t-1}, x_{i0})}{\tilde{p}(\bar{a}_{t-1}, \bar{x}_{1:t-1}, x_{i0}))}\\
    &=\sum_{s=1}^{T}\left\{\prod_{t=s+1}^{T}\frac{\tilde{p}(x_t , \bar{a}_{t-1}, \bar{x}_{1:t-1}, x_{i0})}{\tilde{p}(\bar{a}_{t-1}, \bar{x}_{1:t-1}, x_{i0}))} \prod_{t=1}^{s-1}\frac{\tilde{p}_\epsilon(x_t , \bar{a}_{t-1}, \bar{x}_{1:t-1}, x_{i0})}{\tilde{p}_\epsilon(\bar{a}_{t-1}, \bar{x}_{1:t-1}, x_{i0}))} \frac{\epsilon}{\tilde{p}_\epsilon(\bar{a}_{s-1}, \bar{x}_{1:s-1}, x_{i0}))}\right.\nonumber\\
    &\qquad\qquad\times\left.\left[\tilde{\delta}^\lambda_{(\bar{a}_{i,s-1}, \bar{x}_{i,s})}) - \frac{\tilde{p}(x_s , \bar{a}_{s-1}, \bar{x}_{1:s-1}, x_{i0})}{\tilde{p}(\bar{a}_{s-1}, \bar{x}_{1:s-1}, x_{i0}))}\tilde{\delta}^\lambda_{(\bar{a}_{i,s-1}, \bar{x}_{i,s-1})}\right]\right\}.
\end{align}
Thus the first term of \Cref{eq:dec-3-0} is
\begin{align*}
    &\int \EE{\tilde{p}_\epsilon}{Y \,|\, \bar{A}=\bar{a}, \bar{X}_{1:T}=\bar{x}_{1:T}, X_0=x_{i0}} \prod_{t=1}^{T} \tilde{p}_\epsilon(x_t \,|\, \bar{a}_{t-1}, \bar{x}_{1:t-1}, x_{i0})\dif \bar{x}\nonumber\\
    & = \int \EE{\tilde{p}_\epsilon}{Y \,|\, \bar{A}=\bar{a}, \bar{X}_{1:T}=\bar{x}_{1:T}, X_0=x_{i0}} \prod_{t=1}^{T} \tilde{p}(x_t \,|\, \bar{a}_{t-1}, \bar{x}_{1:t-1}, x_{i0})\dif \bar{x}\\
    &\qquad + \int \EE{\tilde{p}_\epsilon}{Y \,|\, \bar{A}=\bar{a}, \bar{X}_{1:T}=\bar{x}_{1:T}, X_0=x_{i0}} \\
    &\qquad\qquad\times\left[\sum_{s=1}^{T}\left\{\prod_{t=s+1}^{T}\frac{\tilde{p}(x_t , \bar{a}_{t-1}, \bar{x}_{1:t-1}, x_{i0})}{\tilde{p}(\bar{a}_{t-1}, \bar{x}_{1:t-1}, x_{i0}))} \prod_{t=1}^{s-1}\frac{\tilde{p}_\epsilon(x_t , \bar{a}_{t-1}, \bar{x}_{1:t-1}, x_{i0})}{\tilde{p}_\epsilon(\bar{a}_{t-1}, \bar{x}_{1:t-1}, x_{i0}))} \right.\right.\nonumber\\
    &\qquad\qquad\times\left.\left.\frac{\epsilon}{\tilde{p}_\epsilon(\bar{a}_{s-1}, \bar{x}_{1:s-1}, x_{i0}))}\left[\tilde{\delta}^\lambda_{(\bar{a}_{i,s-1}, \bar{x}_{i,s})}) - \frac{\tilde{p}(x_s , \bar{a}_{s-1}, \bar{x}_{1:s-1}, x_{i0})}{\tilde{p}(\bar{a}_{s-1}, \bar{x}_{1:s-1}, x_{i0}))}\tilde{\delta}^\lambda_{(\bar{a}_{i,s-1}, \bar{x}_{i,s-1})}\right]\right\}\right]\dif \bar{x}.
\end{align*}
Plugging this calculation into $\frac{\Psi(\tilde{P}^i_\epsilon)-\Psi(\tilde{P})}{\epsilon}:$
\begin{align}
    &\frac{\Psi(\tilde{P}^i_\epsilon)-\Psi(\tilde{P})}{\epsilon} \nonumber\\
    &\textstyle= \left[\int \EE{\tilde{p}_\epsilon}{Y \,|\, \bar{A}=\bar{a}, \bar{X}_{1:T}=\bar{x}_{1:T}, X_0=x_{i0}} \prod_{t=1}^{T} \tilde{p}_\epsilon(x_t \,|\, \bar{a}_{t-1}, \bar{x}_{1:t-1}, x_{i0})\dif \bar{x}  - \Psi(\tilde{P})\right] \nonumber\\
    &\textstyle\quad+ \int\tilde{p}(x_0)\left\{\left(\int y \cdot \frac{\tilde{p}_\epsilon(y , \bar{A}=\bar{a}, \bar{X}=\bar{x})}{\tilde{p}_\epsilon(\bar{A}=\bar{a}, \bar{X}=\bar{x})} \dif y\right)\left[\sum_{s=1}^{T}\left\{\prod_{t=s+1}^{T}\frac{\tilde{p}(x_t , \bar{a}_{t-1}, \bar{x}_{t-1})}{\tilde{p}(\bar{a}_{t-1}, \bar{x}_{t-1}))} \prod_{t=1}^{s-1}\frac{\tilde{p}_\epsilon(x_t , \bar{a}_{t-1}, \bar{x}_{t-1})}{\tilde{p}_\epsilon(\bar{a}_{t-1}, \bar{x}_{t-1}))} \right.\right.\right.\nonumber\\
    &\textstyle\qquad\qquad\times\left.\left.\left.\frac{1}{\tilde{p}_\epsilon(\bar{a}_{s-1}, \bar{x}_{s-1}))}\left[\tilde{\delta}^\lambda_{(\bar{a}_{i,s-1}, \bar{x}_{i,s})}) - \frac{\tilde{p}(x_s , \bar{a}_{s-1}, \bar{x}_{s-1})}{\tilde{p}(\bar{a}_{s-1}, \bar{x}_{s-1}))}\tilde{\delta}^\lambda_{(\bar{a}_{i,s-1}, \bar{x}_{i,s-1})}\right]\right\}\right]\right\} \dif \bar{x}\nonumber\\
    &\textstyle\quad+  \int\tilde{p}(x_0)\left\{\left[\frac{\tilde{\delta}^\lambda_{(\bar{a}_{i},\bar{x}_{i})}}{\tilde{p}_\epsilon(\bar{A}=\bar{a}, \bar{X}=\bar{x})}{}\left[\int y \cdot \tilde{\delta}^\lambda_{y_i} \dif y - \EE{\tilde{p}}{Y\,|\, \bar{A}=\bar{a}, \bar{X}=\bar{x} }\right] \right]\prod_{t=1}^T \frac{\tilde{p}(x_t , \bar{a}_{t-1}, \bar{x}_{t-1})}{\tilde{p}(\bar{a}_{t-1}, \bar{x}_{t-1}))}\right\} \dif \bar{x}%\\
    \end{align}
    \begin{align}
    &\textstyle= \left[\int \EE{\tilde{p}_\epsilon}{Y \,|\, \bar{A}=\bar{a}, \bar{X}_{1:T}=\bar{x}_{1:T}, X_0=x_{i0}} \prod_{t=1}^{T} \tilde{p}(x_t \,|\, \bar{a}_{t-1}, \bar{x}_{1:t-1}, x_{i0})\dif \bar{x}  - \Psi(\tilde{P})\right] \nonumber\\
    &\textstyle\quad+  \int\tilde{p}(x_0)\left\{\left[\frac{\tilde{\delta}^\lambda_{(\bar{a}_{i},\bar{x}_{i})}}{\tilde{p}_\epsilon(\bar{A}=\bar{a}, \bar{X}=\bar{x})}{}\left[\int y \cdot \tilde{\delta}^\lambda_{y_i} \dif y - \EE{\tilde{p}}{Y\,|\, \bar{A}=\bar{a}, \bar{X}=\bar{x} }\right] \right]\prod_{t=1}^T \frac{\tilde{p}(x_t , \bar{a}_{t-1}, \bar{x}_{t-1})}{\tilde{p}(\bar{a}_{t-1}, \bar{x}_{t-1}))}\right\} \dif \bar{x}\nonumber\\
    &\textstyle\quad+ \sum_{s=1}^{T}\int\tilde{p}(x_0)\left\{\left(\int y \cdot \frac{\tilde{p}_\epsilon(y , \bar{A}=\bar{a}, \bar{X}=\bar{x})}{\tilde{p}_\epsilon(\bar{A}=\bar{a}, \bar{X}=\bar{x})} \dif y\right)\left[\left\{\prod_{t=s+1}^{T}\frac{\tilde{p}(x_t , \bar{a}_{t-1}, \bar{x}_{t-1})}{\tilde{p}(\bar{a}_{t-1}, \bar{x}_{t-1}))} \prod_{t=1}^{s-1}\frac{\tilde{p}_\epsilon(x_t , \bar{a}_{t-1}, \bar{x}_{t-1})}{\tilde{p}_\epsilon(\bar{a}_{t-1}, \bar{x}_{t-1}))} \right.\right.\right.\nonumber\\
&\textstyle\qquad\qquad\times\left.\left.\left.\frac{1}{\tilde{p}_\epsilon(\bar{a}_{s-1}, \bar{x}_{s-1}))}\left[\tilde{\delta}^\lambda_{(\bar{a}_{i,s-1}, \bar{x}_{i,s})}) - \frac{\tilde{p}(x_s , \bar{a}_{s-1}, \bar{x}_{s-1})}{\tilde{p}(\bar{a}_{s-1}, \bar{x}_{s-1}))}\tilde{\delta}^\lambda_{(\bar{a}_{i,s-1}, \bar{x}_{i,s-1})}\right]\right\}\right]\right\} \dif \bar{x}\nonumber\\
    &\textstyle\quad + \epsilon\sum_{s=1}^{T}\int \EE{\tilde{p}_\epsilon}{Y \,|\, \bar{A}=\bar{a}, \bar{X}_{1:T}=\bar{x}_{1:T}, X_0=x_{i0}} \\
    &\textstyle\qquad\qquad\times\left[\left\{\prod_{t=s+1}^{T}\frac{\tilde{p}(x_t , \bar{a}_{t-1}, \bar{x}_{1:t-1}, x_{i0})}{\tilde{p}(\bar{a}_{t-1}, \bar{x}_{1:t-1}, x_{i0}))} \prod_{t=1}^{s-1}\frac{\tilde{p}_\epsilon(x_t , \bar{a}_{t-1}, \bar{x}_{1:t-1}, x_{i0})}{\tilde{p}_\epsilon(\bar{a}_{t-1}, \bar{x}_{1:t-1}, x_{i0}))} \right.\right.\nonumber\\
&\textstyle\quad\qquad\qquad\times\left.\left.\frac{1}{\tilde{p}_\epsilon(\bar{a}_{s-1}, \bar{x}_{1:s-1}, x_{i0}))}\left[\tilde{\delta}^\lambda_{(\bar{a}_{i,s-1}, \bar{x}_{i,s})}) - \frac{\tilde{p}(x_s , \bar{a}_{s-1}, \bar{x}_{1:s-1}, x_{i0})}{\tilde{p}(\bar{a}_{s-1}, \bar{x}_{1:s-1}, x_{i0}))}\tilde{\delta}^\lambda_{(\bar{a}_{i,s-1}, \bar{x}_{i,s-1})}\right]\right\}\right]\dif \bar{x}.
\end{align}
\end{proof}

\new{
\begin{proposition}[Finite-difference calculus simplifies dynamic treatment regime example]\label{prop-dtr-fdc}
Suppose $u \subseteq o$; i.e., we evaluate densities on subsets of the complete observation $o$. Then define:%\useshortskip
$$
\textstyle
\Delta_\eps f(u) = \eps^{-1} \{ ( (1-\eps)f(u)+\eps\smoothdel{\tilde u} (u)) -f(u) \} = \delta(u)-f(u).$$
Then 
\begin{align*}
\MoveEqLeft{\frac{\Psi(\tilde{P}^i_\epsilon)-\Psi(\tilde{P})}{\epsilon}} \\  &=\int \int y \cdot \left[ \frac{
\tilde{p}(\bar{a}, \bar{x})\Delta_\epsilon\left(\tilde{p}(y , \bar{a}, \bar{x})\right)
 - \tilde{p}(y , \bar{a}, \bar{x})\Delta_\epsilon\left(\tilde{p}(\bar{a}, \bar{x})\right)
 }{\tilde{p}(\bar{a}, \bar{x})\tilde{p}_\epsilon(\bar{a}, \bar{x})}
\prod_{t=1}^T \frac{\tilde{p}(x_t , \bar{a}_{t-1}, \bar{x}_{t-1})}{\tilde{p}(\bar{a}_{t-1}, \bar{x}_{t-1})} \cdot \tilde{p}(x_0) \right. \\
&\qquad \qquad+ \sum_{k=1}^T \left\{\frac{\tilde{p}(y , \bar{a}, \bar{x})}{\tilde{p}(\bar{a}, \bar{x})}
\frac{\tilde{p}(\bar{a}_{k-1}, \bar{x}_{k-1})\Delta_\epsilon\left(\tilde{p}(x_k , \bar{a}_{k-1}, \bar{x}_{k-1})\right) - 
\tilde{p}(x_k , \bar{a}_{k-1}, \bar{x}_{k-1})
\Delta_\epsilon\left(\tilde{p}(\bar{a}_{k-1}, \bar{x}_{k-1})\right)
}{\tilde{p}(\bar{a}_{k-1}, \bar{x}_{k-1})\tilde{p}_\epsilon(\bar{a}_{k-1}, \bar{x}_{k-1})}\right. 
\\
&\qquad \qquad \qquad \left.\prod_{t\in [T]\backslash k} \frac{\tilde{p}(x_t , \bar{a}_{t-1}, \bar{x}_{t-1})}{\tilde{p}(\bar{a}_{t-1}, \bar{x}_{t-1})} \cdot \tilde{p}(x_0) \right\}\\
&\qquad \qquad\left.+  \frac{\tilde{p}(y , \bar{a}, \bar{x})}{\tilde{p}(\bar{a}, \bar{x})}\prod_{t=1}^T \frac{\tilde{p}(x_t , \bar{a}_{t-1}, \bar{x}_{t-1})}{\tilde{p}(\bar{a}_{t-1}, \bar{x}_{t-1})} \cdot \Delta_\epsilon\left(\tilde{p}(x_0) \right)\right]\dif \bar{x} \dif y.
\end{align*}
\end{proposition}
}
% \if\forjournal 0 
% \begin{proof}[Proof of finite-difference calculus characterization]
% \else
\begin{proof}{Proof of \Cref{prop-dtr-fdc}, finite-difference calculus characterization.}
% \fi
\new{
Below we characterize how this empirical Gateaux derivative at the smoothed distribution differs from the one at the (unsmoothed) estimated distribution. The following result is analogous to \Cref{prop-decomposition-gateaux-derivative-pdf-rep} but is extended to the dynamic treatment regimes:
\begin{align*}
\MoveEqLeft{\frac{\Psi(\tilde{P}^i_\epsilon)-\Psi(\tilde{P})}{\epsilon}} \nonumber\\
&=\int \int y \cdot \left(\frac{\tilde{p}_\epsilon(y , \bar{a}, \bar{x})}{\tilde{p}_\epsilon(\bar{a}, \bar{x})}\prod_{t=1}^T \frac{\tilde{p}_\epsilon(x_t , \bar{a}_{t-1}, \bar{x}_{t-1})}{\tilde{p}_\epsilon(\bar{a}_{t-1}, \bar{x}_{t-1}))} \cdot \tilde{p}_\epsilon(x_0) - \frac{\tilde{p}(y , \bar{a}, \bar{x})}{\tilde{p}(\bar{a}, \bar{x})}\prod_{t=1}^T \frac{\tilde{p}(x_t , \bar{a}_{t-1}, \bar{x}_{t-1})}{\tilde{p}(\bar{a}_{t-1}, \bar{x}_{t-1}))} \cdot \tilde{p}(x_0) \right)\dif \bar{x} \dif y\\
&=\int \int y \cdot \Delta_\epsilon\left( \frac{\tilde{p}(y , \bar{a}, \bar{x})}{\tilde{p}(\bar{a}, \bar{x})}\prod_{t=1}^T \frac{\tilde{p}(x_t , \bar{a}_{t-1}, \bar{x}_{t-1})}{\tilde{p}(\bar{a}_{t-1}, \bar{x}_{t-1}))} \cdot \tilde{p}(x_0) \right)\dif \bar{x} \dif y\\
&=\int \int y \cdot \left[\Delta_\epsilon\left( \frac{\tilde{p}(y , \bar{a}, \bar{x})}{\tilde{p}(\bar{a}, \bar{x})}\right)\prod_{t=1}^T \frac{\tilde{p}(x_t , \bar{a}_{t-1}, \bar{x}_{t-1})}{\tilde{p}(\bar{a}_{t-1}, \bar{x}_{t-1}))} \cdot \tilde{p}(x_0) \right. \\
&\qquad \qquad\left.+ \sum_{k=1}^T \frac{\tilde{p}(y , \bar{a}, \bar{x})}{\tilde{p}(\bar{a}, \bar{x})}
\Delta_\epsilon\left(\frac{\tilde{p}(x_k , \bar{a}_{k-1}, \bar{x}_{k-1})}{\tilde{p}(\bar{a}_{k-1}, \bar{x}_{k-1}))} \right)
 \prod_{t\in [T]\backslash k} \frac{\tilde{p}(x_t , \bar{a}_{t-1}, \bar{x}_{t-1})}{\tilde{p}(\bar{a}_{t-1}, \bar{x}_{t-1}))} \cdot \tilde{p}(x_0) \right.\\
&\qquad \qquad\left.+  \frac{\tilde{p}(y , \bar{a}, \bar{x})}{\tilde{p}(\bar{a}, \bar{x})}\prod_{t=1}^T \frac{\tilde{p}(x_t , \bar{a}_{t-1}, \bar{x}_{t-1})}{\tilde{p}(\bar{a}_{t-1}, \bar{x}_{t-1}))} \cdot \Delta_\epsilon\left(\tilde{p}(x_0) \right)\right]\dif \bar{x} \dif y
\end{align*}
\begin{align*}
&=\int \int y \cdot \left[ \frac{
\tilde{p}(\bar{a}, \bar{x})\Delta_\epsilon\left(\tilde{p}(y , \bar{a}, \bar{x})\right)
 - \tilde{p}(y , \bar{a}, \bar{x})\Delta_\epsilon\left(\tilde{p}(\bar{a}, \bar{x})\right)
 }{\tilde{p}(\bar{a}, \bar{x})\tilde{p}_\epsilon(\bar{a}, \bar{x})}
\prod_{t=1}^T \frac{\tilde{p}(x_t , \bar{a}_{t-1}, \bar{x}_{t-1})}{\tilde{p}(\bar{a}_{t-1}, \bar{x}_{t-1})} \cdot \tilde{p}(x_0) \right. \\
&\qquad \qquad+ \sum_{k=1}^T \left\{\frac{\tilde{p}(y , \bar{a}, \bar{x})}{\tilde{p}(\bar{a}, \bar{x})}
\frac{\tilde{p}(\bar{a}_{k-1}, \bar{x}_{k-1})\Delta_\epsilon\left(\tilde{p}(x_k , \bar{a}_{k-1}, \bar{x}_{k-1})\right) - 
\tilde{p}(x_k , \bar{a}_{k-1}, \bar{x}_{k-1})
\Delta_\epsilon\left(\tilde{p}(\bar{a}_{k-1}, \bar{x}_{k-1})\right)
}{\tilde{p}(\bar{a}_{k-1}, \bar{x}_{k-1})\tilde{p}_\epsilon(\bar{a}_{k-1}, \bar{x}_{k-1})}\right. 
\\
&\qquad \qquad \qquad \left.\prod_{t\in [T]\backslash k} \frac{\tilde{p}(x_t , \bar{a}_{t-1}, \bar{x}_{t-1})}{\tilde{p}(\bar{a}_{t-1}, \bar{x}_{t-1})} \cdot \tilde{p}(x_0) \right\}\\
&\qquad \qquad\left.+  \frac{\tilde{p}(y , \bar{a}, \bar{x})}{\tilde{p}(\bar{a}, \bar{x})}\prod_{t=1}^T \frac{\tilde{p}(x_t , \bar{a}_{t-1}, \bar{x}_{t-1})}{\tilde{p}(\bar{a}_{t-1}, \bar{x}_{t-1})} \cdot \Delta_\epsilon\left(\tilde{p}(x_0) \right)\right]\dif \bar{x} \dif y.
\end{align*}
}
\end{proof}

\section{Proofs for \Cref{sec-stochopt}}
\new{
\subsection{Intermediate results}
The following is a standard result about non-degenerate linear programs. 
\begin{lemma}[Primal and dual basic feasible solution]\label{lemma-primal-dual-bfs}
    If $\beta$ is a basis corresponding to an optimal solution, then $v^\top = c_\beta^\top M_\beta^{-1}$ is also a dual optimal solution. 
\end{lemma}
}
\rev{
\begin{lemma}[Matrix inverse finite-difference approximation stability.]\label{lemma-matrix-inverse-fd-stability}
Suppose that $M$ is an invertible matrix and \Cref{asn-matrixentriesarewellapproxfnls}, i.e. its entries are stable under finite differences. 
\begin{align*} 
   \eps^{-1} ((M^\eps)^{-1}- M^{-1}) = 
     -  M^{-1}   G(M^{-1}) + O(\eps)
\end{align*} 
\end{lemma}
\begin{proof}[Proof of \Cref{lemma-matrix-inverse-fd-stability}]
        % $$\left( (M_\beta^\eps)^{-1}  - (M_\beta)^{-1} \right).$$
% We analyze the term $\E_P[c^\top(u^*_\eps(X)-u^*(X))] $ (as its analysis also implies the second-order remainder, i.e. that $(\E_{P_\eps} - \E_P)[c^\top (u^*_\eps(X)-u^*(X))]$ is second order, $O(\eps^2).$
    Note that for two generic matrices $A,B$, we have the resolvent identity $(A+B)^{-1} = A^{-1} - A^{-1}B(A+B)^{-1}.$ 
    % Applying this recursively we obtain the power series expansion 
    % \begin{equation}
    %     (A+B)^{-1}=A^{-1} - A^{-1} \sum_{k=0}^\infty ( B)^k (A^{-1})^k.
    %     \label{eqn-matrix-perturbation-expansion}
    % \end{equation}
    We use this result to study  $(M^\eps_{\beta})^{-1}  - M^{-1}$ in terms of entrywise differences. 
Under \Cref{asn-matrixentriesarewellapproxfnls} we have
$M^\eps_{\beta}-M_{\beta}=\eps G+O(\eps^2)$.
Using the resolvent identity,
\begin{align}
(M^\eps_{\beta})^{-1}-M_{\beta}^{-1}
&= -\,M_{\beta}^{-1}\,(M^\eps_{\beta}-M_{\beta})\,(M^\eps_{\beta})^{-1}.
\label{eqn-matrix-perturbation-expansion}
\end{align}
Applying the same identity once more (i.e., writing $(M^\eps_{\beta})^{-1}
= M_{\beta}^{-1}-M_{\beta}^{-1}(M^\eps_{\beta}-M_{\beta})(M^\eps_{\beta})^{-1}$ inside
\eqref{eqn-matrix-perturbation-expansion}) yields the exact second-order decomposition
\begin{align}
(M^\eps_{\beta})^{-1}-M_{\beta}^{-1}
&= -\,M_{\beta}^{-1}\,(M^\eps_{\beta}-M_{\beta})\,M_{\beta}^{-1}
\nonumber\\
&\quad\;+\;M_{\beta}^{-1}\,(M^\eps_{\beta}-M_{\beta})\,M_{\beta}^{-1}\,(M^\eps_{\beta}-M_{\beta})\,(M^\eps_{\beta})^{-1}.
\label{eqn-matrix-perturbation-expansion-2ndorder}
\end{align}
Since $M^\eps_{\beta}-M_{\beta}=\eps G+O(\eps^2)$ and $\|M_\beta^{-1}\|,\|(M^\eps_\beta)^{-1}\|$ remain bounded for $\eps$ small,
\[
(M^\eps_{\beta})^{-1}-M_{\beta}^{-1}
= -\,\eps\,M_{\beta}^{-1} G M_{\beta}^{-1} + O(\eps^2).
\]
\end{proof}
}
\subsubsection{Conditions in $\epsilon$ to maintain the same basis }

\rev{
\begin{proposition}[Primitive conditions for same optimal basis]
\label{cor:eps-same-optimal-basis}
Consider the linear program $\min_{u\in\mathbb{R}^n} \{c^\top u \ \text{s.t. } Mu \le b\}$, 
with $M\in\mathbb{R}^{m\times n}$, $b\in\mathbb{R}^m$, $c\in\mathbb{R}^n$. 
Let $\beta$ index the optimal basis, $N:=[n]\setminus\beta$ index the non-basic variables, and define the reduced costs vector $r_N^* := c_N - M_{:,N}^\top v^*$. Note that therefore the basic variables are $u_B^* := M_\beta^{-1} b$ and recall that the dual variables are $ v^* := M_\beta^{-\top} c_\beta.$
% \[
% u_B^* := M_\beta^{-1} b, \qquad
% v^* := M_\beta^{-\top} c_\beta, \qquad
% r_N^* := c_N - M_{:,N}^\top v^*.
% \]
Define the margins of non-degeneracy (\Cref{asn-nondegeneracy}) $\eta := \min_{i\in\beta} (u_B^*)_i > 0, 
\zeta := \min_{j\in N} (r_{N,j}^*) > 0$. 
Consider simultaneous perturbations $(\Delta_M,\Delta_b,\Delta_c)$ to $(M,b,c)$ arising from the perturbed density $P_\varepsilon$.
Suppose that $Mu$, $c^\top u$, and $b^\top v$ are (vector) expectation functionals with integrands bounded by $Y_{\max}$
under both $p$ and $\tilde\delta$, then $
\|\Delta_M\|_\infty,\ \|\Delta_b\|_\infty,\ \|\Delta_c\|_\infty \le 2\varepsilon Y_{\max}$,
where for matrices $\|\cdot\|_\infty$ denotes the induced $\ell_\infty$ operator norm. Define the stability constants $K_M:= ({\,8\,Y_{\max}\,\|M_\beta^{-1}\|_\infty})^{-1},
K_b:= 
3\,\eta
({\,8\,Y_{\max}\,\|M_\beta^{-1}\|_\infty(1+\|u_B^*\|_\infty)})^{-1}$ and 
% In the proposition statement, replace only the definition of K_c by:
$K_c :=
{\zeta}{
\left( 4Y_{\max}\Bigl(
1+\frac43\|M_\beta^{-\top}\|_\infty(\|c_\beta\|_\infty+2Y_{\max})
+\frac43\|M_{:,N}^\top\|_\infty\|M_\beta^{-\top}\|_\infty
\bigl(1+\|M_\beta^{-\top}\|_\infty\|c_\beta\|_\infty\bigr)
\Bigr) \right)}$.
% \]
% $K_c := ({\zeta{\,8\,Y_{\max}\!\big(
% 1+\|v^*\|_\infty
% +\|M_{:,N}^\top\|_\infty\|M_\beta^{-\top}\|_\infty(1+\|c_\beta\|_\infty)
% \big)}})^{-1}$.
% \begin{align*}
% K_M &:= \frac{1}{\,8\,Y_{\max}\,\|M_\beta^{-1}\|_\infty},\\
% K_b &:= \frac{3\,\eta}{\,8\,Y_{\max}\,\|M_\beta^{-1}\|_\infty(1+\|u_B^*\|_\infty)},\\
% K_c &:= \frac{\zeta}{\,8\,Y_{\max}\!\big(
% 1+\|v^*\|_\infty
% +\|M_{:,N}^\top\|_\infty\|M_\beta^{-\top}\|_\infty(1+\|c_\beta\|_\infty)
% \big)}.
% \end{align*}
Then the same basis $\beta$ remains optimal for all $0\le\varepsilon\le \min\{K_M,K_b,K_c\}$.
\end{proposition}
\begin{proof}[Proof of \cref{cor:eps-same-optimal-basis}]
We use the induced $\ell_\infty$ matrix norm, so that $\|Ax\|_\infty\le \|A\|_\infty\|x\|_\infty$
and $\|AB\|_\infty\le \|A\|_\infty\|B\|_\infty$.
A fixed basis $\beta$ remains optimal if (i) the associated basic vector remains componentwise nonnegative
and (ii) all nonbasic reduced costs remain nonnegative.
% \paragraph{Mixture bounds.}
By assumption, $Mu$, $c^\top u$, and $b^\top v$ are expectation functionals with integrands bounded by $Y_{\max}$
under both $p$ and $\tilde\delta$. Along the mixture path
$p_\varepsilon=(1-\varepsilon)p+\varepsilon\tilde\delta$, each coefficient varies linearly in $\varepsilon$,
and hence
\[
\|\Delta_M\|_\infty,\ \|\Delta_b\|_\infty,\ \|\Delta_c\|_\infty \le 2\varepsilon Y_{\max}.
\]
% \paragraph{Invertibility of the basis block.}
Let $M_{\beta,\varepsilon}:=M_\beta+\Delta_{M,\beta}$. Since
$\|M_\beta^{-1}\Delta_{M,\beta}\|_\infty \le 2\varepsilon Y_{\max}\|M_\beta^{-1}\|_\infty$,
the condition $\varepsilon\le K_M$ implies $\|M_\beta^{-1}\Delta_{M,\beta}\|_\infty\le \tfrac14$.
By the Neumann--series bound, $M_{\beta,\varepsilon}$ is invertible and
\[
\|M_{\beta,\varepsilon}^{-1}\|_\infty \le \tfrac{4}{3}\|M_\beta^{-1}\|_\infty,
\qquad
\|M_{\beta,\varepsilon}^{-\top}\|_\infty \le \tfrac{4}{3}\|M_\beta^{-\top}\|_\infty.
\]
The perturbed basic vector satisfies
\[
u_{B,\varepsilon}
= M_{\beta,\varepsilon}^{-1}(b+\Delta_b)
= u_B^* + M_{\beta,\varepsilon}^{-1}(\Delta_b-\Delta_{M,\beta}u_B^*).
\]
Using the mixture bounds and the inverse bound above,
\[
\|u_{B,\varepsilon}-u_B^*\|_\infty
\le \tfrac{4}{3}\|M_\beta^{-1}\|_\infty\cdot 2\varepsilon Y_{\max}(1+\|u_B^*\|_\infty).
\]
If $\varepsilon\le K_b$, the right-hand side is at most $\eta/2$, so
$u_{B,\varepsilon}\ge 0$ componentwise and the same basis remains feasible.
Reduced-cost nonnegativity. Let $v_\varepsilon:=M_{\beta,\varepsilon}^{-\top}(c_\beta+\Delta_{c,\beta})$. Then
\[
r_{N,\varepsilon}
= (c_N+\Delta_{c,N})-(M_{:,N}+\Delta_{M,:,N})^\top v_\varepsilon.
\]
Subtracting $r_N^*$ and applying triangle inequality yields
\[
\|r_{N,\varepsilon}-r_N^*\|_\infty
\le \|\Delta_c\|_\infty
+ \|\Delta_M\|_\infty\|v_\varepsilon\|_\infty
+ \|M_{:,N}^\top\|_\infty\|v_\varepsilon-v^*\|_\infty.
\]
Using the inverse bounds above together with $\|\Delta_c\|_\infty\le 2\varepsilon Y_{\max}$ and
the fact that $0\le \varepsilon\le 1$ along the mixture path gives
\[
\|v_\varepsilon\|_\infty
\le \|M_{\beta,\varepsilon}^{-\top}\|_\infty \|c_\beta+\Delta_{c,\beta}\|_\infty
\le \frac43 \|M_\beta^{-\top}\|_\infty(\|c_\beta\|_\infty+2Y_{\max}).
\]
Also,
\[
v_\varepsilon-v^*
=
(M_{\beta,\varepsilon}^{-\top}-M_\beta^{-\top})c_\beta
+
M_{\beta,\varepsilon}^{-\top}\Delta_{c,\beta},
\]
and
\[
M_{\beta,\varepsilon}^{-\top}-M_\beta^{-\top}
=
-\,M_{\beta,\varepsilon}^{-\top}\Delta_{M,\beta}^\top M_\beta^{-\top},
\]
so
\[
\|v_\varepsilon-v^*\|_\infty
\le
\frac83 \varepsilon Y_{\max}\|M_\beta^{-\top}\|_\infty
\Bigl(1+\|M_\beta^{-\top}\|_\infty\|c_\beta\|_\infty\Bigr).
\]
Combining these bounds,
\[
\|r_{N,\varepsilon}-r_N^*\|_\infty
\le
2\varepsilon Y_{\max}
\Bigl(
1+\frac43\|M_\beta^{-\top}\|_\infty(\|c_\beta\|_\infty+2Y_{\max})
+\frac43\|M_{:,N}^\top\|_\infty\|M_\beta^{-\top}\|_\infty
\bigl(1+\|M_\beta^{-\top}\|_\infty\|c_\beta\|_\infty\bigr)
\Bigr).
\]
Thus, if $\varepsilon\le K_c$, then $\|r_{N,\varepsilon}-r_N^*\|_\infty\le \zeta/2$ and hence
$r_{N,\varepsilon}\ge 0$.
\end{proof}
}

% \if\forjournal 0 
% \begin{proof}[Proof of \Cref{prop-general-LP}]
% \else
\subsubsection{Proof of approximation stability for general marginalized linear programs}
\begin{proof}{Proof of \Cref{prop-general-LP}.}
% \fi 
\new{
The proof applies the product rule to ``closed-form representations" of the functional via basic feasible solutions (and interprets these from the output). 
}

\new{Our proof strategy broadly follows from that of \cite{freund1985postoptimal}, who studies perturbations to linear programs of the constraint coefficient matrix. However our perturbation is entrywise upon the transition probability (a conditional transition probability); hence nonlinear in $\eps$; we compute the entrywise matrix derivative directly using an alternative technique described in the proof of \citet[Thm.1 ]{freund1985postoptimal}. But, the overall argument for sensitivity analysis subject to left-hand-side perturbations based on active set identification is similar, so we recount the argument here as a preface. 
}
\new{
Namely, note that for $\eps$ small enough, the \textit{active basis} remains the same when solving with respect to unperturbed or perturbed distributions, $P_0$ or $P_\eps$. Consider the primal program in standard form, $$d(\eps) = \max\{ c^\top x \colon M^\eps x = b\}.$$ 
\cite{freund1985postoptimal} considers perturbations where $M^\eps=M+\epsilon G$. Letting $\beta$ denote an index vector for the optimal basis, note that $$d(\eps) = c_\beta x^*(\eps) = c_\beta (M^\eps_\beta)^{-1} b_\beta,$$ by nondegeneracy and properties of the basic feasible solution for linear programming. \cite{freund1985postoptimal} notes that the derivative $d'(\eps) = - v Gu$ (where $(v, u)$ are the dual and primal optimal solutions, respectively) can be obtained from the well-known matrix identity: 
\begin{equation}\frac{d M^{-1}(\eps)}{d \eps} = - M^{-1}(\eps) \left(\frac{dM(\eps)}{d \eps}\right)M^{-1}(\eps). \label{eqn-matrixinverse} 
\end{equation} 
}
\new{
We simply compute $\frac{d M^{-1}(\eps)}{d \eps}$ for the perturbations to conditional probabilities, rather than linear perturbations restricted to the form $M+\eps G$, for a perturbation matrix $G$. 
}

 \new{   
Since we have:
$$ 
\Psi\left(P_\epsilon\right)=\left(c^\epsilon\right)^{\top}\left(\left(M_\beta^\epsilon\right)^{-1} b_\beta^\epsilon\right),
$$
then 
\begin{align*} 
\frac{\intd}{\intd\eps} \fnl(P_\eps)\Big\vert  &= \frac{\intd}{\intd\eps}
\left\{ \left(c^\epsilon\right)^{\top}\left(\left(M_\beta^\epsilon\right)^{-1} b_\beta^\epsilon\right)
\right\} 
\\
&= 
 \{\frac{\intd}{\intd\eps} \left(c^\epsilon\right)^{\top}\} \left(M_\beta^{-1} b_\beta\right) +   c^{\top} \left(\left\{\frac{\intd}{\intd\eps}\left(M_\beta^\epsilon\right)^{-1}
 \right\}
 b_\beta\right)
 +
  c^\top
(    M_{\beta}^{-1}
\frac{\intd}{\intd\eps}\{b_{\beta}^\eps(X)\} )  \\
&= 
\{\frac{\intd}{\intd\eps} \left(c^\epsilon\right)^{\top}\} \left(M_\beta^{-1} b_\beta\right) +   
c^{\top} \left(\left\{ - M_\beta^{-1} \left(\frac{ \intd M^\eps_\beta}{\intd \eps}\right)M_\beta^{-1} 
 \right\}
 b_\beta\right)
 +
  c^\top 
(    M_\beta^{-1}
\frac{\intd}{\intd\eps}\{b_{\beta}^\eps(X)\} ) 
%\\
% & = c(X_i)\left(M_\beta^{-1} b_\beta\right)(X_i) - \fnl(P) + c^{\top} \left(\left\{ - M_\beta^{-1} \left(\frac{dM^\eps_\beta}{d \eps}\right)M_\beta^{-1} 
%  \right\}
%  b_\beta\right)
%  +
%   c^\top 
% (    M_\beta^{-1}
% \frac{\intd}{\intd\eps}\{b_{\beta}^\eps(X)\} ).
\end{align*} 
Applying \Cref{lemma-primal-dual-bfs}, we obtain the simplification, 
\begin{align*}
  &   \frac{\intd}{\intd\eps} \fnl(P_\eps) \Big\vert_{\eps=0} \\&
 = \left\{\frac{\intd}{\intd\eps} \left(c^\epsilon\right)^{\top}\Big\vert_{\eps=0}\right\}  u^*
 -(v^*)^\top  \left(\frac{\intd M^\eps_\beta}{\intd \eps}\Big\vert_{\eps=0}\right) u^*
 +
  c^\top 
(    M_\beta^{-1}
\frac{\intd}{\intd\eps}\{b_{\beta}^\eps\}\Big\vert_{\eps=0} ).
\end{align*}
}
\end{proof}

% \if\forjournal 0 
% \begin{proof}[Proof of \Cref{prop-fd-apx-error-lps}]
% \else
\begin{proof}{Proof of \Cref{prop-fd-apx-error-lps}.}
% \fi 
\new{
\begin{align*}
&(\fnl(P_\eps) - \fnl(P)) \\
&= \left(c^\epsilon\right)^{\top}\left(\left(M_\beta^\epsilon\right)^{-1} b_\beta^\epsilon\right)
- c^{\top}M_\beta^{-1} b_\beta
\\
& = \left(c^\epsilon - c\right)^{\top}(M_\beta^\eps)^{-1} b_\beta^\eps
+ c^{\top}\left( (M_\beta^\eps)^{-1}  - (M_\beta)^{-1} \right) b_\beta^\eps
+ c^\top  M_\beta^{-1} (b_\beta^\eps- b_\beta)\\
& = 
\underbrace{\left(c^\epsilon - c\right)^{\top}(M_\beta^\eps)^{-1} b_\beta^\eps}_{T_1}
+ 
\underbrace{c^{\top}\left( (M_\beta^\eps)^{-1}  - (M_\beta)^{-1} \right) b_\beta
}_{T_2}
+ 
\underbrace{c^{\top}\left( (M_\beta^\eps)^{-1}  - (M_\beta)^{-1} \right) (b_\beta^\eps- b_\beta)}_{T_3}
+ 
\underbrace{c^\top  M_\beta^{-1} (b_\beta^\eps- b_\beta)}_{T_4}.
% &  = \left(c^\epsilon - c\right)^{\top}M_\beta^{-1} b_\beta
% + \left(c^\epsilon - c\right)^{\top}\left( (M_\beta^\eps)^{-1} b_\beta^\eps - (M_\beta)^{-1} b_\beta \right) 
% + c^\top ( (M_\beta^\eps)^{-1} b_\beta^\eps- M_\beta^{-1} b_\beta^\eps)
% + c^\top ( M_\beta^{-1} b_\beta^\eps- M_\beta^{-1} b_\beta)
% \\
% &= (\E_{P_\eps} - \E_P)[c^\top u(X)] + (\E_{P_\eps} - \E_P)[c^\top (u^*_\eps(X)-u^*(X))] + \E_P[c^\top(u^*_\eps(X)-u^*(X))] 
\end{align*}
The finite-difference approximation error of $T_1,T_4$ can be directly concluded as follows. Observe that 
\begin{align*} \frac{T_1}{\eps}
%\left(c^\epsilon - c\right)^{\top}(M_\beta^\eps)^{-1} b_\beta^\eps
&= (c - \delta_{X_i})^\top u_\eps^*,
\\
\frac{T_4}{\eps}
% (c^\top  M_\beta^{-1} (b_\beta^\eps- b_\beta))
&= 
 c^\top 
(    M_\beta^{-1}
\frac{\intd}{\intd\eps}\{b_{\beta}^\eps\}\Big\vert_{\eps=0} ).
\end{align*}
Next we bound $T_2, T_3,$ and establish the approximation error of $u_\eps$ by establishing the matrix perturbation error of 
$$\left( (M_\beta^\eps)^{-1}  - (M_\beta)^{-1} \right).$$
Applying \Cref{lemma-matrix-inverse-fd-stability} to $M_\beta^\eps$, we obtain that $ (M^\eps_{\beta})^{-1}  - M_{\beta}^{-1} =  - \eps M_{\beta}^{-1}   G(M_{\beta}^{-1}) + O(\eps^2).$
% \az{refactor this out as a lemma?}
% % We analyze the term $\E_P[c^\top(u^*_\eps(X)-u^*(X))] $ (as its analysis also implies the second-order remainder, i.e. that $(\E_{P_\eps} - \E_P)[c^\top (u^*_\eps(X)-u^*(X))]$ is second order, $O(\eps^2).$
%     Note that for two generic matrices $A,B$, we have $(A+B)^{-1} = A^{-1} - A^{-1}B(A+B)^{-1}.$ Applying this recursively we obtain the power series expansion 
%     \begin{equation}
%         (A+B)^{-1}=A^{-1} - A^{-1} \sum_{k=0}^\infty ( B)^k (A^{-1})^k.
%         \label{eqn-matrix-perturbation-expansion}
%     \end{equation}
%     We use this result to study  $(M^\eps_{\beta})^{-1}  - M_{\beta}^{-1}$ in terms of entrywise differences, which are assumed to be well-behaved under finite-differences (under marginalization) by \Cref{asn-matrixentriesarewellapproxfnls}. 
% Applying \cref{eqn-matrix-perturbation-expansion} under \Cref{asn-matrixentriesarewellapproxfnls} implies that 
% \begin{align} 
%     (M^\eps_{\beta})^{-1}  - M_{\beta}^{-1} &= 
%      - M_{\beta}^{-1} \sum_{k=0}^\infty ( \eps G + O(\eps^2) )^k (M_{\beta}^{-1})^k 
%     \nonumber
%      \\
%      & = 
%      - \eps M_{\beta}^{-1}   G(M_{\beta}^{-1}). + O(\eps^2)
%      % \text{ higher order terms in epsilon}
% % (M^\eps_{\beta})^{-1}  - M_{\beta}^{-1}  = 
%      % - \eps M_{\beta}^{-1}   G(M_{\beta}^{-1}) + \text{ higher order terms in epsilon}
% \end{align} 
% \az{refactoring out the above}
% \flag{notation}
Therefore, 
\begin{align*}
    &\frac{T_2}{\eps}  =  -  c^\top M_{\beta}^{-1}   G(M_{\beta}^{-1})b_\beta = -(v^*)^\top  \left(\frac{dM^\eps_\beta}{d \eps}\Big\vert_{\eps=0}\right) u^* + O(\eps), \qquad 
    % \\ &
        \frac{T_3}{\eps}  =  O(\eps) \\
        &\int u^*(x) - u^*_\eps(x) = O(\eps).
\end{align*}
The claim follows by collecting terms and approximation terms.
}
\end{proof}

\subsection{Proofs for infinite-horizon evaluation (\Cref{ex-infhorizonope})} 
\begin{proof}{Proof of \Cref{prop-infhorizonmdp}}
\textbf{Perturbations for the optimal policy. }
Define the $(s,s')$ entry of the perturbed transition matrix, in the direction of a generic observation $(\td{s},\td{a},\td{s}'),$ as $P^\eps_a(s,s') = \frac{(1-\eps) p(s,a,s') + \eps \indic{s,a,s'} }{ (1-\eps) d(s,a) + \eps\indic{s,a}}.$ 

Define the perturbed version of \Cref{eqn-primalmdplp}:
\begin{equation}
\textstyle \fnl_D(P_\eps) = 
 \underset{V}{\operatorname{min}}\left\{ (1-\gamma)\sum_s \left( (1-\eps) \mu_0(s) + \eps \indic{\td{s}}\right) V_\eps^*(s) \colon \left(I-\gamma P_{a}^\eps\right) {V}-{r}_{a} \geq 0, \quad \forall a \in \mathcal{A} \right\}.
 \label{eqn-perturbedprimalmdplp}
\end{equation}

Let $V_\eps^*(s) \in \arg\min \fnl_D(P_\eps)$ achieve the optimal solution of $\fnl_D(P_\eps)$.
We will repeatedly use the fact that ${\frac{d}{d\eps} \{ (1-\eps) p(o) + \eps\indic{o} \} = \eps( \indic{o} - p(o)).}$
We have:
\begin{align*}
    \frac{d}{d\eps} \fnl_D(P_\eps) \Big\vert_{\eps=0} &= 
    (1-\gamma) \sum_s \frac{d}{d\eps} \left\{ (1-\eps) \mu_0(s) + \eps \indic{\td{s}}  \right\}\Big\vert_{\eps=0} V^*(s) + (1-\gamma) \sum_s \mu_0(s) \frac{d}{d\eps}  \{V_\eps^*(s)\}\Big\vert_{\eps=0} \\
    &= 
    (1-\gamma) \sum_s (\indic{s=\td{s}} - \mu_0(s)) V^*(s) + (1-\gamma) \sum_s \mu_0(s) \frac{d}{d\eps}  \{V_\eps^*(s)\}\Big\vert_{\eps=0}, 
\end{align*}
where $V^*(s)$ is the unperturbed solution. Letting $P_{a^*}$ denote the transition matrix under $a(s) \sim \pi^*$, note that we have $V_\eps^* = (I - \gamma P_{a^*}^\eps)^{-1} r_{a^*}.$

Entrywise, where we overload notation for $a^*=a^*(s)$ (the optimal action at $s$, i.e., under the active basis), and $\indic{\td{s},\td{a},\td{s'}}= \indic{(s,a^*, s')=(\td{s},\td{a},\td{s'})},$ etc:
\begin{align*}
    \frac{d (I-\gamma P^\eps_{a^*})_{(s,s')} }{d \eps}\Big\vert_{\eps=0 } &= -\gamma \; \frac{d}{d\eps}  \left\{ 
    \frac{(1-\eps) p(s,a^*,s') + \eps \indic{\td{s},\td{a},\td{s}'} }{(1-\eps) d(s,a^*) + \eps \indic{\td{s},\td{a} }}
    \right\} \Big\vert_{\eps=0 }\\
     & = -\gamma 
    \left( \frac{\indic{\td{s},\td{a},\td{s}'} - p(s,a^*,s')}{d(s,a^*)} - p(s,a^*,s') \frac{\indic{\td{s}, \td{a}} - d(s,a^*)}{d(s,a^*)^2} \right) \\
        &=-\gamma 
    \left( \frac{\indic{\td{s},\td{a},\td{s}'} }{d(s,a^*)} - p(s'\mid s,a^*) \frac{\indic{\td{s}, \td{a}} }{d(s,a^*)} \right).
\end{align*}
Hence, applying \Cref{eqn-matrixinverse}, 
\begin{align*}
    \frac{d}{d\eps} \fnl_D(P_\eps) \Big\vert_{\eps=0} &= (1-\gamma) \sum_s (\indic{s=\td{s}} - \mu_0(s)) V^*(s) + (1-\gamma) \sum_s \mu_0(s) \frac{d}{d\eps}  \{V_\eps^*(s)\}\Big\vert_{\eps=0} \\
     &= (1-\gamma) \sum_s (\indic{s=\td{s}} - \mu_0(s)) V^*(s)\\
     &\qquad - (1-\gamma){\mu_0} (I-\gamma P_{a^*})^{-1}  \left(  \frac{d (I-\gamma P^\eps_{a^*}) }{d \eps}\Big\vert_{\eps=0 } \right) (I-\gamma P_{a^*})^{-1} {r}.
\end{align*}
% \az{go through and check dim and transposes}
Note that, as in \cite{freund1985postoptimal}, $(1-\gamma){\mu_0} (I-\gamma P_{a^*})^{-1} = \mu^* $ is the dual (unperturbed) optimal solution of \Cref{eqn-dualmdplp} and $ (I-\gamma P_{a^*})^{-1} {r}$ is the primal (unperturbed) optimal solution $V^*$. 
Therefore, 
\begin{align}
    \MoveEqLeft{\frac{d}{d\eps} \fnl_D(P_\eps) \Big\vert_{\eps=0}} \\
    &= 
    (1-\gamma)(V^*(\td{s}) - \sum_s \mu_0(s) V^*(s) )\nonumber\\
    & \qquad -  (-1) \cdot \gamma \sum_s \mu^*(s,a) \sum_{s'} V^*(s')     \left( \frac{\indic{\td{s},\td{a},\td{s}'} - p(s,a^*,s')}{d(s,a^*)} - p(s,a^*,s') \frac{\indic{\td{s}, \td{a}} - d(s,a^*)}{d(s,a^*)^2} \right) \label{eqn-infhorizonmdp-fd-sub} \\
    &  =   (1-\gamma)(V^*(\td{s}) - \sum_s \mu_0(s) V^*(s) ) + \gamma \frac{\mu^*(\td{s},\td{a})}{d(\td{s},\td{a})}
    \left(  V^*(\td{s}') - (P^\pi V^*)(\td{s}) \right),\nonumber
\end{align}
where $(P^\pi V^*)(\td{s})= \sum_{s'} p(s' \mid \td{s},a) V^*(\td{s})$ is the Bellman operator evaluated starting from the state $\td{s}$. 

Further note that $\gamma(P^\pi V^*)(\td{s}) = (V^*(\td{s}) - r(\td{s},\td{a}))$ (i.e., a Bellman residual term), and $\fnl_D(P)={(1-\gamma)\sum_s \mu_0(s) V^*(s) )}.$

Hence
\begin{equation}
    \frac{d}{d\eps} \fnl_D(P_\eps) \Big\vert_{\eps=0} =  (1-\gamma)V^*(\td{s})
    +\frac{\mu^*(\td{s},\td{a})}{d(\td{s},\td{a})}
    \left( r(\td{s},\td{a}) + \gamma V^*(\td{s}') - V^*(\td{s}) \right) - \fnl_D(P).
\end{equation}

\end{proof}
% \if\forjournal 0 
% \begin{proof}[Proof of \Cref{prop-mdp-finitedifferenceapprox}]
% \else
\begin{proof}{Proof of \Cref{prop-mdp-finitedifferenceapprox}.}
% \fi 
We analyze the error introduced by finite differencing. 
Consider the dual formulation of \Cref{eqn-primalmdplp}, and the perturbed formulation of \Cref{eqn-perturbedprimalmdplp}. Fix $\eps$ small enough such that the active basis remains the same under $P$ and $P_\eps$; let $P^\pi$ denote the submatrix of the transition matrix $P$ (stacked for the optimization problem) corresponding to the active basis (equivalently, optimal policy).

Define the entries of $G_\eps$ such that $P_\eps = P+\eps G_\eps$ (analogous to \Cref{lemma-numerical-derivative-decomposition}):
$$(G_\eps^a)_{s,s'} = \frac{\indic{\td{s},\td{a},\td{s}'}}{d_\eps(s,a)} - \frac{\indic{\td{s},\td{a}}P(s'\mid s,a)}{d_\eps(s,a)}. $$
Let $V_\eps^\pi(s), V^\pi(s)$ be the active subvector of the optimal solution achieving the optimal values of $\fnl(P_\eps), \fnl(P)$, respectively. 

Then 
\begin{align}
\eps^{-1}(\fnl_D(P_\eps) - \fnl_D(P))&
\textstyle =
\eps^{-1}(1-\gamma) \left( 
\sum_s \left( (1-\eps) \mu_0(s) + \eps \indic{\td{s}}\right) V_\eps^\pi(s) - \sum_s  \mu_0(s) V^\pi(s)
\right) \nonumber\\
&\textstyle =\eps^{-1}(1-\gamma) \left( 
\sum_s \eps\left(\indic{\td{s}}- \mu_0(s)  \right) V_\eps^\pi(s) + \sum_s  \mu_0(s)(V_\eps^\pi(s)- V^\pi(s))
\right) \nonumber\\
&\textstyle =(1-\gamma) \left( 
\sum_s \left(\indic{\td{s}}- \mu_0(s)  \right) V_\eps^\pi(s) + \eps^{-1}\sum_s  \mu_0(s)(V_\eps^\pi(s)- V^\pi(s))
\right). \label{eqn-infhorizonmdp-expansion}
\end{align}

The optimal solutions $V_\eps^\pi, V^\pi$ differ in the perturbation matrix; inverting the basic feasible solution, where invertibility is assured by nondegeneracy, gives that: \begin{align*}V_\eps^\pi=(I-\gamma P^\pi_\eps)^{-1} r_\pi, \qquad  V^\pi=(I-\gamma P^\pi)^{-1} r_\pi.\end{align*} 

We analyze $V_\eps^\pi-V^\pi= ((I-\gamma P^\pi_\eps)^{-1}-(I-\gamma P^\pi)^{-1}) r_\pi$ via \cref{eqn-matrix-perturbation-expansion}.

% via a general result for perturbed matrices. Note that for two matrices $A,B$, we have $(A+B)^{-1} = A^{-1} - A^{-1}B(A+B)^{-1}.$ Applying this recursively we obtain the power series expansion $$(A+B)^{-1}=A^{-1} - A^{-1} \sum_{k=0}^\infty ( B)^k (A^{-1})^k. $$ 

Define the approximating matrix: 
$$(G^a)_{s,s'} = \frac{\indic{\td{s},\td{a},\td{s}'}}{d(s,a)} - \frac{\indic{\td{s},\td{a}}P(s'\mid s,a)}{d(s,a)}. $$
Applying this result to our setting, with $A = (I-\gamma P^\pi)$ and $B = -\gamma \eps G_\eps$, we obtain
\begin{align}V_\eps^\pi-V^\pi &= 
%= ((I-\gamma P^\pi_\eps)^{-1}-(I-\gamma P^\pi)^{-1}) 
\eps \gamma (I-\gamma P^\pi)^{-1}  (G^\pi + (G_\eps-G^\pi) ) (I-\gamma P^\pi)^{-1} r_\pi + O(\eps^2) 
\label{eqn-valuefn-eps-nuisance}\\
&=\eps \gamma (I-\gamma P^\pi)^{-1}  (G^\pi  ) (I-\gamma P^\pi)^{-1} r_\pi  + O(\eps^2). \nonumber
\end{align} 
The statement follows by substituting in the above expression, evaluating based on definition of the entries of $G^\pi$, and substituting the above and \Cref{eqn-infhorizonmdp-expansion} into \Cref{eqn-infhorizonmdp-fd-sub} of the proof of \Cref{prop-infhorizonmdp} and following that proof onwards. 
\end{proof}

\subsection{Proofs for orthogonalized sensitivity analysis (\Cref{ex-sensitivityanalysis})}

\subsubsection{Intermediate results}

\paragraph{Sensitivity results for optimal solution to linear program. }

% \begin{lemma}[Perturbed optimal solution]\label{lemma-perturbedoptimalsolution}
% Recall that 
% $$W^*_\eps(x,y) \in \arg\max_{W} \left\{ 
%   \E_{P_\eps}[Y  W \mid A=1,X=x] \colon \E_{P_\eps}[ W \mid A=1,X=x] = 1, \forall x; \;\;  \alpha(x) \leq W(x,y)\leq  \beta(x), \forall  y
%     \right\} $$
% Consider perturbations in the direction of an (unsmoothed) Dirac delta function $\delta_{o}$at the generic observation $o=(X_i,\td{a},Y_i)$. 
% Then $W_\eps^*(x,y)-W^*(x,y)= 0$ if $x\neq X_i.$
% \end{lemma}
\new{
\begin{lemma}[Quantile function influence function (see, e.g., \citep{van2000asymptotic})]\label{lemma-quantile-if}
$$\frac{d}{d\eps}q_\tau(P_\eps)\mid_{\eps=0} =  \frac{ (\indic{y \geq q_\tau} - \tau)}{p(q_\tau)} $$
\end{lemma}
\begin{lemma}[Quantile finite-difference approximation error]\label{prop-sensitivityanalysis-quantile-fd}
% Let \begin{align*}
%     W^*(x) &\in \arg\max \{ \E[YW\mid A=1,X=x] \colon \E[W\mid A=1,X=x]=1, 0 \leq W \leq \alpha^{-1} \} && (\text{Primal}) \\  
%         v^*(x) &\in \arg\min \{ \alpha^{-1} \E[(Y-\lambda)_+\mid A=1,X=x] + \lambda^*\} && (\text{Dual}) \\  
% \end{align*}
$$ \Delta q_\tau(P) - \frac{d}{d\eps}q_\tau(P_\eps)\mid_{\eps=0} = O(\eps) $$
\end{lemma}
}

\begin{lemma}[Derivative of integral finite difference]\label{lemma-leibniz-fd} 
 \new{
    \begin{align*} &{\frac{\mathrm{d}}{\mathrm{d}\eps} 
\left\{\int_{\lambda_\eps(x)}^\infty y p(y\mid A=1,x) \intd y\right\}     \Big\vert_{\eps = 0}
} - \eps^{-1} \left( \int_{\lambda_\eps(x)}^\infty y p(y\mid A=1,x) \intd y 
- 
\int_{\lambda}^\infty y p(y\mid A=1,x) \intd y 
\right) \\&= O(\eps).
\end{align*} 
}

\end{lemma}
% \begin{proposition}[Sensitivity of optimal solution (under sufficient smoothness)]\label{lemma-sensitivityanalysis-genericmatrixderivative}
% \end{proposition}
% \begin{proposition}[Finite-dimensional analogue of optimal solution (under sufficient smoothness)]\label{lemma-sensitivityanalysis-genericmatrixderivative-fd}
% \end{proposition}
\new{
\begin{lemma}[Conditional quantile interpretation of optimal $\lambda^*$]\label{lemma-sensitivityanalysis-whichquantile}
$\lambda^*(x)$ is a conditional quantile of $Y\mid(A=1,X=x)$ at level
$\tau(x)
=1-\frac{1-\alpha(x)}{\beta(x)-\alpha(x)}
=\frac{\beta(x)-1}{\beta(x)-\alpha(x)}.
$
% $\lambda^*(x)$ is the $(1-(e_1(a(x)-b(x))^{-1})-$(conditional) quantile under the distribution $P(Y\mid A=1,X)$. 
\end{lemma}
}

In contrast to the results in the main text, the next influence function is derived without perturbing the optimal solution.
\begin{theorem}[Proof of optimization influence function (without perturbed solution) ]\label{thm-analytical-optif}
\new{
\begin{align*}
&{\left.\frac{\mathrm{d}}{\mathrm{d} \epsilon}\left\{\mathbb{E}_{P_\epsilon}\left[\mathbb{E}_{P_\epsilon}\left[Y W^* \mid A=1, X\right]\right]\right\}\right|_{\epsilon=0}}\\
&  = 
\frac{(Y_i-\lambda^*)W^*(X_i,Y_i)\indic{A=1} - \E_{P} [(Y-\lambda^*)W^* \indic{A=1}\mid X_i]}{p(A=1\mid X_i)} + \E_{P} [(Y-\lambda)W^*\indic{A=1}\mid X_i] - \fnl(P).
\end{align*}}
\end{theorem}

\subsubsection{Proofs of orthogonalized sensitivity analysis results.}

% \if\forjournal 0 
% \begin{proof}[Proof of \Cref{proposition-if-closedform}]
% \else
\begin{proof}{Proof of \Cref{proposition-if-closedform}.}
% \fi 
\new{
We can also define the perturbed optimal primal and dual closed-form solutions, $W_\eps(x,y),$ and $\lambda^*_\eps(x)$, which solve the primal and dual optimization problems under the perturbed distributions:  
\begin{align*} 
W^*_\eps(x,y) &\in \arg\max_{W} \left\{ 
  \E_{P_\eps} [Y  W \mid A=1,X=x] \colon \E_{P_\eps}[ W \mid A=1,X=x] = 1, \forall x; \;\; \alpha(x) \leq W(x,y)\leq  \beta(x), \forall  y
    \right\} \\
    \lambda^*_\eps(x) &\in \arg\min_{\lambda \in\mathbb{R}} \left\{ 
\left(\alpha(x)\E_{P_\eps}[ (\lambda-Y)_-\mid  A=1,X=x])+ \beta(x) \E_{P_\eps}[(Y -\lambda)_+ \mid A=1,X=x]\right) + \lambda 
    \right\} \\
% \end{align*} 
% \begin{align*}
\phi(o) &= \frac{d}{d\eps} \left\{ 
\E_{P_\eps}[ \alpha(x)(\lambda_\eps -Y)_- + \beta(x) (Y -\lambda_\eps)_+ \mid  A=1,X=x]+ \lambda_\eps 
    \right\}
    \mid_{\eps =0 }
\end{align*}
Then: 
\begin{align}
 \textstyle   \frac{\intd}{\intd \eps}  
\{\fnl(P_\eps) \}\vert_{\eps = 0}= &\textstyle 
    \frac{\mathrm{d}}{\mathrm{d}\eps} \left\{
    \int y W_\eps^* p_\eps(y\mid A=1,x) p_\eps (x) \intd x  \right\}\Big\vert_{\eps = 0} = 
    \int        \frac{\mathrm{d}}{\mathrm{d}\eps}  \left\{y W_\eps^* p_\eps(y\mid A=1,x) p_\eps (x) \right\}\Big\vert_{\eps = 0} \intd x  \label{eqn-sensitivity-productruledecomp-1}\\
    &= \textstyle 
\E_P\left[\E_P\left[     Y   \frac{\mathrm{d}}{\mathrm{d}\eps}\left\{W_\eps^*\right\}\Big\vert_{\eps = 0}\mid A=1,X \right]\right] 
+  
% \underbrace
{\frac{\mathrm{d}}{\mathrm{d}\eps} \left\{ \E_{P_\eps}\left[\E_{P_\eps}\left[     Y W^*\mid A=1,X \right]\right] \right\}\vert_{\eps = 0}}%^_{\frac{\intd}{\intd \eps}   \{  \fnl^{opt}(P_\eps) \}\vert_{\eps = 0}}
% \label{eqn-sensitivity-productruledecomp-2}
\\
  &= \textstyle 
\underbrace{\textstyle\E_P\left[\frac{\mathrm{d}}{\mathrm{d}\eps}\left\{\E_P\left[     Y   W_\eps^*\mid A=1,X \right]\right\}\Big\vert_{\eps = 0}\;\right] }_{T_1}
+  
\underbrace{\textstyle
\frac{\mathrm{d}}{\mathrm{d}\eps} \left\{ \E_{P_\eps}\left[\E_{P_\eps}\left[     Y W^*\mid A=1,X \right]\right] \right\}\vert_{\eps = 0}
}_{
T_2
}.
% \label{eqn-sensitivity-productruledecomp-2}
\end{align}
}
\new{
$T_2$ is given by \Cref{thm-analytical-optif}.
We further study $T_1$. By the saddle point characterization, 
$$
W_\eps(x,y) = \alpha(x) \indic{Y - \lambda_\eps(x)} + \beta(x) \indic{Y - \lambda_\eps(x)},
$$
so that 
\begin{align}
    \frac{\mathrm{d}}{\mathrm{d}\eps} \left\{ \E_{P}\left[\E_P\left[     Y   W_\eps^*\mid A=1,X \right]\right] \right\}\vert_{\eps = 0} &= 
    \frac{\mathrm{d}}{\mathrm{d}\eps}
\left\{  \E_{P}\left[\E_P\left[ \alpha(x)(\lambda_\eps -Y)_- + \beta(x) (Y -\lambda_\eps)_+ \mid A=1,X \right]\right]  \right\}
    \vert_{\eps = 0} \nonumber
    \\ 
    & = \frac{\mathrm{d}}{\mathrm{d}\eps}
\left\{  \alpha \mu(s, a)+(1-\alpha) \frac{1}{\tau} \int_{\lambda_\eps(x)}^\infty y p(y\mid A=1,x) \intd y \right\}
    \Big\vert_{\eps = 0}, \label{eqn-perturbedW-leibniz} 
\end{align}
where the second equality follows because for any threshold $t,$ since $\mathbb{E}[Y \mid X]=\mathbb{E}\left[Y \mathbb{I}\left(Y>t\right) \mid X\right]+\mathbb{E}\left[Y \mathbb{I}\left(Y \leq t\right) \mid X\right],$ and $\tau(\beta-\alpha)= 1-\alpha,$ we have that $
\E_P\left[ \alpha(x)(\lambda_\eps -Y)_- + \beta(x) (Y -\lambda_\eps)_+ \mid A=1,X \right]
= \alpha \mu(s, a)+(1-\alpha) \frac{1}{\tau} \int_{\lambda_\eps(x)}^\infty y p(y\mid A=1,x) \intd y.$
}
\new{
Omitting functional dependence of the quantile on $x$ for brevity, the Leibniz integral rule gives that 
\begin{equation}{\frac{\mathrm{d}}{\mathrm{d}\eps} 
\left\{\int_{\lambda_\eps(x)}^\infty y p(y\mid A=1,x) \intd y\right\}     \Big\vert_{\eps = 0}
}
  = - \lambda \cdot p(\lambda\mid A=1,x) \cdot \frac{d}{d \eps } \lambda_\eps(x). \label{eqn-quantile-if}
  \end{equation}
  Therefore, putting the above together with the quantile influence function of \Cref{lemma-quantile-if}, perturbing in the direction of $O_i = (X_i,A_i,Y_i)$, we obtain that: 
  \begin{align*}
   &  \frac{\mathrm{d}}{\mathrm{d}\eps}
\left\{  \E_P\left[ \alpha(x)(\lambda_\eps -Y)_- + \beta(x) (Y -\lambda_\eps)_+ \mid A=1,X \right] \right\}\Big\vert_{\eps=0} \\
&\qquad = - \lambda \cdot p(\lambda\mid A=1,X_i)  \cdot \frac{(1-\alpha)\indic{A_i=1}}{\tau} \cdot- \frac{ (\indic{Y_i\geq \lambda} - \tau)}{p(\lambda\mid A=1,X_i)}\end{align*}
The result follows by simplifying and combining with \Cref{thm-analytical-optif}.
}
\end{proof}
% \if\forjournal 0 
% \begin{proof}[Proof of \Cref{prop-sensanalysis-finitediff}]
% \else
\begin{proof}{Proof of \Cref{prop-sensanalysis-finitediff}.}
% \fi 
\new{
We first simplify the finite-difference in the integrand. First note that by adding and subtracting $\E_P[YW^*_\eps\mid A=1,X],$ we obtain:
\begin{align}
&\E_{P_\eps
  }[Y  W_\eps^* \mid A=1,X] - \E_{P
  }[Y  W \mid A=1,X] \\
  & = \left(\E_{P_\eps}-\E_P)[Y W_\eps^*\mid A=1,X] + \E_P[Y(W_\eps^*-W^*)\mid A=1,X] \right). 
  \label{eqn-sensanalysis-interiordecomp}
\end{align}
% Consider the decomposition, where for brevity we denote the conditional distribution $\E_{P \mid A=1,X},\E_{P_\eps \mid A=1,X}$ 
We next show that
\begin{align*}
  &  \E_{P_\eps}[\E_{P_\eps
  }[Y  W_\eps^* \mid A=1,X]] - \E_{P}[\E_{P
  }[Y  W \mid A=1,X]] \\
  &= (\E_{P_\eps}-\E_P) \{ \E_{P_\eps}[YW\mid A=1,X] \}\\
  &\qquad + \E_P\{ \left(\E_{P_\eps}-\E_P)[Y W_\eps^*\mid A=1,X] + \E_P[Y(W_\eps^*-W^*)\mid A=1,X] \right)  \} + O(\eps)
\end{align*}
via the following expansions: 
\begin{align}
    &\E_{P_\eps}[\E_{P_\eps
  }[Y  W_\eps^* \mid A=1,X]] - \E_{P}[\E_{P
  }[Y  W \mid A=1,X]]  
 \nonumber \\
  &=(\E_{P_\eps}-\E_P) \{ \E_{P_\eps}[YW_\eps^*\mid A=1,X] \}+ \E_P\{ \E_{P_\eps
  }[Y  W_\eps^* \mid A=1,X] - \E_{P
  }[Y  W \mid A=1,X] \} \label{eqn-sensanalysis-fd-decomp-1} \\ %&& ( by \pm )
    &=(\E_{P_\eps}-\E_P) \{ \E_{P_\eps}[YW\mid A=1,X] \}
    + (\E_{P_\eps}-\E_P) \{ \E_{P_\eps}[Y(W_\eps^*-W)\mid A=1,X] \} \nonumber
    \\ &\qquad+ \E_P\{ \E_{P_\eps
  }[Y  W_\eps^* \mid A=1,X] - \E_{P
  }[Y  W \mid A=1,X] \}  \label{eqn-sensanalysis-fd-decomp-2}\\
      &=(\E_{P_\eps}-\E_P) \{ \E_{P}[YW\mid A=1,X] \}
    + (\E_{P_\eps}-\E_P) \{ \E_{P}[Y(W_\eps^*-W)\mid A=1,X] \} \nonumber 
    \\ &\qquad+ \E_P\{ \E_{P_\eps
  }[Y  W_\eps^* \mid A=1,X] - \E_{P
  }[Y  W \mid A=1,X] \} \nonumber
 \\
& \qquad + (\E_{P_\eps}-\E_P) \{ (\E_{P_\eps}-\E_P)[YW\mid A=1,X] \}
 + (\E_{P_\eps}-\E_P) \{ (\E_{P_\eps} - \E_P)[Y(W_\eps^*-W)\mid A=1,X] \}  \label{eqn-sensanalysis-fd-decomp-3}\\
  &=(\E_{P_\eps}-\E_P) \{ \E_{P}[YW\mid A=1,X] \}+ \E_P\{ \E_{P_\eps
  }[Y  W_\eps^* \mid A=1,X] - \E_{P
  }[Y  W \mid A=1,X] \} + O(\eps) \label{eqn-sensanalysis-fd-decomp-4} \\
  &=\underbrace{(\E_{P_\eps}-\E_P) \{ \E_{P}[YW\mid A=1,X] \}+ \E_P\{ (\E_{P_\eps}-\E_P)[Y W_\eps^*\mid A=1,X]\} }_{T_{1}} \\ &\qquad+ 
  \underbrace{\E_P[\E_P[Y(W_\eps^*-W^*)\mid A=1,X]}_{T_{2}} + O(\eps), \label{eqn-sensanalysis-fd-decomp-5}
\end{align}
where in \cref{eqn-sensanalysis-fd-decomp-1}, we added and subtracted $\E_P\E_{P_\eps}[YW_\eps\mid A=1,X]]$, in \cref{eqn-sensanalysis-fd-decomp-2}, we added and subtracted $(\E_{P_\eps}-\E_P) \{ \E_{P_\eps}[YW\mid A=1,X] \}$, in \cref{eqn-sensanalysis-fd-decomp-3}, $(\E_{P_\eps}-\E_P)\left\{ \E_P[Y(W^*_\eps - W^*)\mid A=1,X]\right\}$, in \cref{eqn-sensanalysis-fd-decomp-4}, we simplify away higher-order terms, and in \cref{eqn-sensanalysis-fd-decomp-5}, we apply \cref{eqn-sensanalysis-interiordecomp}.
}
\new{
Term $T_1$ can be studied by the same analysis as previously for the ATE, while term $T_2,$ $ \E_P[Y(W_\eps^*-W^*)\mid A=1,X],$ is a finite-difference analogue of the additional perturbed solution term in analysis of \Cref{proposition-if-closedform}. The analysis of term $T_{2}$ follows by applying the finite-difference approximation characterization of \Cref{lemma-leibniz-fd} (quantile approximation error via implicit discrete differentiation) to the analysis of \Cref{proposition-if-closedform}:
\begin{align}
 \Delta\left\{ \E_{P}\left[\E_P\left[     Y   W_\eps^*\mid A=1,X \right]\right] \right\}
    & =  \Delta
\left\{  \alpha \mu(s, a)+(1-\alpha) \frac{1}{\tau} \int_{\lambda_\eps(x)}^\infty y p(y\mid A=1,x) \intd y \right\}
   \label{eqn-perturbedW-leibniz-fd} \\
   & = - (1-\alpha) \frac{1}{\tau} \lambda \cdot p(\lambda\mid A=1,x) \cdot  \frac{ (\indic{Y_i \geq \lambda} - \tau)}{p(\lambda\mid A=1,{X_i})} + O(\eps) \label{eqn-perturbed-cvar-fd}
\end{align}
}
% {\frac{\mathrm{d}}{\mathrm{d}\eps} 
% \left\{\int_{\lambda_\eps(x)}^\infty y p(y\mid A=1,x) dy\right\}     \Big\vert_{\eps = 0}
% }
\new{
Here, \cref{eqn-perturbedW-leibniz-fd} follows from \cref{eqn-perturbedW-leibniz} and \cref{eqn-perturbed-cvar-fd} follows from \Cref{lemma-leibniz-fd} (quantile and CVaR influence function and finite-difference approximation error characterizations). 
}
% \az{check sign} 

% We can possibly make this cleaner with the following: 

% \begin{align*}
%     &\E_{P_\eps}[\E_{P_\eps^{Y\mid 1,X}
%   }[Y  W_\eps^* ]] - \E_{P}[\E_{P^{Y\mid 1,X}}[Y  W ]] \\
%   &(\E_{P_\eps}-\E_P) \{ \E_{P_\eps^{Y\mid 1,X}} [YW] \}+ \E_P\{ \E_{P_\eps^{Y\mid 1,X}}[Y  W_\eps^* ] - \E_{P^{Y\mid 1,X}}
%   [Y  W ] \} 
% \end{align*}
\end{proof}

\subsubsection{Proofs of intermediate results.}
% \begin{proof}[Proof of \Cref{lemma-perturbedoptimalsolution}]
% The result follows because $W^*(x,y)$ is the solution to a pointwise covariate-conditional optimization program. Because $p_\eps(y\mid a=1,x)$ evaluates to $\frac{(1-\eps)p(y,1,x)}{(1-\eps)p(1,x)}$ for $x \neq X_i.$, then
% \begin{align*} W^*_\eps(x,y)& \in \arg\max_{W} \left\{ 
%   \E_{P}[Y  W \mid A=1,X=x] \colon \E_{P}[ W \mid A=1,X=x] = 1, \forall x; \;\; \alpha(x) \leq W(x,y)\leq  \beta (x), \forall  y
%     \right\} \\
%      & = W^*(x,y).\end{align*} 
     
% \end{proof}
% \if\forjournal 0 
% \begin{proof}[Proof of \Cref{lemma-quantile-if}]
% \else
\begin{proof}{Proof of \Cref{lemma-quantile-if}.}
% \fi 
\new{
The quantile function solves $\tau = \int_{q_\tau(P)}^\infty dP(x).$ Under the perturbed distribution $P_\eps$, we have that $\tau =\int_{q_\tau(P_\eps)}^\infty dP_\eps(x).$ Therefore, differentiating both sides of the equation: 
\begin{align*} 
0 &= \frac{d}{d\eps} \left\{  \int_{q_\tau(P_\eps)}^\infty dP_\eps(x)\right\} \Big\vert_{\eps=0} \\ 
0 & \textstyle = -\phi(y; q_\tau, P) \cdot p(q_\tau) + \int_{q_\tau}^\infty (\delta(x) - P(x)) 
% \\
% &
=-\phi(y; q_\tau, P) \cdot p(q_\tau) + (\indic{y \geq q_\tau} - \tau).
\end{align*} 
Therefore $$\phi(y; q_\tau, P) = \frac{ (\indic{y \geq q_\tau} - \tau)}{p(q_\tau)}.$$
}
\end{proof}

% \if\forjournal 0 
% \begin{proof}[Proof of \Cref{prop-sensitivityanalysis-quantile-fd}]
% \else
\begin{proof}{Proof of \Cref{prop-sensitivityanalysis-quantile-fd}.}
% \fi 
\new{
Following the finite-difference analogue of the typical implicit differentiation argument: 
    \begin{align} 
% 0 & \textstyle = \Delta \left\{  \int_{q_\tau(P_\eps)}^\infty dP_\eps(x)\right\}  \nonumber \\ 
0 &\textstyle= \int_{q_\tau(P_\eps)}^\infty dP_\eps(x) - \int_{q_\tau(P)}^\infty dP(x) \nonumber \\
0 &\textstyle= \int_{q_\tau(P_\eps)}^\infty dP_\eps(x) - \int_{q_\tau(P)}^\infty dP_\eps(x)+ \int_{q_\tau(P)}^\infty dP_\eps(x) - \int_{q_\tau(P)}^\infty dP(x) \nonumber \\
0 &\textstyle= { -( q_\tau(P) - q_\tau(P_\eps))}{}  p_\eps(\lambda \mid A=1,x)  + \eps\int_{q_\tau}^\infty (\delta(x) - dP(x)) + O(\eps^2) \label{eqn-apx-leibnizrule}  
\end{align}
In \Cref{eqn-apx-leibnizrule}, we approximate the integral with a left Riemann sum, which incurs approximation error on the order of $\eps^2,$ since the approximation error of a left Riemann sum of $\int_a^b f(x) \intd x$ generically is $\frac{(b- a)^2 M}{2}$ where $M$ is an upper bound on the function derivative value on the interval. 
Therefore, dividing by $\eps$:
$$
\textstyle= \frac{ -( q_\tau(P) - q_\tau(P_\eps))}{\eps} \cdot  p_\eps(q_\tau \mid A=1,x)  + (\indic{y \geq q_\tau} - \tau) + O(\eps). \nonumber 
$$
Hence, $$\frac{ (q_\tau(P_\eps) - q_\tau(P))}{\eps} = \frac{(\indic{y \geq q_\tau} - \tau)}{ p_\eps(q_\tau \mid A=1,x) } + O(\eps). $$
% \az{check sign}
}
\end{proof}
% \if\forjournal 0 
% \begin{proof}[Proof of \Cref{lemma-leibniz-fd}]
% \else
\begin{proof}{Proof of \Cref{lemma-leibniz-fd}.}
% \fi 
\new{
We have:
\begin{align} 
\MoveEqLeft{ \textstyle \eps^{-1} \left( \int_{\lambda_\eps(x)}^\infty y p(y\mid A=1,x) \intd y 
- 
\int_{\lambda}^\infty y p(y\mid A=1,x) \intd y
\right)} %\nonumber
% \\
% &
=  \eps^{-1}\int_{\lambda_\eps}^\lambda y p(y\mid A=1,x) \intd y\nonumber \\
&= \frac{ -(\lambda_\eps - \lambda)}{\eps} \lambda p(y\mid A=1,x) + O(\eps^2) \label{eqn-quantilefd-leftriemann} \\ 
& =\frac{d}{d\eps} \{ \lambda_\eps \}\mid_{\eps=0} \cdot  \lambda p(y\mid A=1,x) + O(\eps). \label{eqn-quantilefd}
\end{align}
In \cref{eqn-quantilefd-leftriemann} we apply similar left Riemannian sum arguments as in \Cref{eqn-apx-leibnizrule}.
In the last line we apply \Cref{prop-sensitivityanalysis-quantile-fd}.
}
    
\end{proof}
% \if\forjournal 0 
% \begin{proof}[Proof of \Cref{lemma-sensitivityanalysis-whichquantile}]
% \else
\begin{proof}{Proof of \Cref{lemma-sensitivityanalysis-whichquantile}.}
% \fi 
\new{Recalling that
\begin{equation}\sup \left\{ 
  \\E[Y  W \mid A=1,X=x ] \colon \E[ W \mid A=1,X=x] = 1; \;\; \alpha(x)  \leq W(x,y)\leq  \beta(x), \forall  y
    \right\}, \label{eqn-sens-primalprogram-apx}
    \end{equation}
    we reparametrize with respect to the affine transformation $Z$ such that $W(x,y) = Z(x,y) + \alpha(x).$ Then the above equals:
     $$\alpha(x) \E[Y\mid A=1,X] + \sup_{Z\geq 0} \left\{ 
  \E_{P(Y\mid 1,x)}[Y  Z] \colon \E_{P(Y\mid 1,x)}[ Z] = 1-\alpha(x); \;\ Z(x,y)\in [0,\beta(x)-\alpha(x)], \forall  y
    \right\}.$$
    We reformulate this in terms of standard representations of conditional value at risk~\citep[see, e.g.,][ex. 6.19]{shapiro2021lectures} by adjusting some constant factors; finally, homogeneity of coherent risk measures will allow for the final identification. Namely, it is a standard result that the optimal solution of the dual problem 
  $\min_{\lambda\in\mathbb{R}} \{ \alpha^{-1} \E[(Z-\lambda)_+] + \lambda $
    is the interval with endpoints the left- and right-hand side $(1-\alpha)$ quantiles of the c.d.f. $H_Z(t) = Pr(Z\leq t).$
    Next, taking the Lagrangian, we obtain:
    $$ =\alpha(x) \E_{P(Y\mid 1,x)}[Y] + \inf_{\lambda \in \mathbb{R}}\sup_{Z\in [0,\beta(x)-\alpha(x)]} \left\{ 
  \E_{P(Y\mid 1,x)}[Y  Z]+(\lambda\left((1-e_1(x) a(x)) - \E_{P(Y\mid 1,x)}[ Z]\right)
    \right\}.$$
    We further simplify the optimization on the right hand side: 
    \begin{align*}
     & = \inf_{\lambda \in \mathbb{R}} \left\{ 
      (\beta(x)-\alpha(x))
  \E_{P(Y\mid 1,x)}[(Y-\lambda)_+]+\lambda\left((1-\alpha(x)) \right)
    \right\}  \\
    &=(1-\alpha(x))^{-1}\left[ \inf_{\lambda \in \mathbb{R}} \left\{ 
     { (\beta(x)-\alpha(x))}
  \E_{P(Y\mid 1,x)}[(1-\alpha(x))(Y-\lambda)_+]+\lambda\left((1-\alpha(x)) \right)
    \right\}  \right]\\
       &= (1-\alpha(x))^{-1} \inf_{\lambda' \in \mathbb{R}} \left\{ 
     { (\beta(x)-\alpha(x))}
  \E_{P(Y\mid 1,x)}[((1-\alpha(x))Y-\lambda')_+]+\lambda'\right\} \qquad \text{ (reparametrize } \lambda' = (1-\alpha(x))) \\ 
  &= (1-\alpha(x)) \,
\mathrm{CVaR}_{\kappa(x)}(Y \mid A=1,x),
\qquad
\kappa(x)=\frac{1-\alpha(x)}{\beta(x)-\alpha(x)}.
  % & = (1-\alpha(x))^{-1} \op{CVaR}_{(\beta(x)-\alpha(x))^{-1}}(((1-\alpha(x))Y\mid A=1,x) \\
  % & = \op{CVaR}_{(\beta(x)-\alpha(x))^{-1}}(Y\mid A=1,x).
    \end{align*} 
    In the third line we reparametrize the optimization in braces over $\lambda' = (1-\alpha(x))\lambda$ so long as $(1-\alpha(x))$ is nonnegative. 
    In the second to last line we use the reformulation of CVaR, which also allows us to recognize that $\lambda^*(x)$ is a conditional quantile of 
$Y \mid (A=1,X=x)$ at level
$\tau(x)
= 1 - \frac{1-\alpha(x)}{\beta(x)-\alpha(x)}
= \frac{\beta(x)-1}{\beta(x)-\alpha(x)}$. In the last line we use the homogeneity property of coherent risk measures. 
    % Hence \Cref{eqn-sens-primalprogram-apx} is equal to 
    % $$ \alpha(x) \E_{P(Y\mid 1,x)}[Y] 
    % + \op{CVaR}_{(\beta(x)-\alpha(x))^{-1}}(Y\mid A=1,x).
    % $$
    }
\end{proof}

\subsubsection{Proofs of results} 
% \if\forjournal 0 
% \begin{proof}[Proof of \Cref{thm-analytical-optif}]
% \else
\begin{proof}{Proof of \Cref{thm-analytical-optif}.}
% \fi 
\new{
For notational clarity we consider perturbations with respect to a Dirac delta function (unsmoothed). We define the integrand and its perturbation:
\begin{align*}
    g(x) & = \sup_{W\colon \alpha(x) \leq W(x,y)\leq  \beta(x)} \left\{ 
  \E_P[Y  W \mid A=1,X=x] \colon \E_P[ W \mid A=1,X=x] = 1
    \right\}\\
        \td{g}_\eps(x) &= \sup_{W\colon \alpha(x) \leq W(x,y)\leq  \beta(x)} \left\{ 
  \E_{P_\eps
%   (1-\eps)P+\eps\delta{(\tilde{o})}
  }[Y  W \mid A=1,X=x] \colon \E_{
  P_\eps
%   (1-\eps)P+\eps\delta{(\tilde{o})}
  }[ W \mid A=1,X=x] = 1
    \right\}.
\end{align*}
Interchanging the marginalization (integral) and the derivative, and applying the product rule: 
$$ \textstyle 
 \frac{\intd}{\intd \eps} \E_{P_\eps}[\td{g}_\eps(x)] = 
\int (\{  \frac{\intd}{\intd \eps} \td{g}_\eps(x) \vert_{\eps = 0} \}  p(x) + g(x)
\{ \frac{\intd}{\intd \eps}  p_\eps(x)\} )\intd x. $$
For the second term, we can apply our previous characterization and note that 
\begin{align}
  \textstyle \int    g(x) \{ \frac{\intd}{\intd \eps}  p_\eps(x)\} )\intd x = \int g(x) (\indic{X_i} - p(x)) \intd x = g(X_i) - \fnl(P). \label{eqn-sensanalyticalderiv-eq2}
\end{align}
}
% For the first term, \cite[6.5.3]{shapiro2009lectures} gives some arguments for von Mises expansions of coherent risk measure functionals. We apply similar arguments conditionally, for the case where propensities are assumed to be known. 
% Passing to the Lagrangian, and using the saddle point property that the optimal dual variables are optimal with respect to the optimal primal variable $W^*$, we obtain that:
% \begin{align*}
%   g_\eps(x) = \inf_{\lambda \in \mathbb{R}} \{ \E_{P_\eps} [YW^*\mid 1,x] + \lambda( 1- \E_{P_\eps}[W^*\mid 1,x] \} 
% \end{align*}
\new{
Therefore, 
\begin{align*}
  \frac{d}{d\eps}g_\eps(x) &=    \frac{d}{d\eps}\left\{ \int  (Y-\lambda)W^* \frac{p_{\eps}(y, 1,x)}{p_{\eps}( 1, x)} \intd y + \lambda^* \right\} \\
  &=   \int (Y-\lambda)W^* \frac{d}{d\eps}\left\{   \frac{p_{\eps}(y, 1,x)}{p_{\eps}( 1, x)} \right\}\intd y   \\
  &= \int (Y-\lambda)W^*    \frac{(\indic{o}-p(y,1,x)) p(1,x) - (\indic{A=1,x}-p(1,x))p(y,1,x) }{p( 1, x)^2} \intd y \\
  &= \indic{A=1}
  \left( 
  \frac{(Y_i-\lambda)W^*(X_i,Y_i)}{p(A=1,X_i)} - \frac{\E_{P} [(Y-\lambda)W^*\mid A=1,x] }{p(A=1,x)}
  \right) , 
\end{align*}
interchanging differentiation and integration. 
}
\new{
Therefore, 
$$ \int (\{  \frac{\intd}{\intd \eps} g_\eps(x) \vert_{\eps = 0} \}  p(x) =
\frac{\indic{A=1}
\left( (Y_i-\lambda)W^*(X_i,Y_i) - \E_{P} [(Y-\lambda)W^*\mid A=1,X=X_i]
\right) 
}{p(A=1\mid X_i)}, $$
and the claim follows from combining the above with \cref{eqn-sensanalyticalderiv-eq2}.
}

\end{proof}

\section{Additional discussion}\label{apx-practicalconsiderations}

\subsection{Additional related work on numerical differentiation}\label{apx-relwork-numdiff}
\paragraph{Numerical derivatives, statistical machine learning, and optimization.} Independently of causal inference, other works study the use of numerical gradients in the context of machine learning and stochastic gradients \citep{oktay2020randomized}, and statistical estimation and inference \citep{hong2015extremum,hong2018numerical}. Analysis of finite-difference estimators has been of interest in (stochastic) zeroth-order optimization and derivative-free optimization \citep{flaxman2004online,duchi2015optimal}.   %We defer detailed discussion to the appendix. %We focus on data-driven versions of finite-difference evaluations of Gateaux derivatives with smoothing; and on applications to optimization-based functionals. 

\subsection{Discussion on pathwise differentiability}\label{apx-discussion-pathwisediff}
Assuming pathwise differentiability is a fundamental limitation of attempts to ``automate" semiparametrics because it is a technical condition that requires a similar amount of specialized knowledge to understand and check as to derive an influence function itself. 

% We can  provide an example of a broad class of problems where this assumption is satisfied in general%, even though analysts can specify arbitrary functionals within this class
% : strongly convex stochastic programs. We find it useful to explicitly contrast these two different types of structural assumptions, because a reasonably informed analyst could check the condition of strong convexity, while checking Fr\'echet vs. Gateaux differentiability can be more difficult. 
% Regarding strongly convex stochastic programs, these provide an example where asymptotic linearity is established for the class as a whole but analysts can specify arbitrary functionals within this class. 
In the main text, we referenced sufficient conditions that Fr\'echet differentiability of the functional can imply asymptotic linearity and that the remainder is second-order. \cite{kallus2019localized} study a different problem, also make a stronger assumption of Fr\'echet differentiability, and hence provide some concrete examples (such as estimating equations with incomplete data) where this is satisfied. Therefore, we think that previously articulated sufficient conditions of Fr\'echet differentiability can be made even more useful in the context of additional results that establish Fr\'echet differentiability of functionals, based on the composition of simpler functionals, akin to identification of convex structure via composition rules and structure of known atoms \citep{grant2008graph}. We think these are promising directions for future study, though beyond the scope of this paper.

This assumption can be particularly difficult to satisfy for optimization problems in general, which are generically directionally Hadamard differentiable. We assume asymptotic linearity, which can arise from studying specific functionals like the infinite-horizon MDP (see \citet{kallus2022efficiently}) or the CVaR functional (see \citet{jeong2020robust}). \citet[Prop. 1]{duchi2021asymptotic} establish asymptotic linearity under primitive conditions of strong convexity (as they mention, these results are known in the literature on perturbation analysis of stochastic programs; see \cite{bonnans1998optimization}).

\subsection{Kernel density estimation practical considerations} 

\paragraph{Monte Carlo sampling for uniform kernel for smoothed perturbation.}

For computational purposes, it is helpful to evaluate the final integral over a uniform kernel in $x$ by Monte Carlo sampling. 
 \if\forjournal 0
\footnote{In computational experiments not reported here, we found scipy.integrate.quad to be quite sensitive in evaluating the smoothed perturbation. We leave more extensive numerical evaluation for further work.} 
\fi
In particular, for perturbations in the direction of $(x_i,a_i,y_i)$, we evaluate the integral using $N_{MC}$ Monte Carlo samples: \begin{align}\textstyle \hat{\phi}^{MC}_{\epsilon,\lambda}(o_i) = \frac{1}{N_{MC}} \sum_k \frac{\left((1-\epsilon) \left(\sum_{j:A_j=1} K(X_j-\tilde{x}_k)Y_j\right)P(A=1)+\epsilon y_i  \cdot\indic{a_i=1} \cdot 1 \right) }{(1-\epsilon)p(A=1,\tilde{x}_k)+\epsilon \mathbb{I}[a_i=1] } \nonumber \\
\textstyle+
(1-\epsilon) \frac{1}{N_{MC}} \sum_k
 \frac{\tilde{p}(\tilde{x}_k) }{   \tilde{p}_\epsilon (A=1,\tilde{x}_k) } \mathbb{I}({a_i = 1})\left\{
  y_i - 
  \mathbb{E}_{\tilde P}[Y\mid A=1,\tilde{x}_k]
        \right\}. \label{eq-mcunifkernel}
        \end{align} 
\paragraph{Bandwidth for kernel density estimation of functionals.}
\cite{bickel2003nonparametric} discuss optimal bandwidth selection for estimation of functionals via plug-in of nonparametric density estimates. They show that asymptotic linearity/root-n consistency of $\hat \beta$ is possible under the optimal bandwidth for density estimation when the number of derivatives of the influence function is more than half the dimension of observation $o$; requiring a higher order kernel. (See \cite{ichimura2015influence} for additional discussion). 
\paragraph{Other density estimation approaches.}
 Although we have discussed estimation of the components of $\tP$ by nonparametric Nadaraya-Watson/kernel-density estimation for simplicity, other estimates of probability densities or generative models can be similarly used; we describe a few below. These approaches may work well empirically; we leave a thorough evaluation for future work. 

\begin{example}[Bayesian Dirichlet process mixture models]
\cite{hannah2011dirichlet} develop Dirichlet process mixtures of generalized linear models with covariate density given by a mixture of exponential-family
distributions, and conditional response density modeled via a generalized linear model (with a set of parameters for each mixture component). This specification of the probability densities implies a regression estimate. 
\end{example}
\begin{example}[Generative modeling with Gaussian mixture models]
\cite{jesson2021quantifying} use generative modeling with Gaussian mixture networks, and sampling from the generative model, to approximate the integral in order to scale-up to high-dimensional settings. %Similar approaches may be used to compute the density estimates in our setting.  
\end{example}

\section{Details on empirics}
    % return (-5*A*X+ (2*A-1)*3)*(X<0.25)+ (5*A*X+3)*(X>0.25)*(X<0.5) + -5*A*X*(X>0.5)*(X<0.75)+ 5*A*X*(X>0.75)*(X<1)
    \subsection{ATE}
    We consider a piecewise linear outcome mean specification which is hard for nonparametric estimation. 
\begin{align*}\E[Y\mid A,X] &= \indic{X<0.25}\cdot (-5AX + 3(2A-1)) + \indic{X\in[0.25,0.5]} \cdot (5AX+3) \\& \qquad + \indic{X\in[0.5,0.75]} \cdot -5AX +  \indic{X\in[0.75,1]} \cdot 5AX
\end{align*}
The propensity score is drawn as $A\sim \op{Bern}(\sin(20X)+0.5).$

\subsection{Sensitivity analysis}\label{apx-experimental-details}

\paragraph{One-dimensional example - data-generating process.}
For the \textit{data-generating process}, we generate a one-dimensional covariate $X \sim \mathrm{Unif}[-2,2]$ and transform it to $\tilde x \in [0,1]$ via
$\tilde x \;=\; \mathrm{clip}\!\left(2\big(\mathrm{expit}(0.2X)-\tfrac12\big),\,0,\,1\right),
$ for use in both the propensity score and the outcome model.
For the \textit{data-generating process}, treatment is assigned according to a logistic propensity score $e_1(X)=\mathrm{expit}(-1.4 + 2.8\,\tilde x)$.
For $a \in \{0,1\}$, outcomes are conditionally Gaussian where $\{Y\mid X,A=a\} \sim \mathcal{N}(\mu_a(\tilde x),\sigma_a^2(\tilde x))$, with
\begin{align*}
    \mu_a(\tilde x)
&= (1 + a) + (0.5 + a)\,\tilde x + (0.3 + 0.2a)\,\sin(3\pi \tilde x)\\
\sigma_a(\tilde x)
&= (0.8 - 0.3a) + (1.5 + a)\,\tilde x.
\end{align*}

\paragraph{Moderate-dimensional example - data-generating process.}
In the high-dimensional setting, we take $X \sim \mathcal{N}(0,I_{10})$, transform each coordinate to $\tilde X_j \in [0,1]$ via the same bounded logistic map as above, form the nonlinear single index $\tilde x = 10^{-1}\sum_{j=1}^{10} \tilde X_j$, assign treatment using the logistic propensity $e_1(X) = \mathrm{expit}(-1.4 + 2.8\,\tilde x)$, and draw $\{Y\mid X,A=a \}\sim \mathcal{N}(\mu_a(\tilde x),\sigma_a^2(\tilde x))$ with $\mu_a$ and $\sigma_a$ as in the one-dimensional DGP.

\paragraph{Details on computation}
Experiments were conducted on a MacBook Pro with 16g RAM.

\end{document}